\documentclass[12pt,a4paper]{article}
\usepackage[utf8]{inputenc}
\usepackage[left=2.75cm,right=2.75cm,top=3cm,bottom=3cm]{geometry}
\pdfoutput=1

\usepackage{amsmath}
\usepackage{amsfonts}
\usepackage{amssymb}
\usepackage{amsthm}
\usepackage{bbm}
\usepackage{bm}
\usepackage{mathtools}
\usepackage{graphicx}
\usepackage{caption}
\usepackage[flushleft]{threeparttable}
\usepackage[dvipsnames]{xcolor}
\usepackage{booktabs}
\usepackage{rotating}
\usepackage{eurosym}
	\DeclareUnicodeCharacter{20AC}{\euro}
\usepackage{multirow}
\usepackage{longtable}
\usepackage{setspace}
\usepackage{relsize}
\usepackage{graphicx,subcaption}
\usepackage{pdflscape} 
\usepackage{xurl}
\usepackage{ragged2e}
\usepackage{float}
\usepackage{hyperref}
	\hypersetup{colorlinks, citecolor=BrickRed, anchorcolor=Black, urlcolor=BrickRed, linkcolor=BrickRed}
\usepackage[nottoc]{tocbibind}
\usepackage{comment}
\usepackage[super]{nth}

\usepackage[backend=biber, style=apa, bibstyle=authoryear, dashed=false, maxcitenames=2, maxbibnames=6, giveninits=true, sorting=nyt, date=year, eprint = false, url=false, doi=false, isbn=false]{biblatex} 
\newcommand\fnote[1]{\captionsetup{font=small}\caption*{#1}}

\title{From Rules to Regs:\\
A Structural Topic Model of Collusion  
Research\footnote{I am grateful for valuable feedback from Oliver Budzinski, Kai Fischer, Jerg Gutmann, Justus Haucap, Paul H\"unermund, Anselm K\"usters, Nima Moshgbar, Hans-Theo Normann, and Chris Snyder as well as feedback from participants of the DICE PhD research workshop, the NOUS workshop at Albert-Ludwigs-University Freiburg, 
 and the \nth{56} Hohenheim workshop. 
All errors are my own. 
Funding by the German Research Foundation (DFG) is gratefully acknowledged (Funding No.~\#235577387/GRK 1974). 
Replication code will be available on github soon.
}}

\author{W. Benedikt Schmal \thanks{Düsseldorf Institute for Competition Economics (DICE) at Heinrich Heine University. Universitätsstraße 1, 40225 Düsseldorf, Germany. Email: \href{mailto:schmal@dice.hhu.de}{schmal@dice.hhu.de}}}

\date{\today}
\begin{document}
\begin{onehalfspace}
\maketitle
\thispagestyle{empty}

\vspace{-5mm}

\begin{center}
\textbf{\textit{Preliminary draft}}\\

\vspace{5mm}

\end{center}


\noindent 

Collusive practices of firms continue to be a major threat to competition and consumer welfare. Academic research on this topic aims at understanding the economic drivers and behavioral patterns of cartels, among others, to guide competition authorities on how to tackle them. Utilizing topical machine learning techniques in the domain of natural language processing enables me to analyze the publications on this issue over more than 20 years in a novel way. Coming from a stylized oligopoly-game theory focus, researchers recently turned toward empirical case studies of bygone cartels. Uni- and multivariate time series analyses reveal that the latter did not supersede the former but filled a gap the decline in rule-based reasoning has left. Together with a tendency towards monocultures in topics covered and an endogenous constriction of the topic variety, the course of cartel research has changed notably: The variety of subjects included has grown, but the pluralism in economic questions addressed is in descent. It remains to be seen whether this will benefit or harm the cartel detection capabilities of authorities in the future.

\vspace{5mm}
\vspace{5mm}
\end{onehalfspace}
\newpage

\begin{onehalfspace}
\section{Introduction}

Currently, two issues are predominant in antitrust: Common ownership and rising mark-ups as well as the emergence of enormous digital platform conglomerates. 
In contrast, cartels and collusive behavior seem to have gone off the radar to some extent. Still, the economic harm caused by it is non-negligible. \textcite{Levenstein.2006} report, for example, an 80\% price increase caused by the international tea cartel, and monopoly pricing of the US beer and the British ocean shipping cartel. Meta studies find an average price overcharge of 30-50\% and a median overcharge of 15-25\% by cartels in general  \parencite{Bolotova.2009a, Boyer.2015, Connor.2008}. A proxy for the impact of collusion are the fines imposed: From 2000 - 2022, the European Union (EU) handed out penalties of some \$ 32 billion.\footnote{See \url{https://competition-policy.ec.europa.eu/document/download/b19175c3-c693-410b-b669-27d4360d359c_en?filename=cartels_cases_statistics.pdf} (up to July 2022).} Moreover, the EU plans to lose restrictions on anti-competitive behavior to promote the dissemination of environment-friendly technology \parencite{Schinkel.2021}. 
Thus, collusion continues to be a major issue of antitrust policy. 

This paper investigates research on collusion during the past two decades. It reveals important changes over time and hints at developments in the literature that are likely to affect competition authorities. Methodologically, it provides a novel way of understanding how collusion is addressed and identifies the centers of academic attention. This is based on latent topics computed with a structural topic model, a state-of-the-art natural language processing technique that relies on unsupervised machine learning algorithms. 
The paper utilizes methods to the best of my knowledge hitherto unused in the `economics of science' literature and provides a novel quantitative comprehension of the development of research on cartels for economists as well as antitrust scholars. As industrial organization research is a major source of inspiration for policymakers, legislators, and antitrust authorities, understanding the patterns of collusion research may have a direct impact on the evolution of related public policy.

The machine-learning-based topic modeling is the foundation of the analysis. I build upon nearly 800 publications in leading economics and antitrust journals addressing collusion. Among this arguably full body of research on this topic, I identify 21 latent topics by analyzing the content of the papers with a structural topic model. Based on that, I cluster these topics and assign them to categories. At first sight, a cluster based on methods (theory, empirics, experiments) might come to mind. However, topic modeling allows for more dimensions due to the multinomial probability computation as well as a more content-based analysis than a methodological categorization. 


Relying on this structure, the first part of the subsequent analysis studies the evolution of topic categories over time. In order to achieve this, I extract the predicted probability for every topic to occur in a specific paper. Aggregating topic probabilities by category, it becomes visible that rule-based approaches that study the behavior of firms, in general, are in retreat. Especially in recent years, empirical case studies of disclosed cartels and their court cases have seen a large boost. Furthermore, there exists some empirical evidence for further specialization, i.e., the focus of a paper seems to be narrowed to one or a few topics. 

The case study-related topics are especially correlated with the leading so-called `top 5' journals in economics. While other general interest and broader field outlets cover a variety of topics, specific IO journals also seem to narrow their focus to a smaller set of topics. This may end up in an endogenous process of streamlining research. 
Conducting a repeated cross-section regression analysis on the relationship between citations and topics, I do not find any significantly positive relation between the number of citations to specific topics. Moreover, I only find four topics to be negatively correlated with citations per year. Remarkably, these topics are in decline over time in terms of the predicted probability to occur in a paper. This content-wise decline is not matched by a subject-wise decrease: Applying the machine learning technique of neural networks, I can show that the superficial \emph{subjects} of papers become more diverse over time while the underlying actual considerations are more closely tied together.

In the last step, I conduct a multivariate time-series analysis based on a vector error correction and a vector autoregression model to understand how the decline in rule-based thinking and the upswing in case studies are intertemporally related. In fact, the case studies do not supersede game theoretic modeling but fill the gap the decline in the latter leaves. It seems to be the case that the discipline tends to have lost some interest in the stylized models of markets and rather turned toward backward-looking case studies. This serves as additional evidence for endogenous constriction of academic research on collusion, which may weaken the ability of competition authorities to sufficiently chase and disclose collusive structures in the future.



As this paper is a discourse analysis of industrial organization research, it contributes to the economic strand of the science of science literature that aims at understanding how the process of creating and disseminating scientific progress works. Furthermore, it is a reflection of the rules and habits of the discipline of economic research as well. Earlier research by \textcite{Einav.2010} has discussed the developments in empirical IO, while \textcite{Hovenkamp.2018, Hovenkamp.2021} has recently criticized the current state of antitrust research. \textcite{NBERw30326} find evidence for a decline in antitrust enforcement in the US, but hardly any evidence for antitrust research playing a role in that. \textcite{Shapiro.2020} raises the concern that antitrust economics has focused on technical aspects in markets but rather miss out on the broader economic issues affected by, e.g., higher prices or less innovation. \textcite{Jones.2020} discuss the problem of antitrust authorities dealing with the market power of digital platforms .

More broadly, there is an ongoing debate over the scope and method of economic research to which this paper shall contribute. \textcite{Akerlof.2020} warns that economics has focused to much on `hard' econometric analyses but omits `softer', i.e., less well quantifiable topics, which -- in his view -- leads to a `bias against new ideas' and overspecialization. \textcite{Zingales.2020} demands more intellectual diversity. In a similar direction goes the critique of \textcite{Colenander.2018, Johnson.2020}, who suggest applied economics to reduce econometric overthinking and to understand it more as an `art' in the sense of weaving in more economic reasoning and what they call `engineering.' This term is also used by \textcite{Banerjee.2002, Duflo.2017}, who also use the term `plumber' to suggest that economists should concentrate more on real world aspects and making models work there than just in a stylized theoretical environment. Closer related to economic publishing is the reproach of a `tyranny of the top 5' [journals]  \textcite{Heckman.2020} that arguably leads to a narrower focus of economics as academic discipline.

This paper aims at contributing a deeper understanding of research on cartels. It follows \textcite{Ambrosino.2018} who already clustered all economic research via topic modeling. They use the latent Dirichlet allocation (LDA) algorithm. It is easy to use but structural topic modeling (STM) allows for a more granular analysis with less restrictive model assumptions. 
Therefore, I choose the latter technique. \textcite{Blei.2007} analyze scientific research by looking at publications in the natural sciences journal \emph{Nature}. 
In general, natural language processing recently gained popularity in economic research: \textcite{Larsen.2021} uses it to understand uncertainty in markets, \textcite{Bandiera.2020} classify CEO behavior with it. Lots of well published research investigates central bank communication and its macroeconomic effects \parencite{Hansen.2016, Hansen.2018, Kusters.2022}. The nexus between science of science and economic questions using NLP is, for example, addressed in work on the change in topics in macroeconomics after the financial crisis \parencite{Levy.2022} or else a major amendment of competition law \parencite{Schmal.2022b}. Eventually, this methodological approach has already been reviewed in the \emph{Journal of Economic Literature} by \textcite{Gentzkow.2019}.

The remainder of the paper is structured as follows. Section \ref{sec.emp} describes the data and the methodology of the machine learning based structured topic modeling. Later on, it also describes the econometrics behind the analysis. Section \ref{sec.iad} sketches how the computed topics are clustered, which forms the cornerstone of the following quantitative study. Section \ref{sec.res} presents and discusses the paper's findings, Section \ref{sec.conc} concludes. 

\section{Empirical Model} 
\label{sec.emp}

\subsection{Data}


The cornerstone of this paper is the natural language analysis, which is based on the abstracts of papers on cartels and collusive behavior published in leading IO and antitrust outlets and its related metadata. While a full-text analysis would encompass more information, abstracts are focused on the core techniques and messages a paper applies and wants to communicate to the reader. Due to its brevity, an abstract usually is not decorated with illustrative examples that are not used for deeper analysis as often seen in introduction sections of full papers. This helps separate important from less important cases in the text analysis. \textcite{Sybrandt.2018} find that full texts may provide more information but in turn more `intruder' terms that obfuscate the topic modeling and subsequent interpretation.

I received the data from the publication database Scopus via the Python API \emph{pybliometrics} \parencite{Rose.2019}. In total, I gather 34,564 publications in 35 journals from 2000 up to 2021. To isolate papers that actually address collusion among firms, I select all those which use at least one of the operators `collusion', `collusive', `cartel', `bidding ring' or its declensions within its title, abstract, or keywords.\footnote{Using JEL codes to identify specific groups of contentwise related papers would be attractive, however, the Scopus database does not collect them. Moreover, antitrust journals do not necessarily use these codes. Keywords would be another starting point for structuring research, but latent topics go into much more detail and avoid the subjective self-assignment of keywords by the authors.} Doing that, the number of relevant papers collapses to 777.\footnote{I also technically exclude those papers without any  abstract listed in Scopus.} Table \ref{tab.papcount} in the appendix shows the distribution of papers across journals. 

The text data from the abstracts is preprocessed to ease the analysis. This includes the removal of punctuation, the transformation of all words into lowercase terms, and the removal of stopwords. These are words such as conjunctions or articles that have no real meaning but are necessary for humans to read and understand a text.\footnote{I use the widely applied list created and published by \textcite{Lewis.2004}. The list is publicly available at MIT's Computer Science and Artificial Intelligence Laboratory website: \url{http://www.ai.mit.edu/projects/jmlr/papers/volume5/lewis04a/a11-smart-stop-list/english.stop}} Additionally, I manually remove `boilerplate' words \parencite[i.e., repetitive words or code, see][]{Lammel.2003} that are technically part of the abstracts but do not contain any useful information, such as copyrights of the publishers. Furthermore, I remove the operators `cartel' and `collusion' as these terms by definition should be part of every abstract and, by that, add no information on underlying topics \emph{within} the subset of papers about cartels and collusion.\footnote{A list containing these additional stopwords can be found in Table \ref{tab.stopword} in the appendix.} 

Last, I stem the words. Hence, plural forms or the past tense for verbs disappear as only the stem of each word is used. For example, `cooperation', `cooperate', and `cooperative' collapse to `cooper'. This bears the loss of some information. However, it comes along with a huge gain in analytic power as otherwise every version of `cooper' (or any other word) would be considered by the algorithm as an independent word.

\begin{figure}[H] 
\begin{center}
\includegraphics[width=\linewidth]{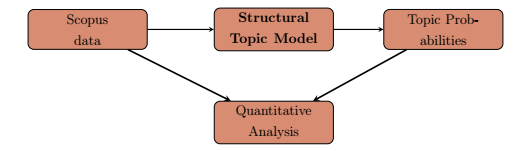}
\caption{Data used in the quantitative analysis}
\label{fig.datasketch}
\end{center}
\vspace{-5mm}
\end{figure}

While the publication records drawn from the Scopus database already exist, a core function of the structural topic model -- explained in detail later on -- is the computation not only of latent topics but also their probability for each publication in the dataset. This is a crucial component of the analysis later on.

\subsection{Structural Topic Modeling}

Based on this body of papers, I conduct natural language processing via text mining. Essentially, I work with the papers' abstracts to understand deeper relations and developments in collusion research over time. To do that, I apply the technique of topic modeling. This method was made popular by \textcite{Blei.2003}.\footnote{As noted by \textcite{Grimmer.2022}, the authors build upon, e.g., \textcite{Hofmann.1999}. The same methodology had been independently invented by \textcite{Pritchard.2000} as well.} The core idea is as follows: 
There exists a number of $D$ text elements (usually called documents, in my case paper abstracts) that contain $K$ so called topics, i.e., a cluster of words that belong together.\footnote{I use the letter $K$ for the choice of the optimal number of topics as done in the literature. However, later one, I label topics with the Greek letter $\tau_k$, which abbreviates topic $k \le K$.} Each document $d$ consists of a number of $W$ words. Consider as an example the `document' $d_1=$``Hello world'', that contains of $w_1=$ hello and $w_2=$ world and, say, one topic $k_1=$ greeting. 

The ultimate goal is the identification of a pre-defined number of $K$ latent topics without a prior which words belong to it. To do this, a data-generating process for each document is assumed in which every document consists of several topics (and of course several actual words), which themselves consist of a set of words, which have their own probability to belong to a particular topic. In the following, I sketch the mechanism of this unsupervised machine-learning algorithm based on the work of \textcite{Blei.2003} and \parencite{Roberts.2013}. Note that I mostly follow the notation of the latter even though I adjust parts to better address an economics readership.

In this paper, I apply one of the most recent advances of topic model, the \emph{structural} topic model \parencite[STM,][]{Roberts.2013}.\footnote{To apply the techniques, I make use of the R package `stm' developed by \textcite{Roberts.2019}.} The STM is particularly useful as it allows the incorporation of covariates as independent variables for the prevalence of topics, which in turn allows me to estimate regressions on relationships between specific topics and covariates $X_d$. Every document has a set of topics in it. The STM assumes that there is a individual prior for the particular set of topics for each document, which depends on a vector of covariates. Put differently, the covaratias are assumed to determine which topics to appear in a particular paper. Thinking of academic publications, a paper's outlet is likely to be related to the topics as the \emph{Journal of Economic Theory} may cover different topics than \emph{Empirical Economics}. 

Second, there exists a vector $\theta_d \sim logit \mathcal{N}(X_d\gamma_k, \sigma)$ for each document that contains the document-specific probability of each topic to occur there $\gamma_k$ is a vector of covariates for $X_d$, with $\gamma \sim \mathcal{N}(0,\sigma_k^2)$, such that we assume as a prior the document-covariates to be uncorrelated with the topics and only deviate from this assumption for a strong correlation between the two. Within each document, there exists a document-specific probability for each word to occur there. This depends on the overall distribution of words ($m$) and the covariates ($X_d$). It is captured by $\beta(X_d,m)$ and, hence, describes the probability of a particular word to occur in a particular document given this document contains some particular topic. $\beta$ is a matrix of size $K \times W$, i.e. it contains for every word ($w_1...w_N$) and every possible topic ($k_1...k_N$) the probability to occur. 

Third, based on $\theta_d$ we draw the topic ($\hat k_w$) that is most likely for every word $w$ in every document using a multinomial logit model, i.e., $\hat k_w \sim Multi(\theta_d)$. Last, we draw an actual word $\hat w$ that has the highest probability to occur given $\hat k_w$ and $\beta$, i.e. $\hat w = max\{p(w_n|\hat k_w,\beta)\}$ while we assume the actual words to be distributed $w_{d,n} \sim Multi(\beta_d^{k=\hat k_w}))$. One can easily see that the predicted word $\hat w$ should converge towards the actual word $w$. As the actual topics within each document are latent, structural topic modeling uses an expectation-maximization algorithm that iteratively tries to find a local maximum likelihood given the prevalence of latent topics. 

It is a crucial task to evaluate how many underlying topics $K$ exist in a body of documents. There is no `one size fits all' approach that provides a unique set of metrics to elicit the optimal $K^*$ \parencite[see, e.g.,][]{Grimmer.2013, Wehrheim.2019}. Nevertheless, there exist certain key measures that contain crucial information on the quality of the topics dependent on the number of topics. One key measure is semantic coherence that states that words that occur very frequent within topic $k_i$ should also occur in a document given this document contains $k_i$ \parencite{Mimno.2011}. While this is a convincing concept, it suffers from the problem that for low $K$ semantic coherence is high by construction. 

As an antagonistic metric, I use exclusivity as proposed by \textcite{Bischof.2012}. The core idea behind that is that words that are only in one topic very frequent are `exclusive' while words with relatively equal frequencies across topics are rather non-exclusive. Consider in the given context the word `collusion' that should occur in many topics. As this is such a big keyword -- it is part of nearly all abstracts by construction -- I exclude it from the documents. Words that occur in many topics should lead to a higher semantic coherence, because in this case it is more likely that the topics cover many words in the actual abstracts. In contrast, these topics should have a low exclusivity, such that the trade-off between these two measures should provide a sufficient choice of $K$. In \href{app.B}{Appendix B}, I sketch in more detail which measures I compute to come to the ideal choice of $K$. Overall, I come to the conclusion that $K^*=21$ is optimal for the purpose of this paper. 

As I use a structural topic model, it is possible to add covariates to the topic regressions, which in general are fitted as linear models with the expected topic prevalence as dependent variable. I include time measured in years to account for changes over time within the two decades covered and I add the journal as major covariate that should be related to a paper's particular content. To draw more general conclusions and increase statistical power, I group the  journals into five categories as shown in Table \ref{tab.papcount} in the appendix, because there exist clusters of journals that share similarities in scope and methods. These categories are: Top 5 journals, general interest (without top 5), microeconomic field journals, specific IO journals, and antitrust journals.\footnote{The so-called top 5 journals in economics are the American Economic Review, Econometrica, the Journal of Political Economy, the Quarterly Journal of Economics, and the Review of Economic Studies (listed in alphabetical order). These journals are considered leading by far and are usually considered as gatekeepers to academic jobs in economics as a publication in one of these five journals often is an ex- or implicit condition for appointments \parencite{Heckman.2020}.} While IO journal nearly exclusively focus on economic elaborations of (in ouzr case) collusion, antitrust journals also encompass legal and practitioners' perspectives, in particular competition law. Further variables could be added. However, given the structure of the model, only highly relevant covariates get assigned a value different from zero given the construction of the covariates with mean zero.   

While most parts of this paper rely on a structural topic model, I amend my analysis with a machine learning algorithm that relies on neural networks. Other than the statistical expectation-maximization approach taken in the STM setting, neural networks set up a plethora of nodes that are interrelated with each other, equivalent to connected neurons in the human brain. By that, a neural network based model draws relations between variables or topics from the relations between nodes and the particular weighting of these relations \parencite[see, e.g.,][]{Amari.1995}. In the present case of natural language processing, I use the \emph{doc2vec} neural network algorithm developed by \textcite{Le.2014}. The unsupervised algorithm autonomously learns relations between words in a text, in our case in academic publications, and translates the processed text into a lower-dimensional vector for each document that captures these relations. Based on that, I compute the inner product of these vectors for each pair of publications within a year to measure the similarity of these papers. I average over all similarities per paper and in a second step per year to compare changes over time.

\subsection{Quantitative Estimation}
\label{sec.emp.metrics}

 The main building block of the analysis are the predicted topic prevalences that are computed based on the structured topic model and its unsupervised machine learning approach. Furthermore, the topic prevalences and correlations are based on linear regression models within the STM framework. As sketched in Figure \ref{fig.datasketch}, these probabilities are amended with additional publication data from the Scopus database. 

To analyze the development of topic prevalences over time, I construct time series based on yearly averages of each topic. Then one can investigate the means over time measured in years. As visual inspections are valuable but limited in explanatory power, I compute a variety of statistical tests for the stationarity of these time series. In particular, it shall be examined which topics are fluctuating around (and, hence, are mean-reverting) and have finite and time-independent covariances and which include a unit root that makes them non-stationary. 

To do this, I conduct the widely applied augmented Dickey-Fuller (ADF) test that tests the null hypothesis of the existence of a unit root against the alternative of stationarity. The original test has been developed by \textcite{Dickey.1979} and has been augmented by \textcite{Said.1984} to account for more specifications than just an autoregressive process with one lag. Additionally, I conduct the Phillips-Perron (PP) test for unit roots \parencite{Phillips.1988}. It adjusts the Dickey-Fuller test statistics to account for autocorrelation in the error terms. I report both as neither ADF nor PP tests are strictly preferable \parencite{Leybourne.1999}. Both procedures share the weakness of non-stationarity being the null hypothesis, 
\textcite{Kwiatkowski.1992} suggest a test that puts non-stationarity as alternative hypothesis, such that a rejection of the null should be clearer evidence for a unit root in the time series (KPSS test). In total, I use all three test procedures to rule ensure valid findings. 

ADF and PP test come in three different types, namely a linear model with no linear time trend and no drift (type 1), a linear model with no linear time trend but a drift (non-stationarity in the mean; type 2), and eventually a model with a linear time trend plus a drift (non-stationarity in mean and variance; type 3). Additionally, the number of lags in the auto-regressive process can be specified. While annual data usually apply one lag as  it mostly relies on the previous year ($t-1$), I opt for a lag of $n=2$. This has an economic and a technical foundation that coincide: Economic publishing faces not a severe publication lag between the submission of a paper and its publication.\footnote{See, e.g., \textcite{Hadavand.2022, Bjork.2013} for more details.} Thus, a two-year time lag is more reasonable, in which the topic prevalence of year $t$ is a function of $t-2$. This is backed by the lag heuristic as part of the computation approach for a heteroskedasticity and autocorrelation consistent covariance matrix as suggested by \textcite{Newey.1994}. It sets $n = 4 \cdot \left(\frac{max(t)}{100}\right)^{(2/9)}$. Given the structure of the data, we need $n \: \in \: \mathbb{N}$, such that we set $n^*=\lfloor n \rfloor$ as done by \textcite{Schwert.1989}. For the range of my data from 2000 until 2021, we have $max(t)=22$ which leads just as the theoretical considerations to $n^*=2$. For the unit root tests, I also rely on the test statistics of the second lag. 

Next, I conduct a repeated cross-sectional analysis (as every paper can only appear once) of the relationship between latent topics and the reception of papers with these topics in the literature. As my dataset allows me to look at the number of citations per paper, I am able to study the relationship with the presence of certain topics. 
I use the number of citations per year of a specific paper $i$ as dependent variable and the expected topic probability $T$ as an explanatory variable. This looks as follows:
\begin{equation}\label{eq.panel}
f(c/y)_i = \beta_T T_{ij} + \beta_{OA} \mathbbm{1}_{OA} + \beta_A A_i + Y\times J + \epsilon_i \:,
\end{equation}
whereby $\beta$ captures the estimated coefficients of the covariates. Citations per year on the LHS are transformed to their logarithmic form. Figure \ref{fig.citeyear} in the appendix shows that this measure is approximately normally distributed. However, since there are papers without citations, some 13\% of the observations get lost. Hence, I apply two transformations to avoid reducing the already rather small sample size. First, I use $log(c/y+1)$ to avoid the dropout of zero values. Second I apply the widely used hyperbolic sine transformation, i.e., $log(c/y+\sqrt{c/y^2+1})$. On the right hand side, I include the logarithm of topic $j$ into my regression to understand how the prevalence of this particular topic corresponds to citations. 

Additionally, I include two highly relevant control variables: I add the corresponding author ($A_i$) to account for the fact that a paper's visibility and reputation certainly depend on the author.\footnote{The use of the corresponding author makes the regression more convenient than using a factor variable for each group of authors. However, my dataset misses 131 cases of corresponding authors as there are in many cases no corresponding authors mentioned, especially in older publications. In these cases, I replace the missing values with the author teams.} Second, I use a binary dummy for open access ($\mathbbm{1}_{OA}$) to capture the fact that publications made open access tends to affect citations \parencite{McCabe.2021, McCabe.2014}, even though the evidence is mixed \parencite{Craig.2007, Gaule.2011, Norris.2008, Tang.2017}. I include year$\times$journal fixed effects to account for changes over time as well as journal reputation as it is a crucial factor for visibility, credibility, and citations. 

Last, I exploit the fact that we have several topics co-moving, such that one can set up a multivariate time series analysis. Both vector autoregression (VAR) and vector error correction models (VEC) are easily conceivable. While the former relies on no cointegration of the different time series included, the latter allows and corrects for it. Technically, that means that for non-stationary data with at least one unit root, there exists a linear combination that has at least one unit root less or else a lower order of integration. I estimate a model including several groups of topics $\mathcal{T}^i$, which I name \emph{categories},\footnote{Analyzing single topics would be interesting, however, given the comparatively large amount of topics and the limited length of the time series, categories as clustered in Table \ref{tab.mapping} are more suitable.} and usually apply a lag length of two periods. This is reasoned by the publication lag in economics as well as econometric tests for the optimal VAR/VEC length. Details are specified and explained in the results section for each estimation. A vector autoregression model with three categories and two lags would look like
\begin{equation}\label{eq.var}
\begin{bmatrix}
\mathcal{T}^a_t\\
\mathcal{T}^b_t\\
\mathcal{T}^c_t
\end{bmatrix}
= 
\begin{bmatrix}
\delta^a \\
\delta^b \\
\delta^c
\end{bmatrix}
+
\begin{bmatrix}
\theta^{aa}_{-1}  & \theta^{ab}_{t-1} & \theta^{ac}_{t-1} \\
\theta^{ba}_{-1} & \theta^{bb}_{t-1} & \theta^{bc}_{t-1} \\
\theta^{ca}_{-1} & \theta^{cb}_{t-1} & \theta^{cc}_{t-1}
\end{bmatrix}
\begin{bmatrix}
\mathcal{T}^a_{t-1}\\
\mathcal{T}^b_{t-1}\\
\mathcal{T}^c_{t-1}
\end{bmatrix}
+
\begin{bmatrix}
\theta^{aa}_{t-2}  & \theta^{ab}_{t-2} & \theta^{ac}_{t-2} \\
\theta^{ba}_{t-2} & \theta^{bb}_{t-2} & \theta^{bc}_{t-2} \\
\theta^{ca}_{t-2} & \theta^{cb}_{t-2} & \theta^{cc}_{t-2}
\end{bmatrix}
\begin{bmatrix}
\mathcal{T}^a_{t-2}\\
\mathcal{T}^b_{t-2}\\
\mathcal{T}^c_{t-2}
\end{bmatrix}
+
\begin{bmatrix}
\epsilon^a_t \\
\epsilon^b_t \\
\epsilon^c_t
\end{bmatrix},
\end{equation}
where topic category $\mathcal{T}^i$ in time $t$ is regressed on its lagged values as well as lagged values of $\mathcal{T}^{j\neq i}$ up to the maximum lag length selected ex-ante. $\delta^i$ is a constant, $\theta_{-t}^{ij}$ describes the coefficient capturing mutual relationship between category $i$ and $j$ (incl. $i=j$) for lag $t$. $\epsilon^i_t$ captures the error term. A VEC model additionally includes an error correction term as explained in many econometric textbooks, see, e.g., \textcite{Verbeek.2017}. I test for the presence of cointegration using the Engle-Granger (\citeyear{Engle.1987}) two-step test  as well as the \textcite{Johansen.1991}, both well-established. 


Based on all these computations, I obtain important insights into the dispute and contention about collusive behavior within the scientific community, the evolution of approaches over time, and the appeal of particular topics captured by relationships between topics and journal types as well as by the `citability' of particular topics.

\section{Clustering Topics}
\label{sec.iad}

\begin{figure}[htbp] 
\begin{center}
\includegraphics[width=\linewidth]{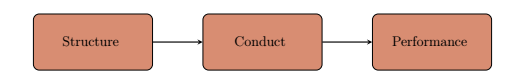}
\caption{Structure-Conduct-Performance Model}
\label{fig.scp}
\end{center}
\vspace{-5mm}
\end{figure}

As mentioned already in the introduction as well as the section on the empirical strategy, I do not use single latent topics computed with a machine-learning ER algorithm. It is from a statistical as well as an economic perspective more useful to cluster and aggregate topics to understand underlying trends, developments and relations. But when clustering any set of topics, the natural question is how to do that.\footnote{For example, in an analysis of central bank communication, \textcite{Kusters.2022} draws from a set of `lessons from the past' in monetary policy as defined by \textcite{Eichengreen.2015}. \textcite{Wehrheim.2021} studies the media reception of a council of economic advisers and uses a combination of the cosine similarity of topics \parencite{Aletras.2014} and human reasoning to group topics to superior meta-topics. This bears the problem of technical similarity but not necessarily content similarity.} In industrial economics, a natural starting point seems to be the structure-conduct-performance (SCP) paradigm. It is sketched in Figure \ref{fig.scp} below and is used to understand firms, industries, their behavior, and the market outcomes. The main idea is that there exists a causal relationship between the structure of a market, the conduct of the participating firms, and, eventually, the performance of firms as main market outcome \parencite[see, for example,][]{Bain.1951, Bester.2017, Ferguson.1994}. 

There have been debates on modifying the framework and especially the relations between the three blocks \parencite{Baumol.1982, Brock.1983}, but this shall not be part of this analysis as I only want to utilize a framework to cluster topics addressed in collusion research. Thus, the SCP framework would suggest to form three categories -- structure, conduct, and performance -- and to class the latent topics. Investigating the computed topics later on, one will see that these topics, although they are far from granular, do not really suit the body of collusion research in the \nth{21} century. Especially the game theoretic modeling of stylized industry relationships does neither fully fit into the structure category nor the conduct of firms. 

An alternative to the SCP paradigm is the the Institutional Analysis and Development (IAD) framework by Elinor Ostrom \parencite[see for an overview for example][]{McGinnis.2011, Ostrom.2011}. Even though she was awarded the Nobel memorial prize in economics, she has always been rather a political scientist. Also the IAD framework is borrowed from political studies. It has been designed to structure and understand jointly used or else managed (scarce) common pool resources, which makes it an economic model or else a model applied to economic questions even though it has been widely neglected by the discipline.\footnote{This holds at least for the leading journals in economics.} 



Applying such a pre-defined model helps in providing a fixed frame for clustering that avoids an ex-post fitting. Nevertheless, assigning the computed topics to a specific component of the framework might provoke the objection of the risk of subjective sorting. I aim at countering this with a rigorous explanation and an extensive definition of the components to avoid arbitrariness. Another objection might be that I do not apply the IAD framework in the sense that I map one or many cartels within the framework and apply it as a whole but rather map publications into single components of the framework. Hence, I map publications \emph{about cartels} and not \emph{cartels themselves} within the framework.\footnote{While in the present paper, the IAD framework is just a means to an end, \textcite{Schmal.2022c} elaborates how this model can be used to better understand the functioning of cartels.} However, this paper aims at clustering and understanding the evolution of collusion research over time and is not a genuine application of the IAD framework. Hence, I am confident one can utilize this framework for that purpose. 

\begin{figure}[htbp]
\centering
\includegraphics[width=\linewidth]{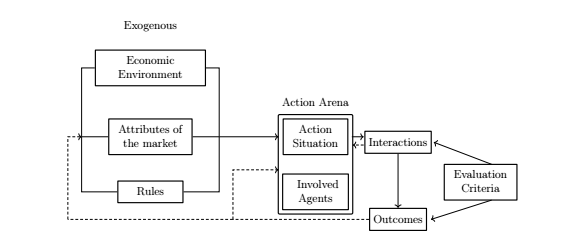}
\fnote{Original taken from \textcite[][p.~15]{Ostrom.2005a}, see Fig.~\ref{fig.iad1.app} in the appendix.}
\caption{The Structure of the IAD Framework adapted for the purpose of clustering}
\label{fig.iad1}
\vspace{-4mm}
\end{figure}

Turning toward the framework shown in Figure \ref{fig.iad1}, one can see that it consists of three major blocks. On the left, one finds the exogenous variables that comprise the social, cultural, and especially in the collusion case the economic environment. On the right, there are `evaluation criteria' devoted to evaluate the interactions of the involved agents as well as the reached outcomes. At the heart of the model lies the action situation. 
Beginning with the exogenous variables, it is obvious that the framework needs an adjustment toward the market environment of firms. Rather than `biophysical' conditions as used in the original (see Fig.~\ref{fig.iad1.app} in the appendix), we should understand it as the overall economic environment. 
Furthermore, I rephrase the `attributes of the community' `attributes of the market' as the community in the case of (colluding) firms is rather their particular market. 

Also the term `rules' needs to be closer defined.
\textcite{Crawford.2005} identify several types of rules. Here, the most applicable types are choice and payoff rules. While the former defines the choice set of the acting individuals, the latter describes what they earn conditional on which action they choose. This can be considered \emph{exogenous}. The computation of Nash equilibria and critical discount factors, which need to be satisfied to stick to a cartel instead of deviating, highlight what would be optimal and what should be done given some specific market conditions. It does not necessarily explain the \emph{actual} behavior of the involved agents. 
Henceforth, papers modeling stylized outcomes of cartel settings might be rather assigned to the exogenous `rules' category.

The last pillar of the framework lies in the center of it. It is called the `action arena' consisting of the involved agents as well as the situation in which they find themselves. I consider this as the part in which no stylized agents or firms act but employees who consider besides incentive constraints for profit maximization of their respective firm their individual constraints. 
Up to now, this is a less addressed issue in industrial organization research and, henceforth, not in the center of the following analysis. In this paper, I attempt to map the underlying topics of existing antitrust and collusion research to the IAD framework to get a first and more detailed understanding, of where research on antitrust and in particular collusion stands and what strands might be less investigated up to now.

\section{Results}
\label{sec.res}

\subsection{Latent Topics and Topic Mapping}
In a first step, I compute the latent topics and their respective content based on the optimal $K$. To elicit the terms related to each topic, I use two metrics. First of all, I show the words that have the highest probability to appear given a document contains a certain topic. Second, I use the FREX measure as proposed by \textcite{Airoldi.2016} that combines the frequency (FR) and the exclusivity (EX) of a specific word in one measure, such that I provide those words with the highest `FREX' value for each topic. It is important to note that the keywords in the topics are stems of words, i.e., the noun `behavior' as well as the verb `behave' would become `behav'. This circumvents the issue of declensions and sets the focus on the core content but of course, misses linguistic details. 

Applying the decision for $K=21$, I obtain a set of 21 latent topics as shown in Table \ref{tab.words_iofull} -- together with the most important key words of each topic.\footnote{As a robustness check, an alternative specification with $\hat K = 10$ topics is shown in Table \ref{tab.words_alt_K10} in Appendix B. It shows a clustering with many similarities to the one chosen, which strengthens the reliability of the setting with $K=21$. However, due to the low number of topics, it lacks many details.}  I assign a name to each topic (besides the number) to make it easier to follow the content of the respective topics. The numbers do not imply any kind of ranking and should be rather considered as some index. 
The main goal of this exploratory study is to map these topics to the IAD framework and its components. This is done following the index of the topics. However, Table \ref{tab.mapping} below summarizes the findings. 

\begin{table}[H]
\centering
\begin{singlespace}
\begin{scriptsize}

\begin{tabular}{l|l}
\toprule
\toprule
\textbf{Topic 1: Antitrust Overview} & \\
 	 Highest Prob:& competit, articl, author, enforc, antitrust, polici, develop   \\
 	 FREX:& author, remedi, jurisdict, articl, sanction, global, review\\
\textbf{Topic 2: Court Cases} & \\
 	 Highest Prob:& court, case, law, competit, decis, articl, damag  \\
 	 FREX:& court, infring, suprem, litig, settlement, judgment, rpm\\
\textbf{Topic 3: Market Shares} & \\
 	 Highest Prob:& market, firm, competit, cooper, share, model, product \\
 	 FREX:& cooper, allianc, market, oligopolist, share, format, forward \\
\textbf{Topic 4: Auction Theory} &\\
 	 Highest Prob:& auction, bidder, bid, first-pric, use, mechan, two  \\
 	 FREX:& bidder, auction, first-pric, ring, second-pric, bid, sealed-bid  \\
\textbf{Topic 5: Information Mechanisms} &\\
 	 Highest Prob:& inform, privat, signal, equilibrium, communic, mechan, game \\
 	 FREX:& signal, inform, privat, payoff, communic, correl, mechan  \\
\textbf{Topic 6: Collusion Cases} & \\
 	 Highest Prob:& antitrust, econom, competit, athlet, analysi, trade, case  \\
 	 FREX:& athlet, ncaa, colleg, restraint, oil, wto, appel \\
\textbf{Topic 7: Principal-Agent Issues} &\\
 	 Highest Prob:& agent, contract, princip, incent, optim, deleg, inform  \\
 	 FREX:& princip, agent, deleg, audit, corrupt, worker, supervisor  \\
\textbf{Topic 8: Law enforcement} &\\
 	 Highest Prob:& law, competit, enforc, state, signific, probabl, firm \\
 	 FREX:& complianc, centuri, freedom, foreign, state, resourc, domest   \\
\textbf{Topic 9: Mark-ups} &\\
 	 Highest Prob:& price, increas, cost, market, firm, find, profit  \\
 	 FREX:& price, announc, discount, transpar, leadership, lower, cycl  \\
\textbf{Topic 10: Capacity Constraints} &   \\
 	 Highest Prob:& firm, market, product, price, sustain, cost, capac \\
 	 FREX:& firm, capac, integr, bertrand, sustain, invest, rival    \\
\textbf{Topic 11: Gasoline Markets} &\\
 	 Highest Prob:&  price, market, retail, use, data, gasolin, cost\\
 	 FREX:& gasolin, ceil, retail, station, adjust, allianc, fuel \\
\textbf{Topic 12: Game Theory} &\\
 	 Highest Prob: & equilibrium, group, incent, game, player, threat, organ  \\
 	 FREX:& group, threat, borrow, ineffici, equilibrium, lend, player  \\
\textbf{Topic 13: Vertical Relations}  &\\
 	 Highest Prob:& competit, retail, supplier, manufactur, suppli, effect, power \\
 	 FREX:& manufactur, supplier, exclus, suppli, retail, chain, downstream  \\
\textbf{Topic 14: Airline Markets }&   \\
 	 Highest Prob:& model, airlin, competit, exchang, price, contact, bargain \\
 	 FREX:& airport, contact, airlin, exchang, bargain, japan, account  \\
\textbf{Topic 15: Merger Analysis} &\\
 	 Highest Prob:& merger, industri, firm, coordin, use, price, effect  \\
 	 FREX:& merger, merg, coordin, margin, activ, multiproduct, screen  \\
\textbf{Topic 16: Leniency} &\\
 	 Highest Prob:& lenienc, effect, fine, program, firm, polici, investig   \\
 	 FREX:& lenienc, program, immun, reduct, fine, programm, commiss \\ 	
\textbf{Topic 17: Antitrust Enforcement} &  \\
 	 Highest Prob:& competit, antitrust, manag, use, effect, crimin, fine  \\
 	 FREX:& pool, crimin, connect, patent, drug, manag, now  \\
\textbf{Topic 18: Procurement Auctions} &   \\
	Highest Prob:& bid, procur, test, use, construct, auction, competit  \\
 	 FREX:& procur, construct, bid-rig, bid, statist, test, round   \\
\textbf{Topic 19: Welfare Analysis }& \\
 	 Highest Prob:&   cost, penalti, welfar, price, trade, higher, profit  \\
 	 FREX:& penalti, differenti, welfar, link, good, cost, revenu  \\ 	  	  
\textbf{Topic 20: Cartel Organization}  &\\
 	 Highest Prob:& price, damag, use, overcharg, guarante, period, predict    \\
 	 FREX:& guarante, overcharg, damag, price-match, subject, period, method  \\ 	 
\textbf{Topic 21: Market Entry} &\\
 	 Highest Prob:&  entri, competit, effect, market, model, chang, increas   \\
 	 FREX:&  entri, entrant, varieti, arrang, barrier, quick, chang  \\ 	 	 	
 	 \bottomrule
 \end{tabular}
 \begin{threeparttable}
\begin{tablenotes}
\item \footnotesize{$K = 21$. Text corpus contains 777 documents, 2590 terms and 37746 tokens. EM-Algorithm iterations: $\le 75$. Spectral initialization. The names are assigned manually.}
\end{tablenotes}
\end{threeparttable}

\caption{Latent Topics in joint 21st Century Antitrust \& Collusion Text Corpus}
\label{tab.words_iofull}
\end{scriptsize}
\end{singlespace}
\end{table}

While topics 1 and 2 address rather review studies and court cases, they cannot be located in the framework. The best fit may be the `evaluative criteria'. Topic 3, however, discusses market shares that arise from competition, cooperation, or else alliances. This can be related to the `outcomes' of the decision situation, i.e., graphically the pillar on the right side in Figure \ref{fig.iad1}. Alternatively, it might be located in the `potential outcome' component of the action situation. However, the main topic keywords rather hint at industry characteristics in a way that it rather fits to actual general outcomes and not necessarily different outcomes depending on the actual behavior within a cartel. Topic 4 on auction theory can be linked to the `rules' within the exogenous variables as the design of procurement auctions shapes the behavior of the participants of auctions that happen later to organize this procurement. Topic 5 on information mechanisms, on the other hand, fits into the action situation and the related information of the participants and how they are exploited. 

Topic 6 is again a rather case study based topic that captures actual cases. In industrial organization, such examples are natural starting points as large blown-up cartels or mergers between leading firms often attract high levels of attention among the informed public. Again, the `evaluative criteria' might capture this topic the most. However, it is as topics 1 and 2 rather are `meta’ topics. In contrast, topic 7 on principal-agent issues fits into the heart of every cartel, namely the action situation, and within that, it fits the participants, their positions, and their actions. The topic of law enforcement (\#8) fits the interactions outside the action arena as it also addresses the outcomes (overall profits depend on potential fines) and also corresponds to exogenous variables such as the legal framework, but also the general attitude towards cartels in society and among politicians. Topic 9 is the already mentioned topic on mark-ups with the highest share among the collusion publications. The following topic on capacity constraints (\#10) seems to be very theory-based and, by that, tends to correspond with the (economic) rules of the game, i.e., a rather exogenous issue. 

The topics on gasoline as well as airlines (\#11 and \#14) again tend to be rather case-study based. This, however, is the case as these two markets have major importance and are distinct as they sell rather homogeneous goods (fuel and transport between two airports), which makes it particularly interesting to study. The topics appear unrelated to the actual IAD framework. The topic of game theory (\#12) is more difficult to allocate. On the one hand, it obviously addresses group behavior and how agents or players interact. However, the equilibrium focus of this topic (and game theory per se) seems to fit into the `rules' of the exogenous variables. One has to keep in mind that these formalized considerations suggest general behavior in the form of strategies to be optimally played. Especially related to sophisticated solution concepts, it is questionable to which extent managers involved in a cartel solve decision situations in such a way when interacting with their colleagues. 

A similar issue is addressed in the following topic on vertical relations, particularly related to grocery retail stores or chains. Again, it addresses rules or `attributes of the community' as it is called by Ostrom in Figure \ref{fig.iad1}. In the case of firms, this might be rather the characteristics of the industry. Merger analysis is another major issue in industrial organization and antitrust in general. It may play a minor role in collusion research but is still non-negligible. This is another topic rather discussing the economic environment and characteristics of some market. 

Topic 16 on leniency addresses potential outcomes within the action situation as leniency is the major `joker' for colluding firms and their involved agents to leave a cartel without being penalized.\footnote{See for an introduction to leniency programs for example \textcite{Hinloopen.2003}.} This changes incentives and, hence, net costs and benefits for the participants. Content-wise closely related to leniency is the topic on antitrust enforcement. Major terms are -- similar to leniency -- `fine', 'criminal', and `manage,' i.e., terms that affect the action situation and potential outcomes, but even more the interactions without the actual action arena. The topic on procurement auctions can  -- as the related topic on auction theory -- be mapped into the exogenous rules of the game. The subsequent topic on welfare issues also rather addresses the `biophysical/material' conditions as welfare analyses take a rather holistic perspective on society as a whole. 

Topic 20 on (internal) cartel organization seems to be one of only a few topics that directly fit into the action arena. Here, terms such as `price,' `guarantee,' `overcharge,' or `period' directly correspond to agents deciding how to price their products and whether to stick to cartel agreements in a certain period or not. The very last topic captures market entry. Again, this is an exogenous issue. It might be endogenously driven by the decisions of the cartel, for example by raising prices so much that an external firm might be inclined towards entering this particular market. However, single participants are likely to be not able to drive the market as a whole in this direction. Hence, a market entry might be again an `attribute' of the market the firms operate in and, by that, an exogenous variable.

\begin{table}[H]
\begin{onehalfspace}
\centering
\begin{tabular}{ll}
\toprule
\toprule
Economic Environment & Welfare Analysis (19)\\
Attributes of the Market & Gasoline (11), Vertical Relations (13), Airlines (14),\\
& Merger Analysis (15), Market Entry (21)\\
Rules & Auction Theory (4), principal-agent issues (7),\\
& Capacity Constraints (10), Game Theory (12),\\
& Procurement Auctions (18)  \\
\midrule
Action Arena & Information Mechanisms (5), Leniency (16),\\
& Cartel Organization (20) \\
\midrule
Interactions & Law Enforcement (8), Antitrust Enforcement (17) \\
\midrule
Outcomes & Market Shares (3),  Mark-Ups (9) \\
\midrule
\midrule
Outside the framework & Antitrust Overview (1), Court Cases (2),\\ 
& Collusion Cases (6) \\
\bottomrule
\end{tabular}
\caption{Clustering of the Topics}
\label{tab.mapping}
\end{onehalfspace}
\end{table}

As mentioned beforehand, three topics do not really fit into the applied cluster, namely the overview topic and those on particular collusion cases. While the overview topic is rather self-explaining, the case-based topics constitute an ex-post analysis of existing cartels. One might question the distinction between the category `rules' and the so-called `action arena' as also topics like leniency can be analyzed theoretically and enter the decision-making process ex-ante. The assignment, however, is made on the consideration that the three topics in the action arena category apply to decision-makers within the actual situation. While incentive constraints for a firm always apply and can be computed in advance, the decision to become a whistleblower or to hold back information is made by being involved in the collusive agreement as a manager. Therefore, it directly relates to the actions of the acting individuals. Last, it is interesting to note that the nowadays ubiquitous topic of algorithmic collusion \parencite[e.g.,][]{Ezrachi.2016, Ezrachi.2017, Calvano.2020, Calvano.2021, Martin.2021} is not present enough for its own latent topic.

\subsection{Topic Prevalence over Time} 
\label{ssec.time}
\begin{figure}[H]
\centering
\includegraphics[scale=.7]{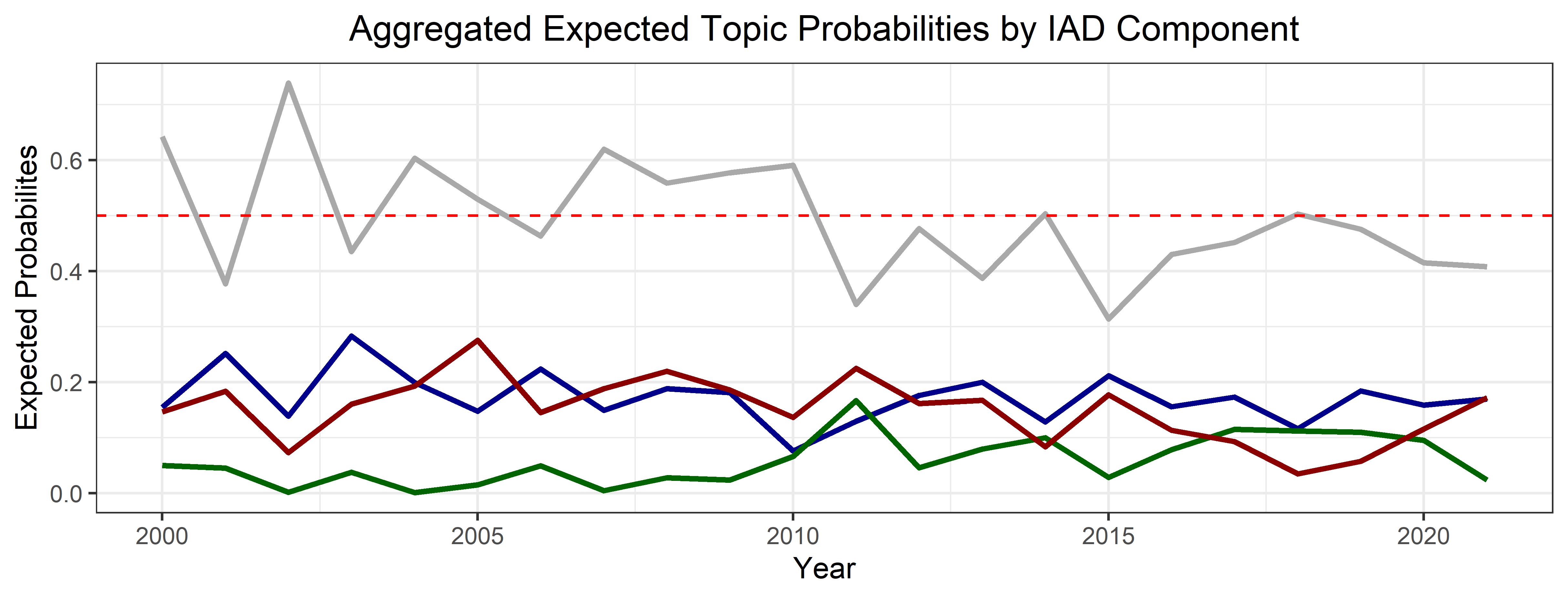}
\fnote{Gray: Exogenous Variables (Material Conditions, Attributes of the Market, Rules); Blue: Action Arena; Green: Interactions; Red: Outcomes. Horizontal red dotted line: Average of exogenous variables. Labels related to the clustering shown in Figure \ref{fig.iad1}.}
\caption{Topic Probabilities over Time}
\label{fig.top_share}
\end{figure}

Taking a look at the topics clustered by the components of the IAD framework in Table \ref{tab.mapping}, one can see a clear trend towards the `exogenous variables.' This becomes even more apparent once one includes the aggregated expected topic probabilities. Figure \ref{fig.top_share} shows the aggregated expected topic probabilities as assigned to the component above. This is done by predicting an expected value for each of the 21 topics per paper and taking the means per year. 

\vspace{-5mm}

One can easily see that topics related to the exogenous variables such as the `rules' for many years count for more than 50\% of the expected topics. Even though there seems to be a slight downward trend, one cannot say that another component would take the lead. In contrast, the action arena or action situation., colored in blue, seems to fluctuate around 20\% with a slight downward move since 2015. Interactions and outcomes seem to be rather stable over time as well. 

Instead of clustering research by latent topics, one can also look at paper types. I distinguish between mainly theoretical, empirical, experimental, and policy papers. If a paper applies two categories, e.g. by developing a theoretical model and testing it empirically, I review the particular publication and weight the contributions to come to a conclusion. I do not cluster the antitrust publications in this step and I also omit policy papers as they only account for a very small fraction of the publications. This leads to the frequencies over time as shown in Figure \ref{fig.types} below.

\begin{figure}[H]
\centering
\includegraphics[scale=.7]{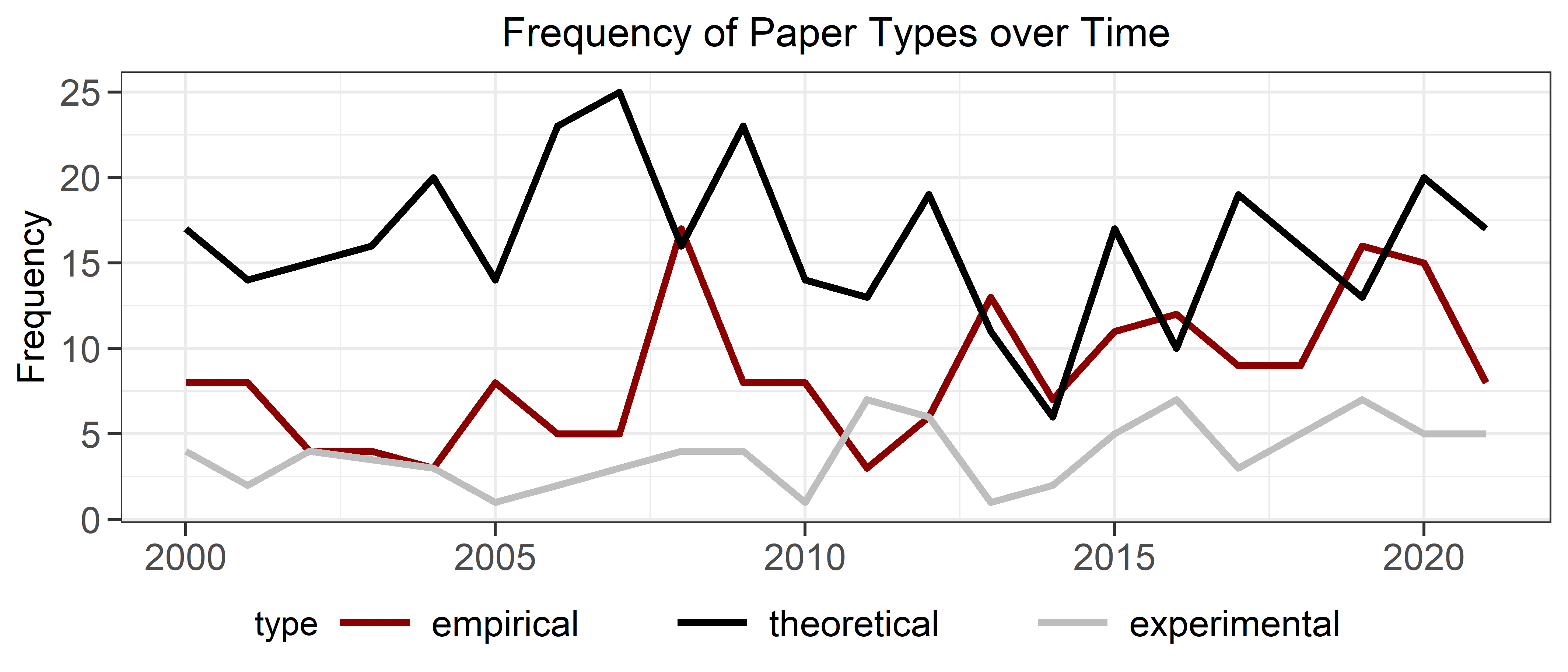}
\caption{Frequency of Paper Types over Time}
\label{fig.types}
\end{figure}

One can see that there is a slight decline in theoretical papers in the years around 2010. However, reaching a minimum in 2014, the number of mainly theoretical contributions has begun to increase again. At the same time, one can observe a minor growth in empirical papers and a rather stable evolution of experimental ones. Even though there might be some changes over time, we mostly observe fluctuations and changes in the topics prevalence may have other reasons than shifts in the methodology applied. 

\begin{figure}[H]
\centering
\includegraphics[scale=.7]{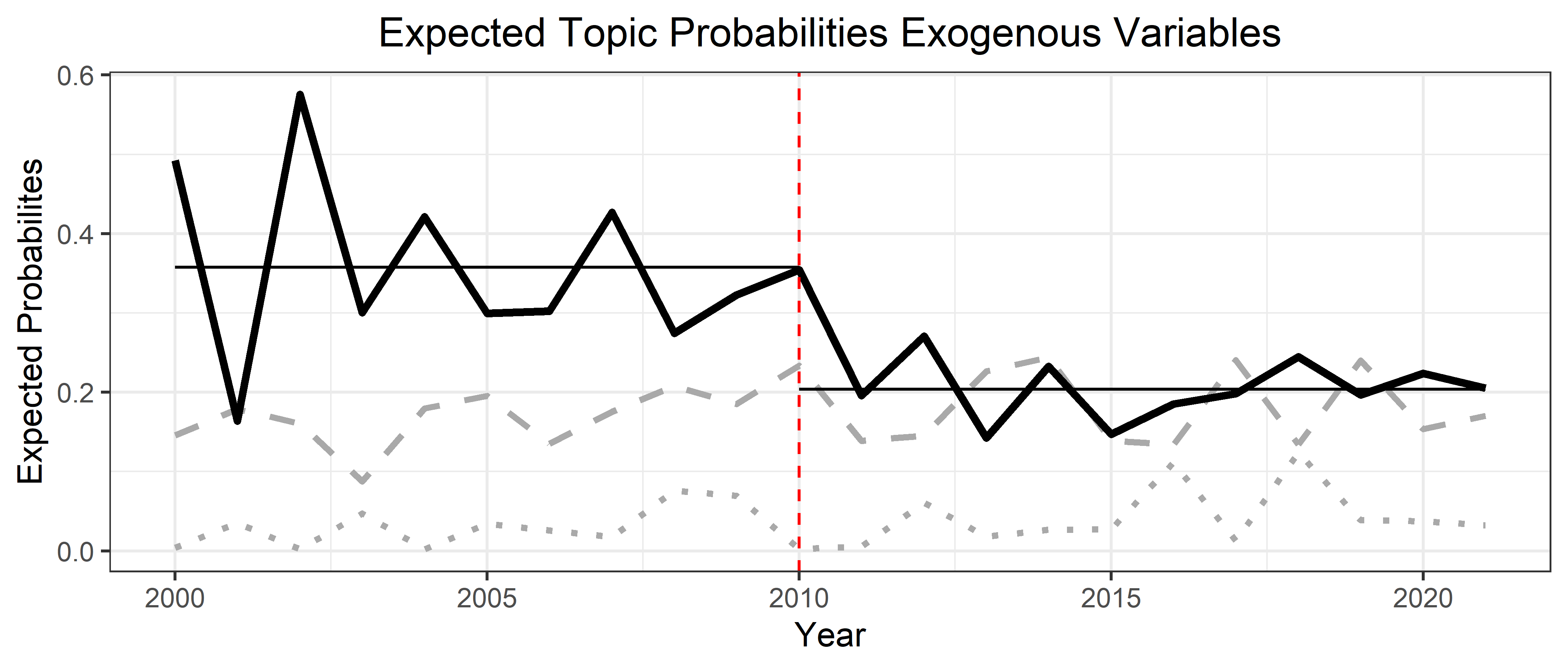}\\
\includegraphics[scale=.7]{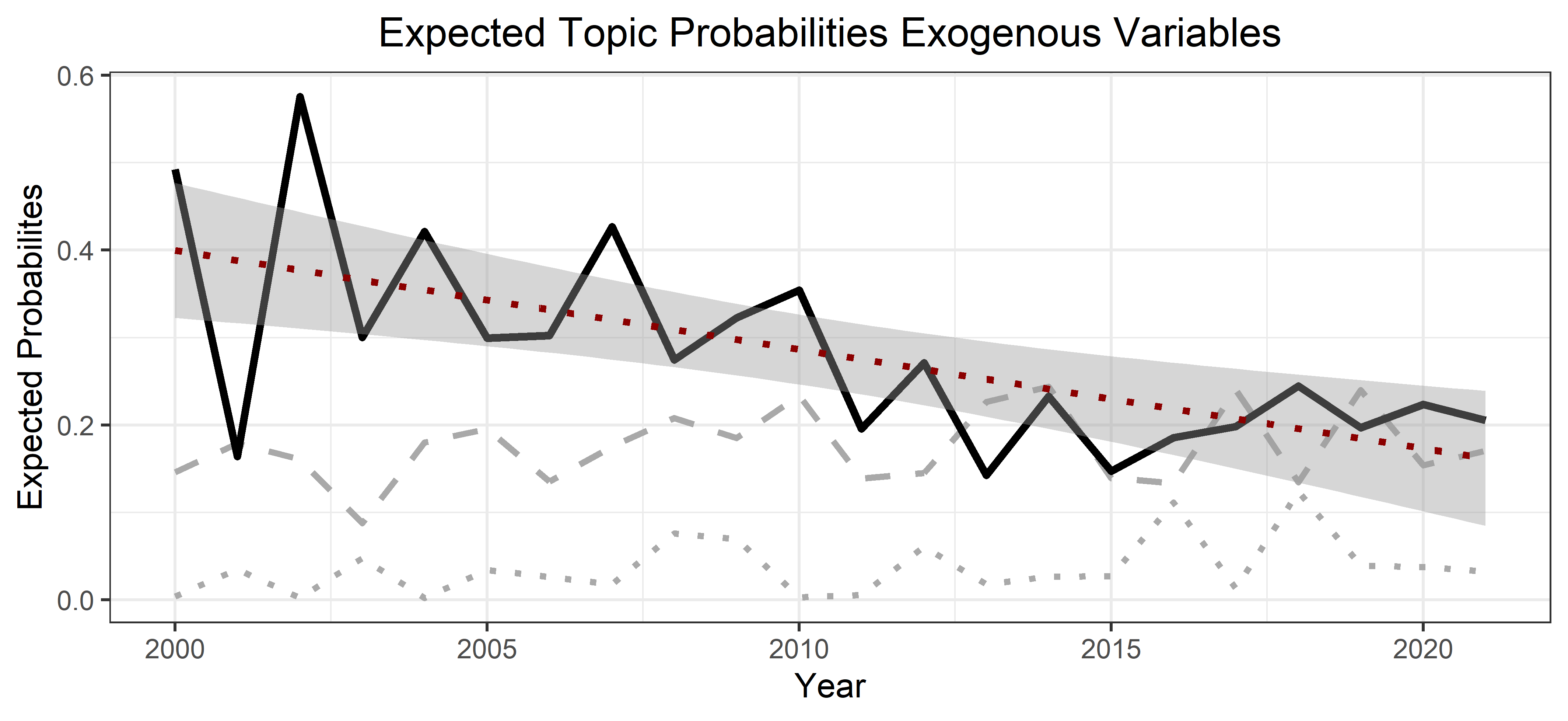}
\fnote{Black solid line: Rules; Gray Dashed line: Attributes of the market; Gray dotted line: Economic environment. Horizontal lines in the upper panel: Mean of the black solid line from 2000-2010 and 2011-2021. Red line in the lower panel: Fitted linear regression line for the `rules' category.}
\caption{Topic Probabilities of the Exogenous Variables}
\label{fig.top_share_exog}
\end{figure}

An important observation is the remarkable drop in the share of the exogenous topics from 2010 to 2011. Figure \ref{fig.top_share_exog} shows this in more detail by decomposing the three subcomponents. One immediately sees that the drop in the overall category is driven by the drop in `rules.' While the mean of the expected topics of the rule-component was $E[Pr(Rules)] = 0.358$ for the years 2000 - 2010, it was $E[Pr(Rules)] = 0.204$ for the subsequent years up to 2021. This implies a fall of 43\%. However, this was accompanied by a drop in the expected probabilities of the market attributes component (gray dashed line) between 2010 and 2011 which does not seem to be part of a major drop but rather a temporal fluctuation. Therefore, the overall decline in the exogenous variables category appears so exaggerated. Hence, fitting a linear regression for $E[Pr(Rules)]$ might be more informative. This is shown in the lower panel of Figure \ref{fig.top_share_exog} and highlights the rather continuous decline over time in that category.

To ensure that this decline is not misinterpreted, or put differently, to ensure there exists significant non-stationarity in the aggregated topics of the `rules' subcluster within the `exogenous variables', I conduct the econometric tests described in subsection \ref{sec.emp.metrics}. The detailed results for all three tests can be found in Table \ref{tab.ur_rules} in the appendix. We can see that the ADF test cannot reject the null hypothesis of non-stationarity on a 5\% significance level for one or two lags for all three types. The PP test, however, rejects the null of non-stationarity for these two types. As \textcite{Leybourne.1999} have shown, the ADF and PP tests are susceptible to diverging especially for $AR(2)$ specifications, the KPSS test supports the economic hypothesis of non-stationarity by clearly rejecting the null of stationarity for a lag parameter of 2. 

\begin{figure}[htbp]
\centering
\includegraphics[scale=.7]{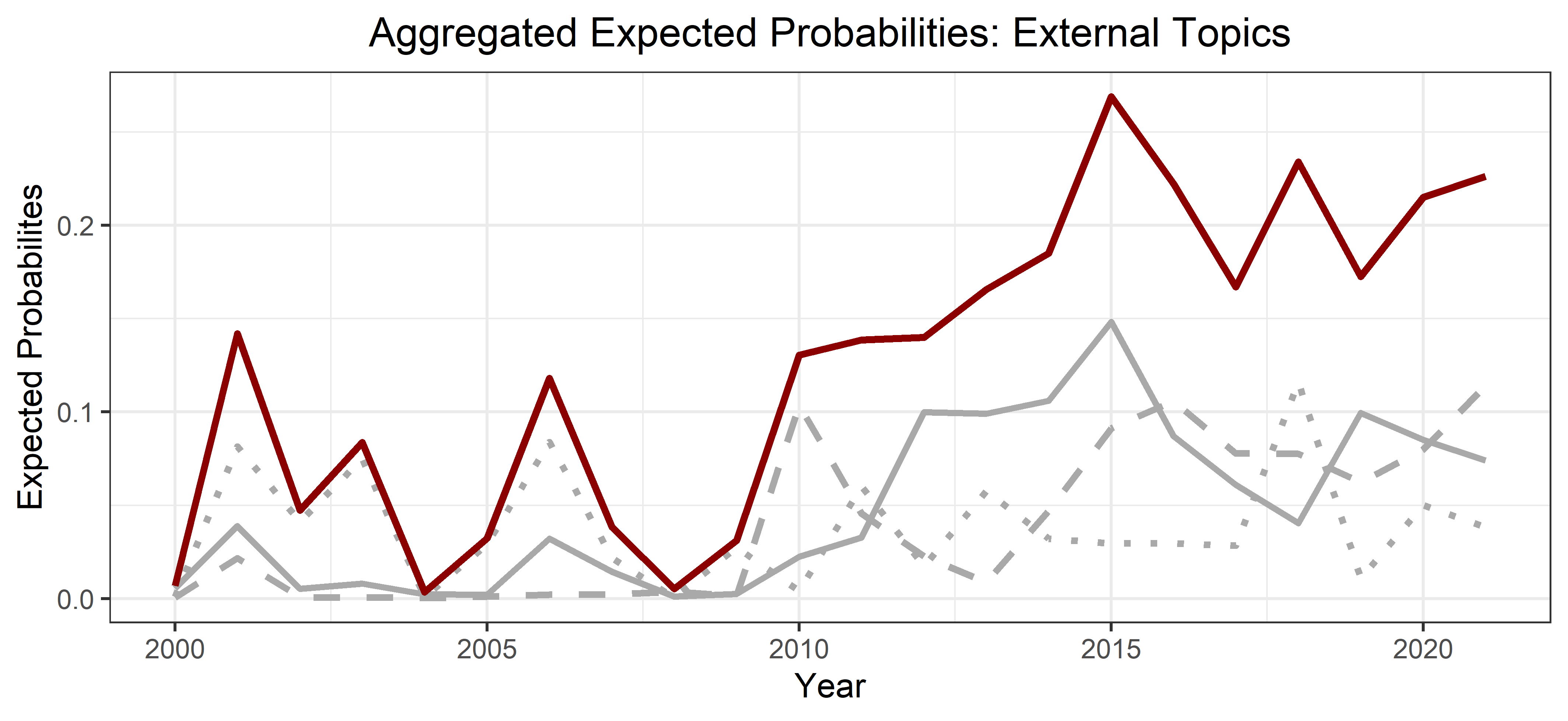}
\fnote{Dark red solid line: Sum of the three topics; Gray solid line: Topic 1 -- Antitrust Overview; Gray Dashed line: Topic 2 -- Court Cases; Gray dotted line: Topic 6 -- Collusion Cases. }
\caption{Topic Probabilities of Topics outside of the IAD framework}
\label{fig.top_share_out}
\end{figure}

In contrast to the `rules'-related topics, one can observe a remarkable surge in the prevalence of case-based topics. This is shown in Figure \ref{fig.top_share_out}. While there seems to be in the aggregate of these topics some major fluctuations in the early 2000s, one can observe a steady rise from 2008, which peaks in 2015 and stabilizes at a high level around 20\% afterward. Furthermore, up to 2008, the prevalence of the aggregated three topics was mainly driven by collusion cases (topic 6), i.e., a topic that mainly addresses leaked out or blown up cartels by authorities and its economic implications. From 2008 on, both rather antitrust-focused topics begin to rise in their respective aggregated prevalence. The statistical unit root tests for non-stationarity are in this case even less ambiguous than for the `rules' topics. As Table \ref{tab.ur_exog} in the appendix shows, neither the ADF nor the PP test rejects the null of a unit root existing for any lag and any type of test. Complimentary, the KPSS test rejects the null of stationarity. Hence, we can be quite sure that this surge is not just an optical illusion but a statistical fact.

This, however, is likely to be driven not only by an intensive margin shift towards these topics but by an extensive margin effect. Especially among antitrust journals are only very few publications addressing collusion in the early 2000s (see Table \ref{tab.ycount} in the appendix). Second, both the number of cases and the number of fines awarded, e.g., by the European Commission only began to rise from 2005 and subsequent years on.\footnote{See, e.g., the EU Summary Report on Cartel Statistics:  \url{https://ec.europa.eu/competition-policy/system/files/2022-07/cartels_cases_statistics.pdf}} Additionally, the EU put a substantially revised leniency regulation in place in 2006 \parencite[see, e.g.,][]{Wils.2007}. Again, this might have triggered new academic inquiries subsequent to disclosed cartels using this leniency scheme. Nevertheless, this extensive effect also affects the intensive margin as the focus on such topics naturally crowds out other topics from a researcher's perspective. But also the attention of the discipline shifts if an increasing share of publications turns towards some specific topics. 

\begin{figure}[H]
\centering
\includegraphics[scale=.7]{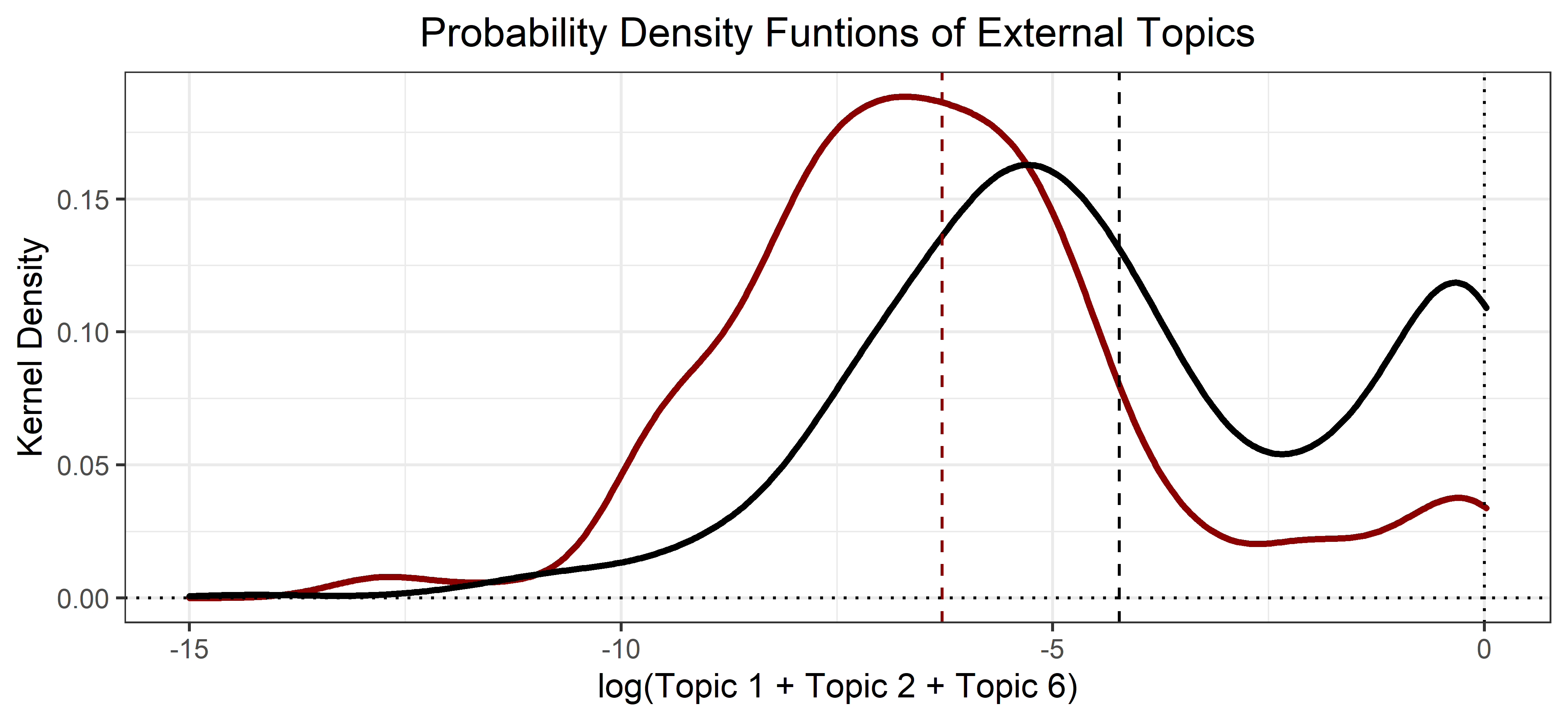}\\
\includegraphics[scale=.7]{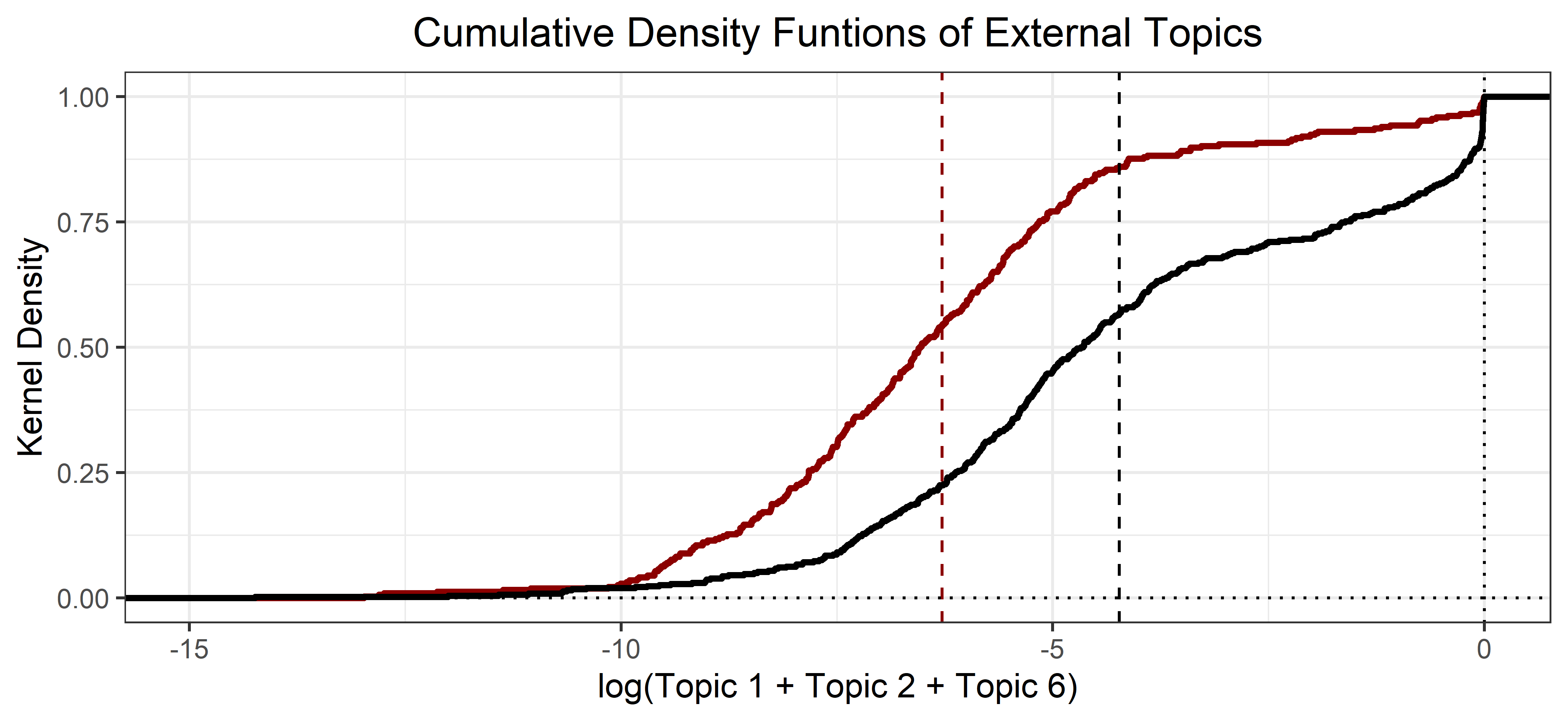}
\fnote{Dark red solid lines: Density functions for the logged sum of external topics (1, 2, 6) up to 2010. Black solid lines: Density functions for the logged sum of external topics 2011 - 2021. Upper panel: Probability density, lower panel: Cumulative density}
\caption{Topic Densities of External Topics prior and post 2010}
\label{fig.topic_dens}
\end{figure}

Looking at the density functions of the logarithmic sum of external topics as described above, one can not only the mean but also the dispersion, i.e., the second moment of the probability distribution. As Figure \ref{fig.top_share_out} highlights, there is a major surge in 2010. Hence, I split the distribution at this point in time, which also separates the time range in half. This leads to $N_{\le 2010}=315$ and $N_{\le 2010}=462$. As Figure \ref{fig.topic_dens} emphasizes, the probability density function (PDF) of the earlier years is shaped close to a normally distributed PDF only with a fatter right tail.  However, the PDF of the later years is clearly bimodal with a second local maximum close to $log\left[\Sigma(\tau_1+\tau_2+\tau_6)\right]=0$, which corresponds to a sum of probabilities close to 1 in levels. Furthermore, we observe a shift to the right around the first local maximum. These characteristics of the PDFs correspond to a first-order stochastic dominance of the probability distribution of the earlier years over the later years, which can be drawn from the comparison of the cumulative density plots (CDF) in the lower panel of Figure \ref{fig.topic_dens}. The corresponding one-sided Kolmogorov-Smirnov test confirms this as it rejects the null of no difference between the CDFs on every significance level (maximum difference $D^+ = 0.35,\: p < 2.2e^{-16}$). This is further empirical evidence for a shift towards this set of topics. Furthermore, the bimodality in recent years also hints at research `monocropping' in the sense that papers not only shift towards these topics but also nearly solely focus on them ignoring most of the other topics.

\pagebreak

\begin{figure}[H]
\centering
\includegraphics[scale=.75]{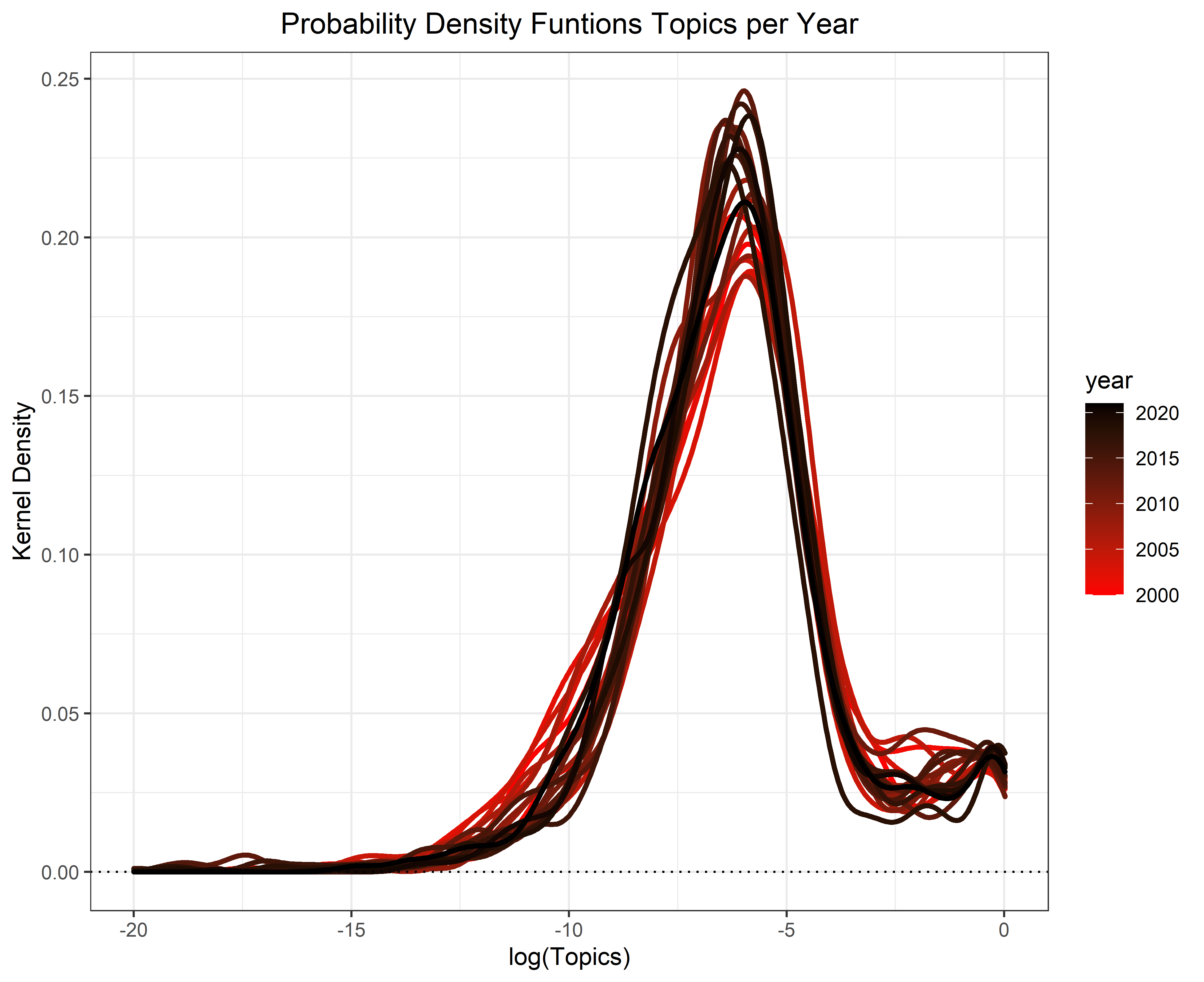}
\caption{Topic Densities per year}
\label{fig.all_dens}
\end{figure}

To investigate this issue further, I look at the annual PDFs across topics, hence, the density of all topic probabilities taken together from 2000 until 2021. This helps understand changes over time. Figure \ref{fig.all_dens} plots these 22 density functions. Even though the joint presentation of 22 graphs in one plot is hard to disentangle, one can grasp from the color gradient interesting developments. Earlier years are colored in brighter red tones while later years turn dark red up to black. One can see that the maximum density is higher and more clinched for later than for earlier years. Nevertheless, it remains ambiguous to which extent monocropping takes place as one also needs to look at the upper threshold of probabilities, i.e., $Pr(\tau_i) \sim 1$. 

\begin{figure}[H]
\centering
\includegraphics[scale=.6]{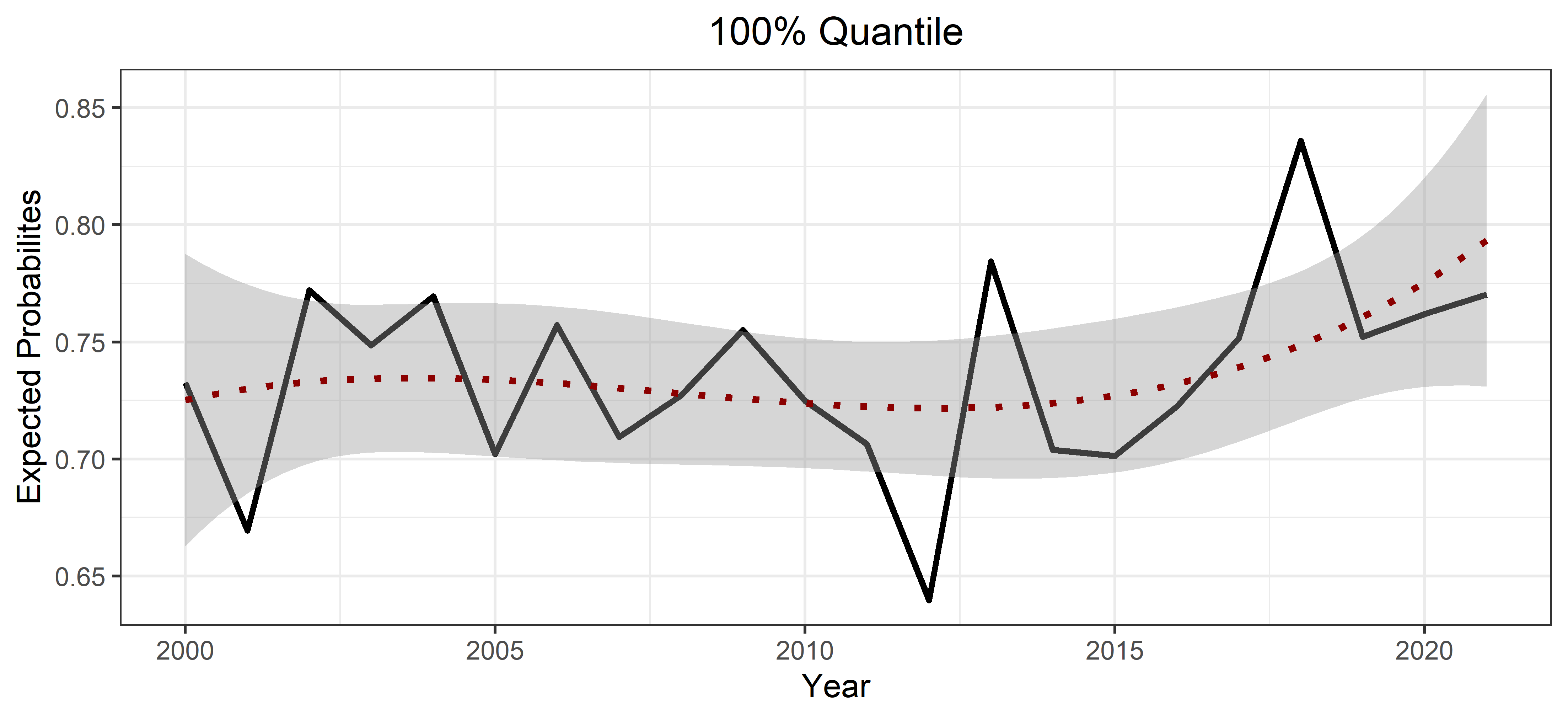}
\caption{Mean of the 100\% quantile per year}
\label{fig.p100}
\end{figure}

Figure \ref{fig.p100} plots the averages over the 100\% quantiles, measured by year, i.e., it takes the highest computed probability among all topics for each publication and averages it over each year. By doing that, one gets a better understanding of monocropping as it looks particularly at the highest single probabilities and how much space the related topic claims in a paper, due to $Pr(\tau_i)\in [0,1]\: \forall \: i \le K$ in combination with the fact that $Pr(\tau_j)=1-\left(\sum^K_{i=1, i\neq j} Pr(\tau_i)\right)$. One can infer from the graph that there seems to be a slight upward trend that has intensified in recent years as the linear model using a third-order polynomial suggests. Figure \ref{fig.p100_b} and Table \ref{tab.ur_p100} in the appendix show a linear time trend and the time series tests conducted to find a unit root. While the ADF test cannot reject its null of non-stationarity, both the PP and the KPSS test suggest the opposite. Hence, it is unclear, whether the recent increase reflects a general upward trend and whether this increase continues in the upcoming years. Overall, there exist several hints for a rise in monocropping, but hard empirical evidence is mixed.

\subsection{Topic correlations with journal types}

Up to now, topic prevalence has been discussed without addressing where these topics are placed. In economics, however, the outlet of a paper has large importance. This holds not only for the reputation of a researcher but also for the audience it reaches. The so-called top 5 and the leading general interest journals are not only widely read but also guide young researchers on where to look and what to investigate.

\begin{figure}[ht]
\centering
\includegraphics[scale=.7]{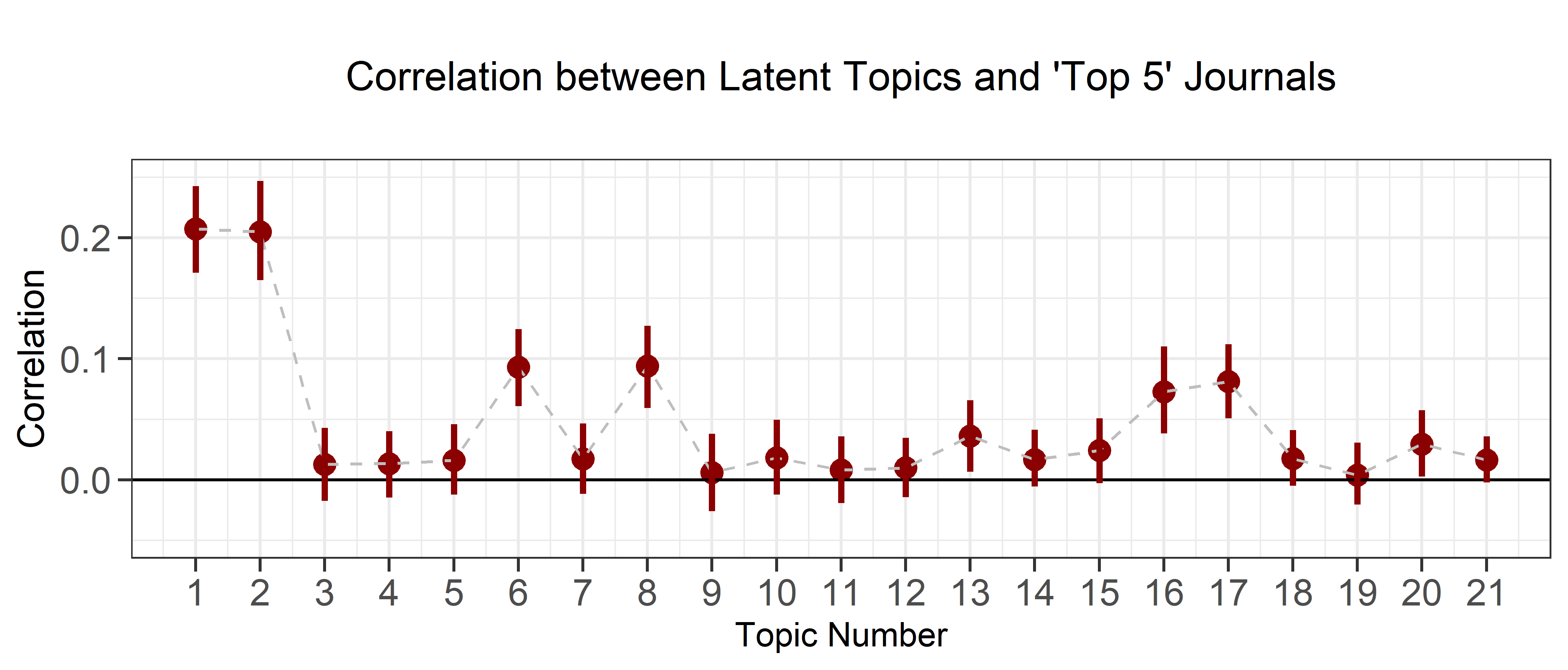}
\includegraphics[scale=.7]{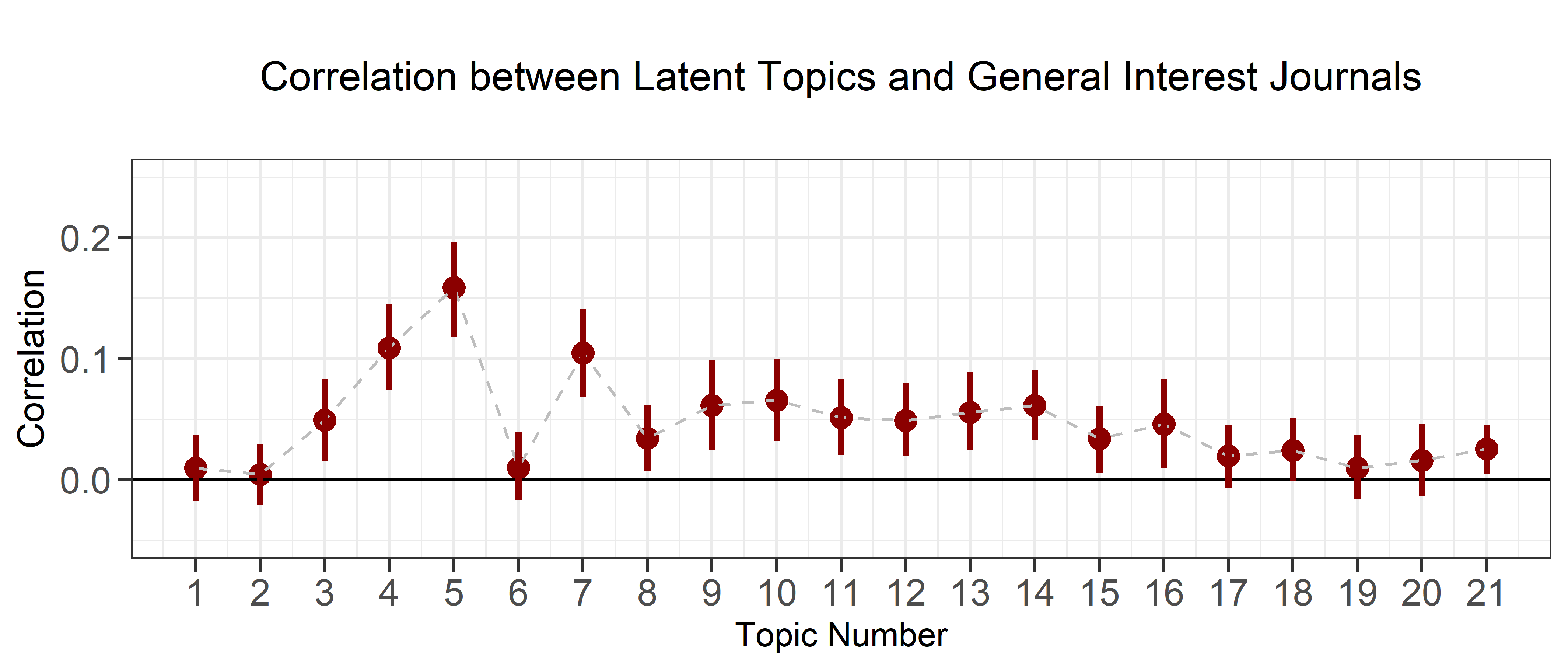}
\caption{Topic Correlation with journal types}
\label{fig.corr1}
\end{figure}

Figure \ref{fig.corr1} shows the correlation between the leading general interest journals and the 21 estimated topics. The upper panel focuses on the `top 5' journals, while the lower panel encompasses the leading `non-top 5' general interest outlets, e.g., the Journal of the European Economic Association or the Economic Journal of the Royal Economic Society. Seven topics are significantly correlated with the top 5 journals, namely 1, 2, 6, 8, 13, 16, and 17. Particularly interesting are the extraordinarily high correlations with the first two topics, i.e., those topics that take a rather general approach and those that discuss specific competition infringement cases. Furthermore, topics 6 and 8 are additional topics that address collusion cases and law enforcement on cartels and anti-competitive institutions. Topics 16 and 17 are case-related as well. Only topic 13 on vertical relations seems to diverge from this leading pattern. 

Among the general interest journals shown in the lower panel of Figure \ref{fig.corr1}, one finds lower levels of correlations with many topics. The three largest connections exist with topics 4, 5, and 7, i.e., on auction theory, information mechanisms, and principal-agent issues, which are also likely to be information-driven. Among field journals, dedicated IO journals, and antitrust journals many topics are significantly correlated (as shown in Figure \ref{fig.a.corr}), but there is not such a clear pattern as for general interest and especially top 5 journals. Especially the pattern for the latter is striking. The correlation with the first two topics is not only significant but significantly different and higher than any correlation with another topic.  Together with topics 6, 8, and 17, these journals seem to focus strongly on past disclosed cartels. Other than topics of the categories `rules' or `action arena', those are rather case study based and also backward-looking.

\subsection{Topic correlation with citations}

In the last step, this analysis looks in more detail at the relationship between topic prevalence and citations. Besides the realization of a paper's placement in a highly ranked journal, the number of citations gained for a paper is a key figure in the measurement of academic impact and eventually success.

\begin{figure}[H]
\centering
\includegraphics[scale=.7]{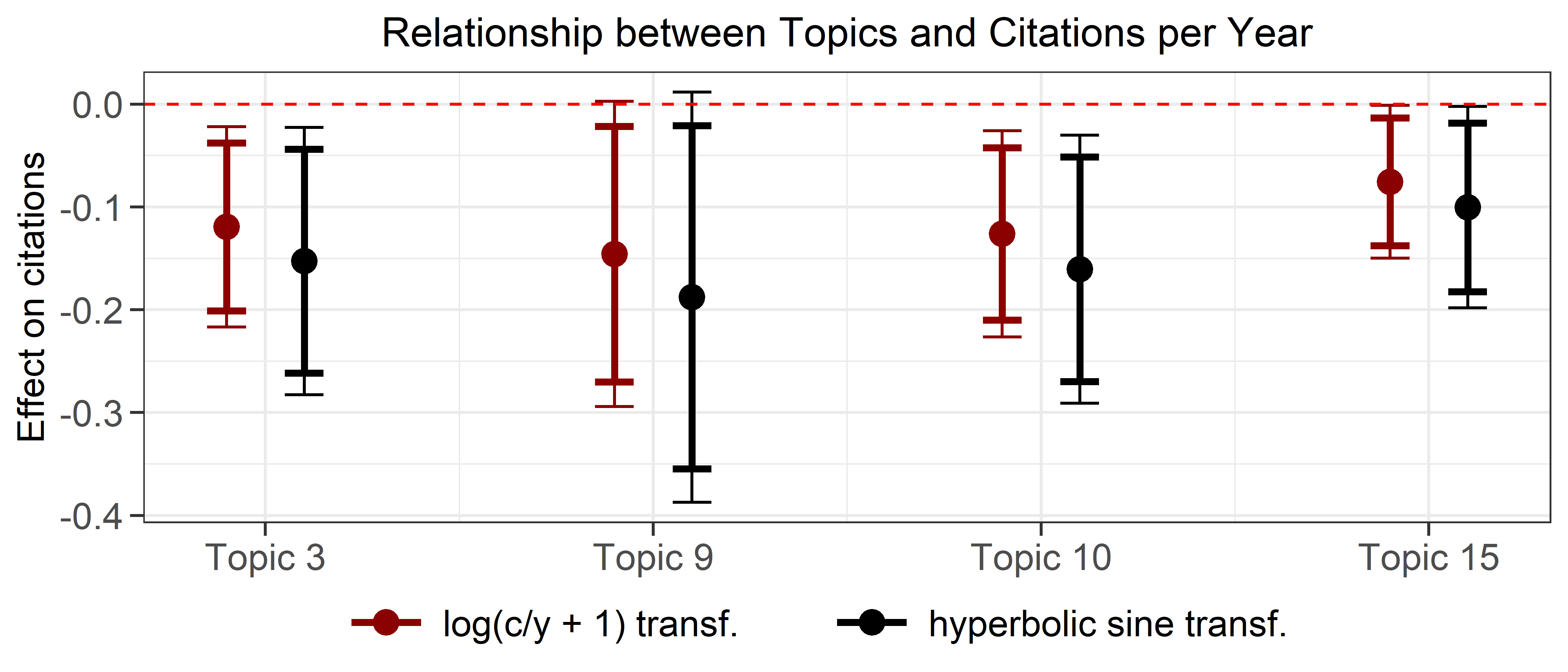}
\fnote{Point estimates for the panel regression as described in eq.~(\ref{eq.panel}). Thick line: 90\% confidence bands, thin line: 95\% confidence bands. }
\caption{Relationship between topic prevalence and citations}
\label{fig.cit1}
\end{figure}

Conducting this regression analysis for all 21 computed latent topics separately, I am able to find significant relations for four of the topics. The results are shown in Figure \ref{fig.cit1}. It shows two estimates for each topic. This stems from the fact that I report both transformations of the citations per year variable as described previously. All regression results including those for $log(c/y)$, which suffer statistically from many missing values, are shown in Table \ref{tab.reg1} in the appendix. 

\vspace{4mm}

The affected topics are those on market shares (3), mark-ups (9), capacity constraints (10), and mergers (15). All four topics are negatively related to the number of annual citations. Given the logarithmic nature of the variables, a 1\% increase in the expected probabilities of one of these topics is related to a decrease by $-0.1\%$ up to $-0.2\%$.  One has to keep in mind that this is related to publications within the subdiscipline of collusion research.

\begin{figure}[H]
\centering
\includegraphics[scale=.7]{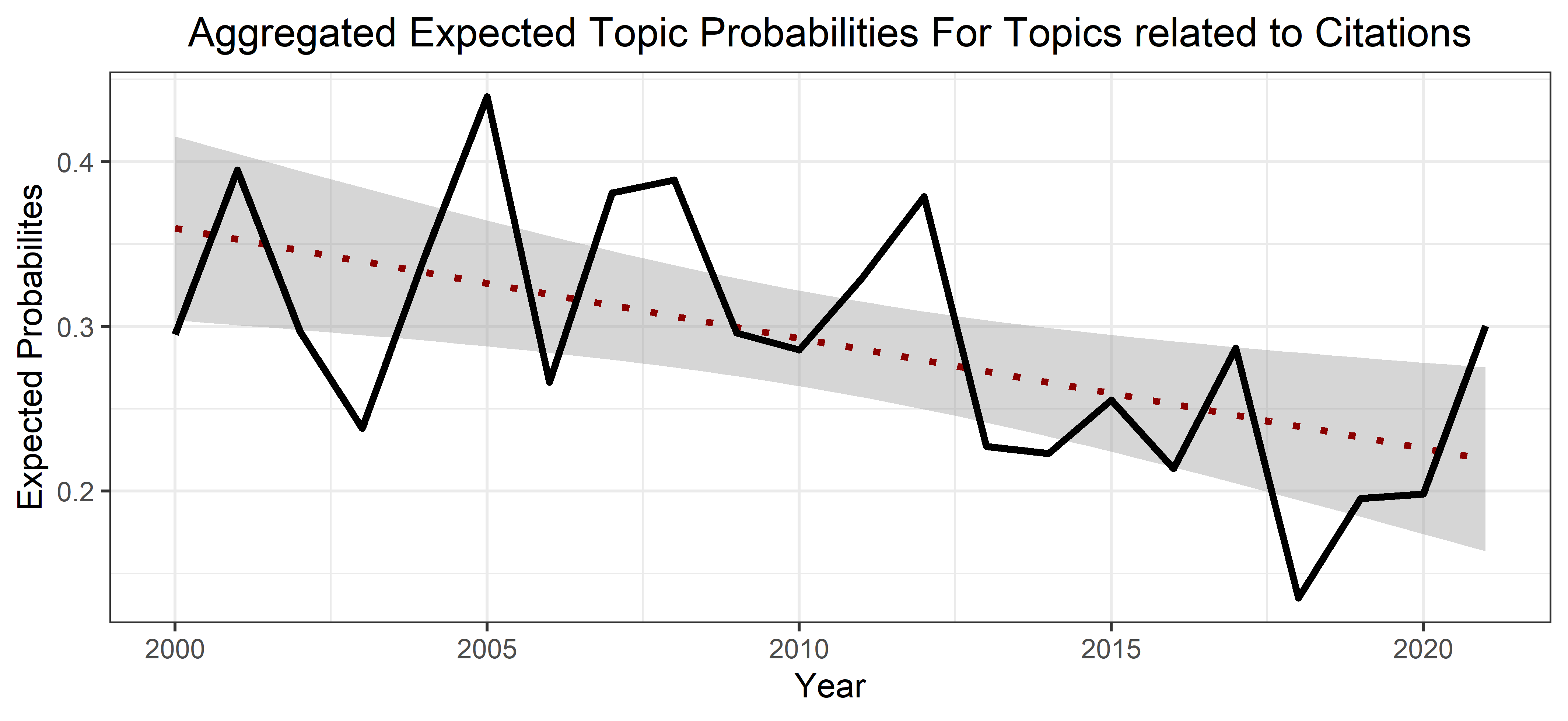}
\fnote{Black solid line: Topics significantly correlated with citations (see Fig.~\ref{fig.cit1}). Red dotted line: Fitted linear regression line.}
\caption{Mean Topic Prevalence of Topics Negatively Correlated with Citations}
\label{fig.top.shares.cit}
\end{figure}

As lower citation rates are obviously an inferior outcome for researchers, this raises the question on the time dimension of the prevalence of these topics. Figure \ref{fig.top.shares.cit} shows indeed that the share of topics with a lower level of citations diminished over time -- even though it is fluctuating.  
Overall, it seems that researchers realize the relative unpopularity of these topics and tend to shift their attention toward other topics. This is backed by the statistical tests for the existence of a unit root as evidence for non-stationarity. Again, the ADF test cannot reject the null of non-stationarity while the PP test rejects it on the 5\% level. As before, the KPSS test rejects its null of a unit root being absent on the 5\% level. An alternative explanation might be that research on these particular topics has come to an end in the sense that there is (currently) nothing significantly new to contribute. Given the fact that economics as a social science is never static and that there is a constant supply of new data, this seems not convincing. 

Hence, I conclude that there is sufficient evidence for non-stationarity being present, such that researchers indeed tend to shift away from topics for which they can observe lower levels of citations. This is in line with a recent NBER working paper studying major grants of the European Research Council (ERC grants). The authors find for rejected funding applicants a cutback in risky research after their repudiation \parencite{NBERw30320}. This is rather related to methodology and the type of questions asked instead of topics but further evidence that researchers quickly adjust their agenda subsequent to signals that their current approach does not pay off.

\begin{figure}[H]
\centering
\includegraphics[scale=.7]{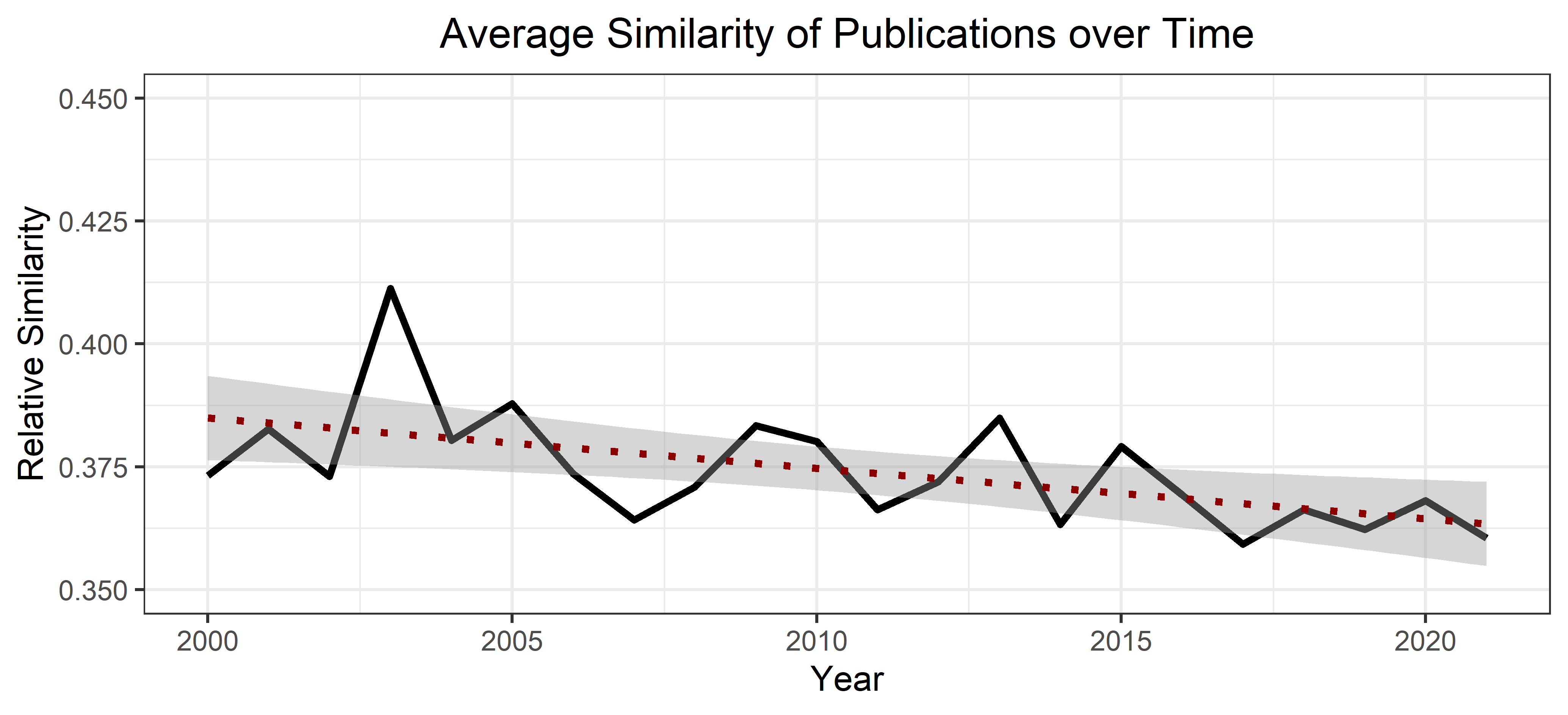}
\fnote{Black solid line: Mean similarity of papers of a specific year with all papers of the same year. Red dotted line: Fitted linear regression line. Neural network specification: paragraph vector dimension: 50. Training iterations: 20. Method: paragraph vector with distributed bag of words \parencite[PV-DBOW, see ][for details]{Le.2014}.}
\caption{Development of text similarity of publications over time}
\label{fig.neural1}
\end{figure}

In contrast to the development of topics, the variety of subjects in papers has grown over time. This can be drawn from the neural network set up to understand similarities between papers year by year. This is done using the \emph{doc2vec} algorithm with 20 iterations of the training model and a $50\times1$ vector. As \textcite{Mikolov.2013} have shown, the dimension of the document vector matters. However, as Figure \ref{fig.app.neural} in the appendix shows, variations with higher dimensions do not vary the result. Figure \ref{fig.neural1} below plots the change over time. One can see a slight decline that is statistically non-stationary as Table \ref{tab.ur_neural} in the appendix proves. Different from the previous analyses, the average similarity of papers \emph{of a particular year} does not suggest an economically reasonable intertemporal relationship, hence a random walk seems to be most suitable. This corresponds to the ADF and PP tests of type 1. Furthermore, adding lags does not seem to be useful either. The results in Table \ref{tab.ur_neural} suggest a non-stationary random walk using all three testing procedures. As we find a small decline in similarity, this implies the variety of subjects actually increases.

Thus, we have two parallel developments: We see growing diversity in the subjects of papers, for example, different industries in which firms collude. However, at the same time, there is evidence for a narrowing in underlying topics.

\subsection{Reciprocal topic prevalence over time}
\label{ssec.var}

While the time series analyses in subsection \ref{ssec.time} already bear important insights how collusion research evolved during the \nth{21} century, a multivariate examination may be able to provide a deeper understanding. Especially interesting is the question how the rise and fall of topic prevalences affect other topics. As causal inference in a microeconometric sense is difficult to obtain given the lack of a regression discontinuity or a major shock that could serve as causal identification for a difference-in-differences set-up, one can utilize the toolbox of macroeconomic multivariate time series analysis. Based on the previous findings, an analysis of the comovement of the group of external topics (outside the IAD framework) as well as the `rules' category appears as particularly interesting. Additionally, there exists lots of interest in the `attributes of the market' category, especially in the gasoline market as famous recent IO literature suggests \parencite[see, e.g.,][]{Assad.2020, Luco.2019, Martin.2020}. Hence, I choose these three categories for a multivariate time series analysis. 

To obtain the lag length, I compute statistics for several selection-order criteria. The results are shown in Table \ref{tab.vec.lags} in the appendix. I compute these statistics for a maximum lag of two and of three. In both cases, the most criteria suggest an optimal lag length of two. Next, I test for cointegration between the variables, i.e., a long-run relationship between some or all of them. The Johansen tests suggests cointegration of order 2, i.e., two cointegrated variable pairs and the Engle-Granger test find evidence for cointegration (the order is not predictable in the latter test; see Table \ref{tab.vec.coint} in the appendix for exact results of both tests). Hence, a vector error correction model would be the right approach to correct for the cointegration. For the VECM with variables in levels, with two lags, and three topic categories (external, market attributes, rules), tests for normality show that there is no reason to suspect any distortion (see Table \ref{tab.vec.norm} in the appendix). Furthermore, there is no autocorrelation in the residuals (see Table \ref{tab.vec.autocorr} in the appendix). Last, the VECM is stable as all eigenvalues (except one, by construction as the VECM specification imposes a unit modulus) have a modulus \textsmaller 1 as one can see in Figure \ref{fig.unitcircle_vec} and Table \ref{tab.vec.ev}, both in the appendix. Hence, the stability condition is satisfied as well. 

\begin{figure}[H]
\includegraphics[width=.49\linewidth]{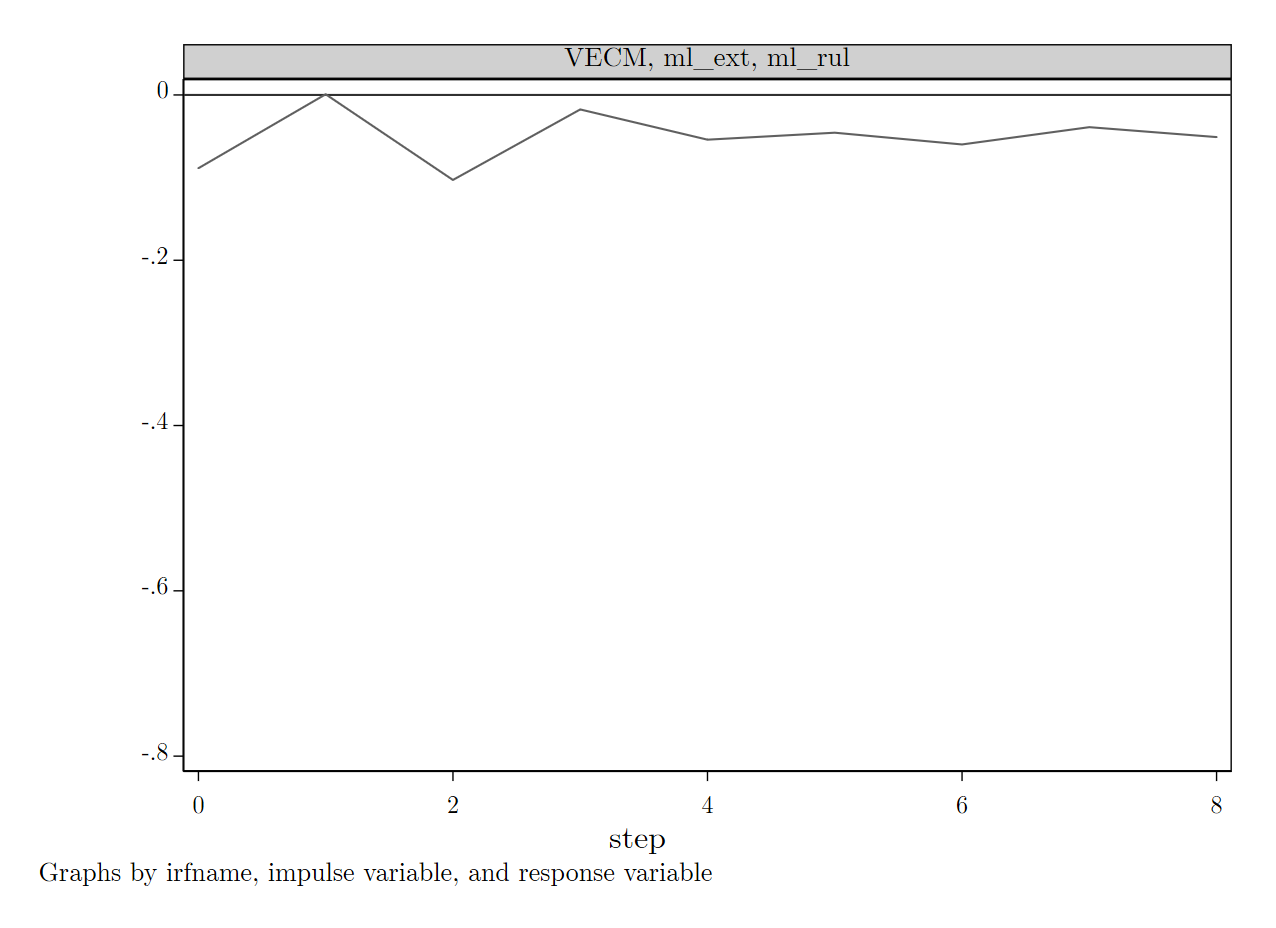}
\includegraphics[width=.49\linewidth]{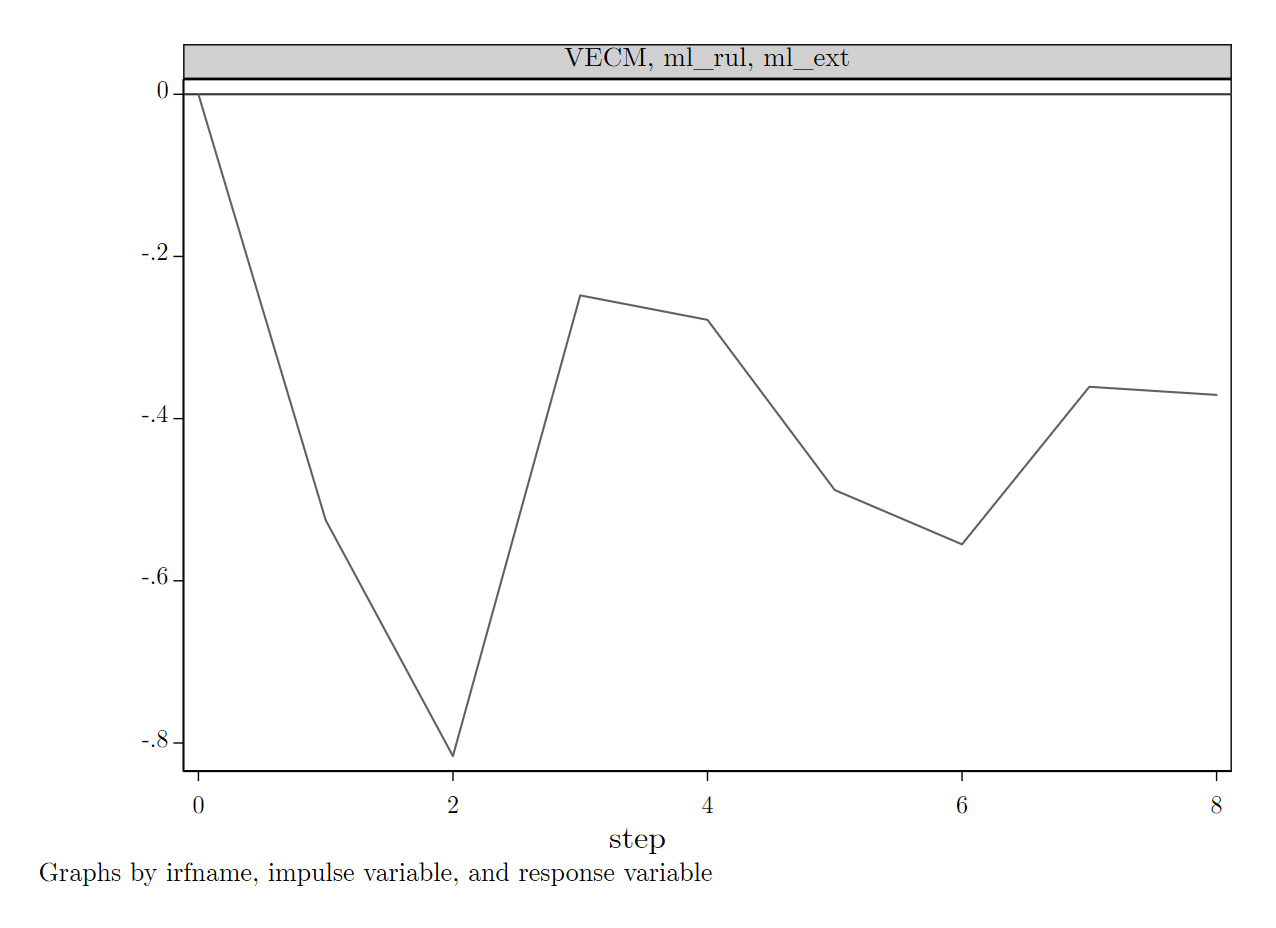}
\caption{Impulse Response Functions VEC model}
\label{fig.vecm.irf}
\end{figure}

In total, these statistical tests suggest a well-fitted model. A major contribution of multivariate models are the impulse response functions (IRFs) that model how a shock to one variable affects another variable. Figure \ref{fig.vecm.irf} shows the IRFs for a shock in the rules category on the prevalence of external variables (left panel) and vice versa (right panel). Both shocks imply a negative relationship, i.e., a jump in the rules category comes along with a seemingly persistent drop in the rules category and vice versa. However, one can see that the effect size largely differs. While the effect of external variables on rules fluctuates around $-0.05$ and reaches at most $-0.1$, the effect of a shock in rules on external variables is much larger. Two periods after the shock, the effect reaches its maximum at $-0.8$. Even though this settles down in a range of approximately $[-0.3,-0.5]$, the effect size is notable. As the whole model is specified in logs, this implies that a 1\% decrease in the rules topics is related to an up to $0.8\% $ increase in the external topics prevalence. Note that confidence intervals cannot be computed due to the error correction part of the model that affects the error estimates. However, it seems that the effect of a shock to the external variables in insignificant while the reverse shock may be significant given the effect size.\footnote{Further IRFs including the `attributes of the market' category can be found in Fig.~\ref{fig.vecm.irf_add} in the appendix. Most relations seem to fluctuate around zero, which suggests no effect at all.}

\vspace{-4mm}
This already serves as early evidence for an important interdependence: As shown earlier, we descriptively observe a surge in the prevalence of external topics and simultaneously a decline of the `rule' based topics. The impulse response functions suggest that there is no displacement of the `rules' by external topics. It rather seems to be the case that  the decline in topics happened in the first place and made place for the empirical studies on antitrust cases. 

To investigate this further, I set up a VAR model. Even though there is strong evidence for cointegration, I utilize the Toda-Yamamoto (\citeyear{Toda.1995}) approach (TY). The authors state that one can set up a VAR in levels (instead of differences) and can test restrictions even if there exists cointegration between variables as long as one adds the order of integration to the lag length of the VAR and the order of cointegration does not exceed the initial leg length. As the `attributes of the market' do not seem to substantially add information, I restrict the VAR to the two main categories of interest. The selection criteria for the optimal lag length suggest two lags (see Table \ref{tab.var.lags} in the appendix) and the Johansen test finds cointegration of order 1, see Table \ref{tab.var.coint} ibid). This leads to a $VAR_{TY}(3)$ specification with three lags. 

The technical evaluation shows that this VAR satisfies the common criteria for a reliable model. All eigenvalues are inside the unit circle and, by that, satisfy the stability condition on vector autoregression models. The residuals are not autocorrelated on a 5\% significance level. Furthermore, the residuals satisfy all conditions for a normally distributed well-behaved distribution. 

Different from a VEC model, a VAR model allows for a Granger causality analysis. Initially proposed already by Granger in \citeyear{Granger.1969}, it tests whether a variable can forecast another variable. In that sense, it only satisfies a rather weak definition of causality.\footnote{See for a more in-depth elaboration on the causality dimension of Granger causality for example \textcite{Granger.1988} or \textcite{Maziarz.2015}.} Table \ref{tab.granger} shows the results for the very small Granger causality Wald tests within the two-variable setting. Essentially, it is tested whether a variable
is influenced by the past values of another variable or else whether a specific
equation is significantly different when a variable is excluded. If so, the exscluded variable `Granger causes' the other variable.

\begin{figure}[H]
\includegraphics[width=.49\linewidth]{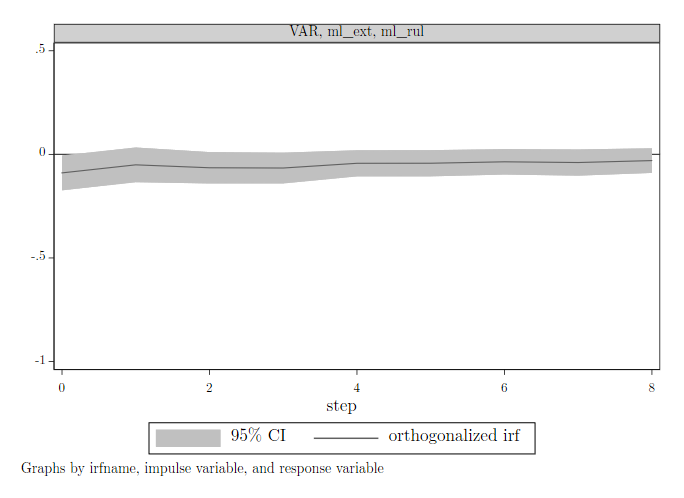}
\includegraphics[width=.49\linewidth]{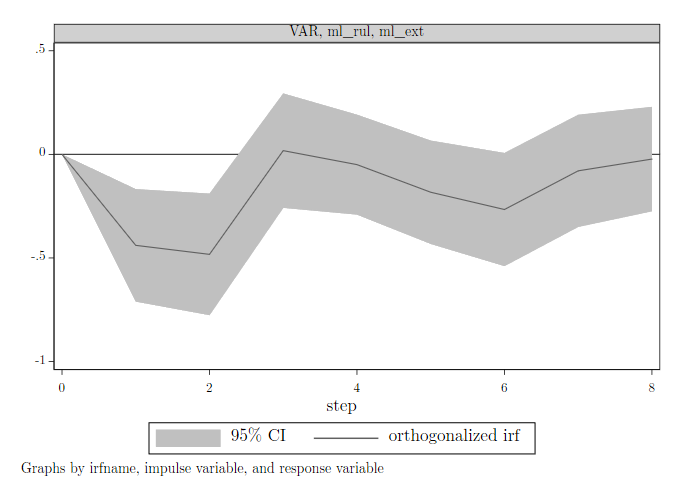}
\caption{Impulse Response Functions -- VAR model}
\label{fig.var.irf}
\end{figure}

\begin{table}[H]
\centering
\def\sym#1{\ifmmode^{#1}\else\(^{#1}\)\fi}
\begin{tabular}{ccccl}
\toprule
\toprule
Dependent Variable & Excluded       & $\chi^2$   & df & Prob \textgreater $\chi^2$ \\
\midrule
external topics  & rules        & 35.325 & 3  & $0.000^{***}$                \\
external topics  & ALL            & 35.325 & 3  & $0.000^{***}$                    \\
rules & external topics       & 1.7052 & 3  & $0.636$                  \\
rules  & ALL            & 1.7052 & 3  & $0.636$  \\
\bottomrule          
\multicolumn{2}{l}{\footnotesize{\emph{Signif. Codes: \sym{***}: $p<0.01$, \sym{**}: $p<0.05$, \sym{*}: $p<0.1$.}}}
\end{tabular}
\caption{Granger causality Wald tests}
\label{tab.granger}
\end{table}
\vspace{-4mm}
One can directly see that past values of the `rules' category influence the prevalence of the `external topics' while the opposite does not hold since the test statistic cannot reject the null of no effect. This is also reflected in the impulse response functions as displayed in Figure \ref{fig.var.irf}. Equivalent to the VEC model, one observes that a shock to the external topics does not affect the `rules'. In contrast, we observe a significant decrease for two-periods in the external topics following a positive shock to the rules-related topics as the right panel in the figure shows. Given the VAR specification, these panels also display the 95\% confidence interval of the estimates. 

\begin{figure}[H]
\includegraphics[width=.49\linewidth]{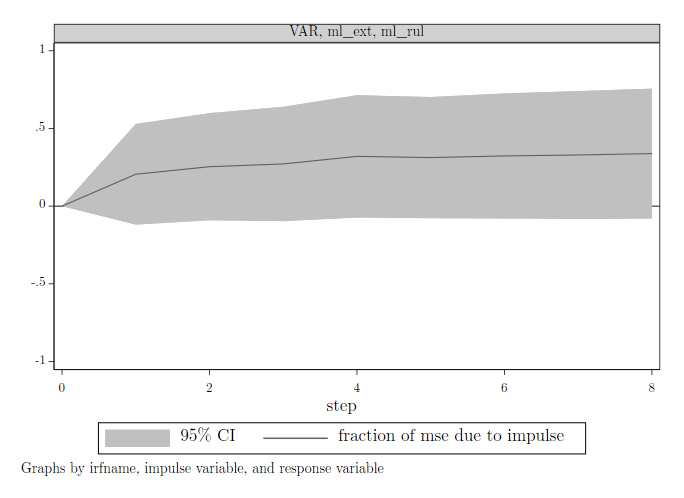}
\includegraphics[width=.49\linewidth]{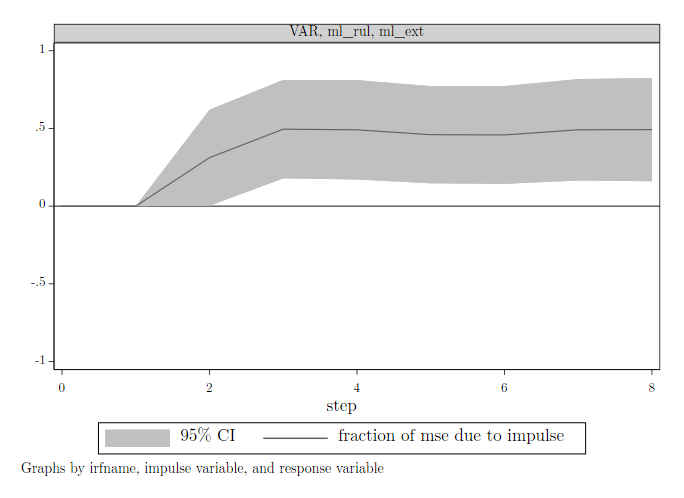}
\caption{Forecast Error Variance Decomposition -- VAR model}
\label{fig.var.fevd}
\end{figure}

\vspace{-4mm}
Figure \ref{fig.var.fevd} adds additional evidence for the previous finding. It shows the forecast error variance decomposition (FEVD) for shocks. In particular, it is scaled between $0$ and $1$ and measures the share a shock to one variable has on the forecast error of the other variable. The left panel of Figure \ref{fig.var.fevd}  shows that a shock to the external variables has no significant effect (95\% CI) on the forecast error of the rules variables. Hence, the rules category evolves independently of the exogenous topics. The opposite is true for the right panel: It takes one period of inertia, but from the second lead on, a shock of the rules category significantly and persistently affects the evolution of the external topic prevalence and effectively predicts a portion of around 1/2 of it (see Table \ref{tab.var.fevd} in the appendix for details). Hence, the external topics are at least partially endogenous in the sense that their evolution are driven by the rules topics but not vice versa.

\subsection{Discussion of results}
\label{ssec.disc}

Considering the empirical findings presented in this section, one can identify several major insights. Using the IAD framework, one can assign the latent topics of the last two decades of collusion research to meaningful categories. The interpretation of a cartel as a common pool resource provides the foundation for that. Inspecting the related time series a notable decline in the `rules of the game' is visible, i.e., a decline in work that studies incentive structures and constraints of firms, mechanisms to run cartels, and to compensate for coordinated adjustments. At the same time, a surge in case studies happened. This means papers that address famous cartels that have broken down -- either through internal collapses, leniency, or the disclosure by competition authorities. While studying these cases, one can gain valuable insights into the mechanics of the cartel organization or the increase in prices, mark-ups, and the decrease in consumer welfare. 

The downside of this development is that such case studies by construction are backward-looking. It allows a much deeper understanding of internal cartel dynamics than the usually highly stylized models that study collusive firm behavior in general. It remains an open question whether this will harm cartel detection soon as rule-based approaches may be more adoptable by competition authorities than granular but case-specific analyses of past cartels. At that point, it should be worrying that these case study topics are the two topics with the highest correlation with the top 5 journals in economics. While in total one-third of the topics are significantly correlated with these prestigious outlets, the first two topics have the highest point estimate for correlation among all topics and journal categories. This implies that the leading figures of the economic discipline particularly value this kind of paper. Naturally, this should foster many follow-up papers of IO researchers on various cartels. Hence, it is likely that these cartel case studies remain a strong pillar of collusion research. 

Taking a closer look at IO field journals such as the Journal of Industrial Economics or the International Journal of Industrial Organization, which constitute the core of collusion research, it turns out that they focus topic-wise indeed still on rules and outcomes. But the `action arena' that should encompass how individuals within a cartel actually behave suffers from several problems. Looking at the number of topics, only three out of 21 can be assigned to this category. Even though this is based on a human assignment of topics to categories, examining the whole list of latent topics reveals that these kinds of issues are rarely addressed in current collusion research. This raises the extensive margin problem that topics related to organizational issues of actual cartels cannot occur without literature studying these issues.

While the topic on information mechanism -- which arguably could also be assigned to the `rules' category -- is correlated with general interest journals, field journals, and antitrust journals, topic 20 on cartel organization is only significantly represented in field journals. The remaining topic in this category is on leniency, which gained attention by the changes in European Union competition law and some famous cartel disclosures. This is captured by the fact that this latent topic gained access to the top 5 and general interest journals in terms of a significant correlation. Nonetheless, it is at least questionable whether this is enough to understand the internal mechanisms of cartels, the organizational ties that affect no stylized agents but involved individuals that may have their personal incentives -- sociological factors that for good reasons do not enter IO models but affect behavior among colluding firms. 

In the next step, I have computed panel regressions that shall investigate the relation between latent topics and citation rates. While there is only a handful of significant correlations, all of them suggest a negative relationship. Not surprisingly, these topics are not correlated with the top 5 journals. Even though I apply year$\times$journal fixed effects to net out the journal effect on citations. But still, these journals are flag-bearers of the discipline, such that topics covered in these journals are likely to be cited more often. Interestingly, we observe a negative intensive margin effect for the topics with a negative relation to citations. As shown in the time series plot on the expected mean topic prevalence, one sees a downward trend. Hence, researchers select these topics, maybe noting that the receptions (captured by citations) are low in the discipline.

Last, I have conducted a detailed multivariate analysis. It reveals that the decline in game theory is not enforced by the new dominance of data heavy case studies but rather fill the gap that the turn away from that stylized form of modeling has left. 


\section{Concluding Remarks}
\label{sec.conc}

This paper applies topical machine learning techniques to more than twenty years of industrial organization and antitrust research that addresses collusion. Using a structural topic model, I am able to identify the underlying latent topics that have driven collusion research in the past two decades. By doing so, I obtain a hitherto not existing clustering of the research in this subdiscipline. It reveals many behavioral and organizational findings that may affect antitrust authorities and practitioners in their ability to detect cartels and collusive behavior in general. There is a strong shift towards backward-looking case studies. At the same time, researchers tend to turn away from game-theoretic rule-based thinking, which may be essential in detecting collusion as authorities usually look for the needle in the haystack, such that rules \emph{`where to look'} may be crucial guidance. Furthermore, the discipline appears to miss out on the question of what actually happens \emph{within} cartels. 

At first sight, this seems to be a rather sociological question. However, personal circumstances as well as individual career paths and the psychological capability to cope with the fact that one is involved in an illegal activity shape perceived utility and, as a consequence, incentive constraints that determine whether a cartel remains stable or not. While it is unreasonable that theoretical partial equilibrium models can encompass all of these issues, modern computing techniques allow for machine learning as well as agent-based modeling approaches that could simulate such scenarios. Laboratory experiments also allow taking a more settled approach in examining the points raised.  

It is notable that the `game-theoretic oligopoly theory' \parencite[][p.~301]{Budzinski.2007} pushed forward in the 1990s and considered a threat to competition economics and antitrust policy (ibid) appears to be in retreat in collusion research. From the availability of large datasets as well as computing capacity rather emerge sophisticated and novel contributions -- as predicted by \textcite{Kovacic.2000}. In contrast, the old Harvard school predominant from the 1930s on and based mainly on industry case studies \parencite{Bresnahan.1987} seems to experience a revival. This might be driven by the persistent trend towards evermore quantitative robustness checks and extensions as outlined by \textcite{Ellison.2002}, who also outlined that this is likely to be not just the zeitgeist but a stable potentially off-equilibrium path in academic publishing. The high demand for validations of the core findings may require such detailed data that can be only drawn from case studies. However, when reviewing the large mobile-phone spectrum auctions at the turn of the millennium, \textcite{Klemperer.2002} noted that it mostly needs``elementary economics'' and what is usually understood as rule and incentive based thinking, namely the avoidance of collusion and misbehavior of the auction participants. 

Already in \citeyear{Caves.2007}, Caves highlighted the benefits of the `old IO' approach based on cross-sectional analyses. While I am far away from criticizing or questioning quantitative rigor, the developments in the IO literature outlined in this paper put in question to which extent competition authorities get equipped with the right tools to investigate markets and detect collusion. This paper also does not make any claims on the usefulness of sophisticated empirical models such as structural estimations in other dimensions such as mark-ups. Nevertheless, in the field of collusion, \textcite{Ghosal.2014} have shown that cartel enforcement in the US has shifted from a large number of cartels detected in the 1980s and 1990s towards an approach that detects a comparatively low number of cartels. Among them are very large cartels and high fines were imposed. This shift from many (small) cartels towards a few `big fishes' might correspond to the shift in the literature. 

\newpage
\begin{singlespace}
\printbibliography
\addcontentsline{toc}{section}{\protect\numberline{}References}
\end{singlespace}
\newpage


\appendix
\begin{singlespace}
\section*{Appendix}
\label{sec.app}
\renewcommand{\thesubsection}{\Alph{subsection}}
\addcontentsline{toc}{section}{\protect\numberline{}Appendix}
\subsection*{Appendix A}
\label{app.a}
\begin{small}

\begin{table}[H]
\centering
\begin{tabular}{lcrrc}
\toprule
\toprule
\multicolumn{1}{c}{Journal} & \#Articles & Share & Cum.   & \emph{Type}  \\
\midrule
International Journal of Ind. Organization                                    & 134                                & 17.25   & 17.25  & IO \\
Review of Industrial Organization                                           & 81                                 & 10.42   & 27.67  & IO \\
Journal of Economic Beh. and Organization                                   & 64                                 & 8.24    & 35.91  & IO \\
RAND Journal of Economics                                                 & 49                                 & 6.31    & 42.21  & IO \\
Journal of Industrial Economics                                          & 48                                 & 6.18    & 48.39  & IO \\
Antitrust Bulletin                                                          & 46                                 & 5.92    & 54.31  & AT \\
Journal of Economic Theory                                                  & 39                                 & 5.02    & 59.33  & FI \\
Journal of Antitrust Enforcement                                            & 35                                 & 4.50    & 63.84  &  AT \\
European Economic Review                                                    & 29                                 & 3.73    & 67.57  & GI \\
World Competition                                                           & 27                                 & 3.47    & 71.04  & AT \\
European Competition Journal                                               & 23                                 & 2.96    & 74.00  & AT \\
Journal of Law and Economics                                                & 22                                 & 2.83    & 76.83  & FI \\
Management Science                                                          & 22                                 & 2.83    & 79.67  & FI \\
Journal of Economics and Mgmt. Strategy                                     & 20                                 & 2.57    & 82.24  & FI \\
American Economic Review                                                  & 15                                 & 1.93    & 84.17  & T5 \\
International Economic Review                                              & 13                                 & 1.67    & 85.84  & GI \\
Review of Economic Studies                                                 & 13                                 & 1.67    & 87.52  & T5 \\
Econometrica                                                              & 12                                 & 1.54    & 89.06  & T5 \\
Economic Journal                                                            & 12                                 & 1.54    & 90.60  & GI \\
Journal of Law, Economics, and Organization                                    & 12                                 & 1.54    & 92.15  & FI  \\
Review of Economics and Statistics                                          & 12                                 & 1.54    & 93.69  & GI \\
American Economic Journal: Microeconomics                                    & 8                                  & 1.03    & 94.72  & FI \\
Journal of International Economic Law                                       & 8                                  & 1.03    & 95.75  & AT \\
Journal of Political Economy                                               & 8                                  & 1.03    & 96.78  & T5 \\
Journal of the European Econ. Association                                     & 7                                  & 0.90    & 97.68  & GI \\
Journal of European Comp. Law and Practice                                     & 5                                  & 0.64    & 98.33  & AT \\
Theoretical Economics                                                       & 4                                  & 0.51    & 98.84  & FI \\
American Economic Journal: Applied Econ.                                     & 2                                  & 0.26    & 99.10  & GI  \\
Economic Policy                                                             & 2                                  & 0.26    & 99.36  & GI \\
Journal of Economic Literature                                              & 2                                  & 0.26    & 99.61  & GI \\
Quarterly Journal of Economics                                            & 2                                  & 0.26    & 99.87  &  T5 \\
Journal of Economic Perspectives                                           & 1                                  & 0.13    & 100 & GI  \\
\midrule
\textbf{Total}                                                                       & \textbf{777}                                &  &     & \\
\bottomrule
\end{tabular}
\fnote{ Journals ordered by the number of articles published in the branch of collusion. Journal types (manually assigned): AT -- Antitrust; FI -- Field; GI -- General Interest (w/o Top 5); IO -- Industrial Organization; T5 -- Top 5 Journals.}
\caption{List of Journals with the number of included articles}
\label{tab.papcount}
\end{table}

\begin{table}[H]
\centering
\begin{tabular}{lcccccrr}
\toprule
\toprule
 & 2000 - 04  & 2005 - 09   & 2010 - 14   & 2015 - 19   & $\ge$ 2020   & Total &  \\
 \midrule
Antitrust     & 1   & 3   & 33  & 72  & 35   & 144   &  \\
Field   & 22  & 24  & 28  & 38  & 15 & 127   &  \\
General Int. & 17  & 19  & 19  & 19  & 6   & 80    &  \\
Ind. Org.  & 69  & 105 & 65  & 92  & 45  & 376   &  \\
Top 5 Journals & 15  & 12  & 7   & 11  & 5 & 50    &  \\
\midrule
Total & 124 & 163 & 152 & 232 & 106 & \textbf{777}   & \\
\bottomrule
\end{tabular}
\caption{Paper distribution by journal type and publication time}
\label{tab.ycount}
\end{table}

\begin{table}[htbp]
\centering
\begin{tabular}{lllll}
\toprule
\toprule
\multicolumn{1}{c}{(1)}& \multicolumn{1}{c}{(2)}& \multicolumn{1}{c}{(3)}& \multicolumn{1}{c}{(4)} & \multicolumn{1}{c}{(5)} \\
\midrule
blackwell       & journal   & academic   & editorial  & cartel    \\
rand            & institute & among      & paper      & cartels   \\
springer        & ltd       & supervisor & section    & collusion \\
north-holland   & llc       & can        & literature & cartels   \\
sciencebusiness & iii       & due        & discuss    & collusive \\
elsevier        & press     &            & may        &           \\
                & eu        &            & show       &           \\
                &           &            & mine       &           \\
                &           &            & will       &           \\
\bottomrule
\end{tabular}
\caption{Custom stopwords additional to the \textcite{Lewis.2004} `SMART' stopwords}
\label{tab.stopword}
\end{table}

\begin{figure}[H]
\centering
\includegraphics[width=\linewidth]{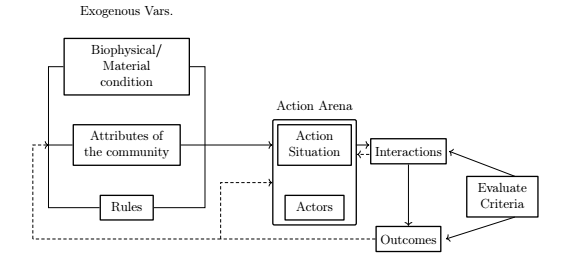}
\fnote{Taken from \textcite[][p. 15]{Ostrom.2005a}.}
\caption{The Main Structure of the IAD Framework}
\label{fig.iad1.app}
\end{figure}

\begin{table}[H]
\centering
\begin{tabular}{ccrr}
  \toprule
  \toprule
$Non$-$stationarity\:type$ & $lag$ & $statistic$ & $p$-$value$ \\ 
  \midrule
  \midrule
  \multicolumn{4}{c}{\emph{Augmented Dickey-Fuller Test}}\\
  \midrule
type 1 & 0 & -1.6836 & 0.0881 \\ 
  type 1 & 1 & -0.7370 & 0.4055 \\ 
  type 1 & 2 & -1.8849 & 0.0593 \\ 
  type 2 & 0 & -4.6838 & 0.0100 \\ 
  type 2 & 1 & -1.0032 & 0.6817 \\ 
  type 2 & 2 & -1.2861 & 0.5841 \\ 
  type 3 & 0 & -8.3101 & 0.0100 \\ 
  type 3 & 1 & -3.0499 & 0.1691 \\ 
  type 3 & 2 & -1.2047 & 0.8741 \\
  \midrule
   \multicolumn{4}{c}{\emph{Phillips-Perron Test}}\\
   \midrule
  type 1 & 2 & -2.4464 & 0.3548 \\ 
  type 2 & 2 & -24.7435 & 0.0100 \\ 
  type 3 & 2 & -35.5442 & 0.0100 \\
  \midrule 
    \multicolumn{4}{c}{\emph{KPSS Test}}\\
    \midrule
  -- & 2 & 0.6927 & 0.0142 \\ 
  \bottomrule
\end{tabular}
\caption{Unit root tests for non-stationarity in rule-based topics}
\label{tab.ur_rules}
\end{table}

\begin{table}[H]
\centering
\begin{tabular}{ccrr}
  \toprule
  \toprule
$Non$-$stationarity\:type$ & $lag$ & $statistic$ & $p$-$value$ \\ 
\midrule
\midrule
  \multicolumn{4}{c}{\emph{Augmented Dickey-Fuller Test}}\\
  \midrule
type 1 & 0 & -0.3874 & 0.5238 \\ 
  type 1 & 1 & 0.1531 & 0.6793 \\ 
  type 1 & 2 & 0.4674 & 0.7698 \\ 
  type 2 & 0 & -1.9613 & 0.3446 \\ 
  type 2 & 1 & -0.7757 & 0.7601 \\ 
  type 2 & 2 & -0.7202 & 0.7793 \\ 
  type 3 & 0 & -3.3657 & 0.0825 \\ 
  type 3 & 1 & -2.8927 & 0.2263 \\ 
  type 3 & 2 & -2.3098 & 0.4383 \\
  \midrule
   \multicolumn{4}{c}{\emph{Phillips-Perron Test}}\\
  \midrule 
  type 1 & 2 & 0.0699 & 0.6945 \\ 
  type 2 & 2 & -5.9502 & 0.3843 \\ 
  type 3 & 2 & -16.7282 & 0.0755 \\ 
 \midrule
    \multicolumn{4}{c}{\emph{KPSS Test}}\\
    \midrule
  -- & 2 & 0.6984 & 0.0137 \\ 
   \bottomrule
\end{tabular}
\caption{Unit root tests for non-stationarity in external topics} 
\label{tab.ur_exog}
\end{table}

\begin{table}[H]
\centering
\begin{tabular}{ccrr}
  \toprule
  \toprule
$Non$-$stationarity\:type$ & $lag$ & $statistic$ & $p$-$value$ \\ 
\midrule
\midrule
  \multicolumn{4}{c}{\emph{Augmented Dickey-Fuller Test}}\\
  \midrule
type 1 & 0 & -0.6638 & 0.4314 \\ 
  type 1 & 1 & -0.7994 & 0.3834 \\ 
  type 1 & 2 & -0.7090 & 0.4154 \\ 
  type 2 & 0 & -3.1228 & 0.0402 \\ 
  type 2 & 1 & -2.3151 & 0.2171 \\ 
  type 2 & 2 & -1.2202 & 0.6068 \\ 
  type 3 & 0 & -4.5066 & 0.0100 \\ 
  type 3 & 1 & -3.3713 & 0.0818 \\ 
  type 3 & 2 & -2.3536 & 0.4223 \\ 
  \midrule
   \multicolumn{4}{c}{\emph{Phillips-Perron Test}}\\
   \midrule
  type 1 & 2 & -0.3350 & 0.6060 \\ 
  type 2 & 2 & -13.7363 & 0.0353 \\ 
  type 3 & 2 & -21.1219 & 0.0183 \\ 
  \midrule 
    \multicolumn{4}{c}{\emph{KPSS Test}}\\
    \midrule
  -- & 2 & 0.5478 & 0.0309 \\ 
   \bottomrule
\end{tabular}
\caption{Unit root tests for non-stationarity in topics with negative citation correlation} 
\label{tab.ur_lost}
\end{table}

\begin{table}[ht]
\centering
\begin{tabular}{ccrr}
  \toprule
  \toprule
$Non$-$stationarity\:type$ & $lag$ & $statistic$ & $p$-$value$ \\ 
\midrule
\midrule
  \multicolumn{4}{c}{\emph{Augmented Dickey-Fuller Test}}\\
  \midrule
  type 1 & 0 & -0.0446 & 0.6224 \\ 
  type 1 & 1 & 0.4223 & 0.7568 \\ 
  type 1 & 2 & 0.2251 & 0.7000 \\ 
  type 2 & 0 & -4.4146 & 0.0100 \\ 
  type 2 & 1 & -2.7745 & 0.0808 \\ 
  type 2 & 2 & -2.0803 & 0.3017 \\ 
  type 3 & 0 & -4.6213 & 0.0100 \\ 
  type 3 & 1 & -2.7805 & 0.2671 \\ 
  type 3 & 2 & -2.2250 & 0.4691 \\ 
  \midrule
   \multicolumn{4}{c}{\emph{Phillips-Perron Test}}\\
   \midrule
  type 1 & 2 & 0.0288 & 0.6855 \\ 
  type 2 & 2 & -22.1887 & 0.0100 \\ 
  type 3 & 2 & -23.5700 & 0.0100 \\ 
  \midrule 
    \multicolumn{4}{c}{\emph{KPSS Test}}\\
    \midrule
  -- & 2 & 0.2160 & 0.1000 \\ 
   \bottomrule
\end{tabular}
\caption{Unit root tests for non-stationarity in the p100-quantile time series} 
\label{tab.ur_p100}
\end{table}

\begin{figure}[H]
\centering
\includegraphics[scale=.5]{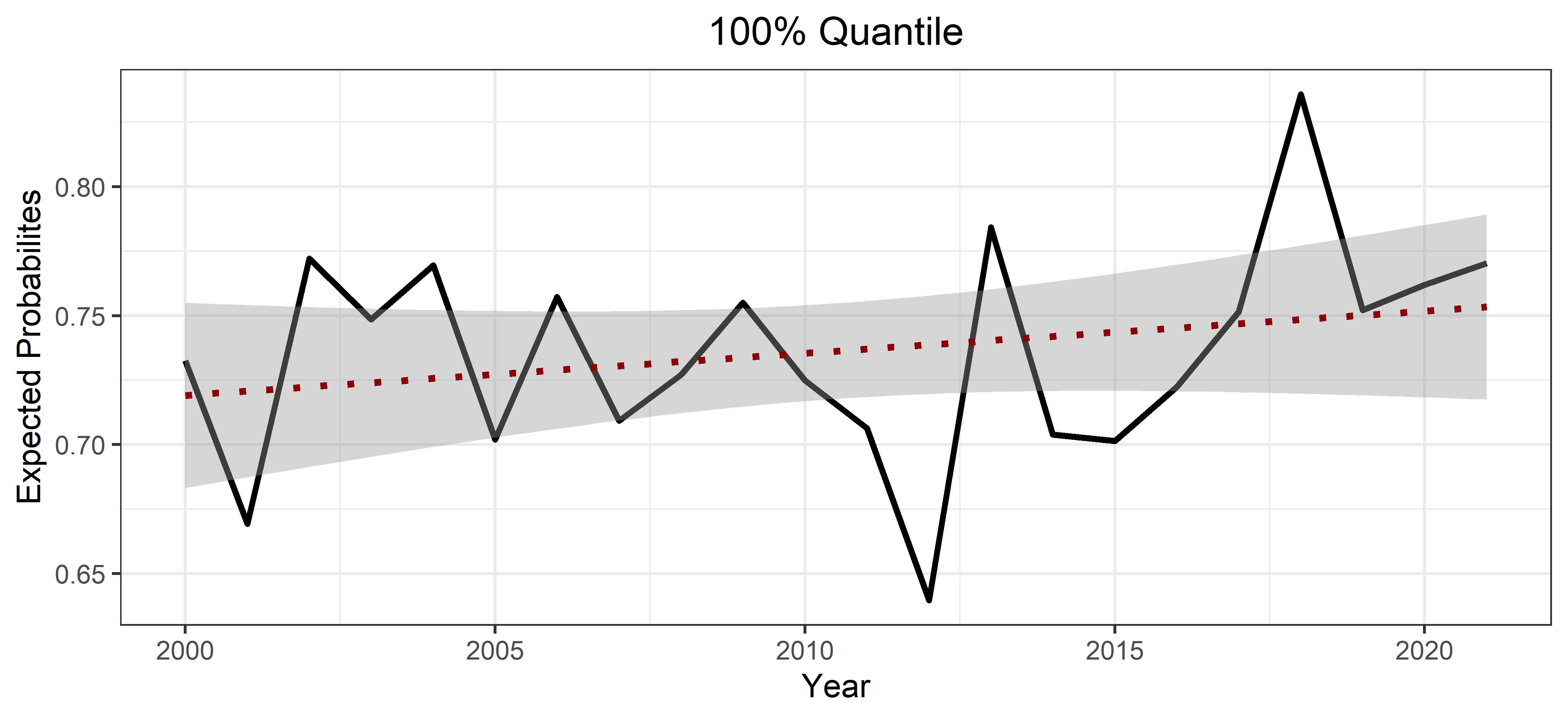}
\caption{Mean of the 100\% quantile per year with first linear order polynomial fitting}
\label{fig.p100_b}
\end{figure}

\begin{figure}[H]
\centering
\includegraphics[scale=.7]{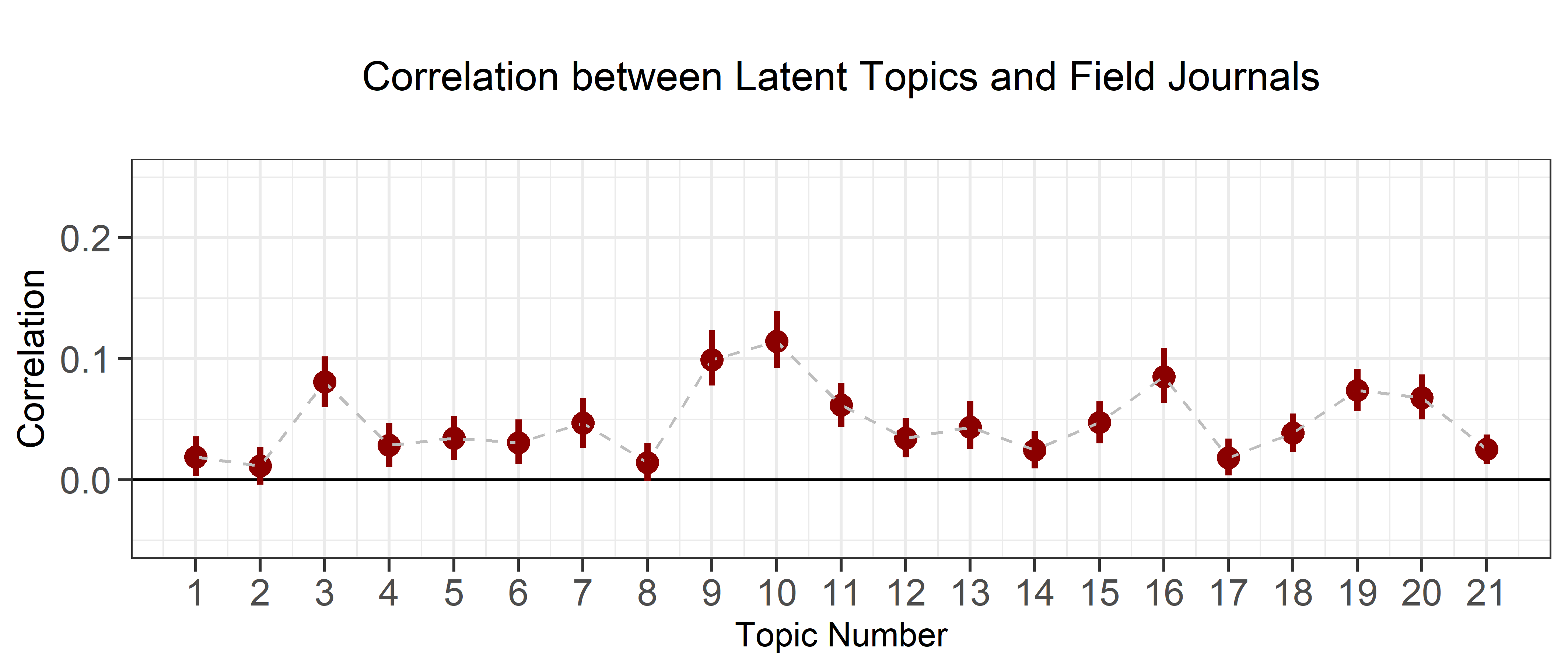}\\
\includegraphics[scale=.7]{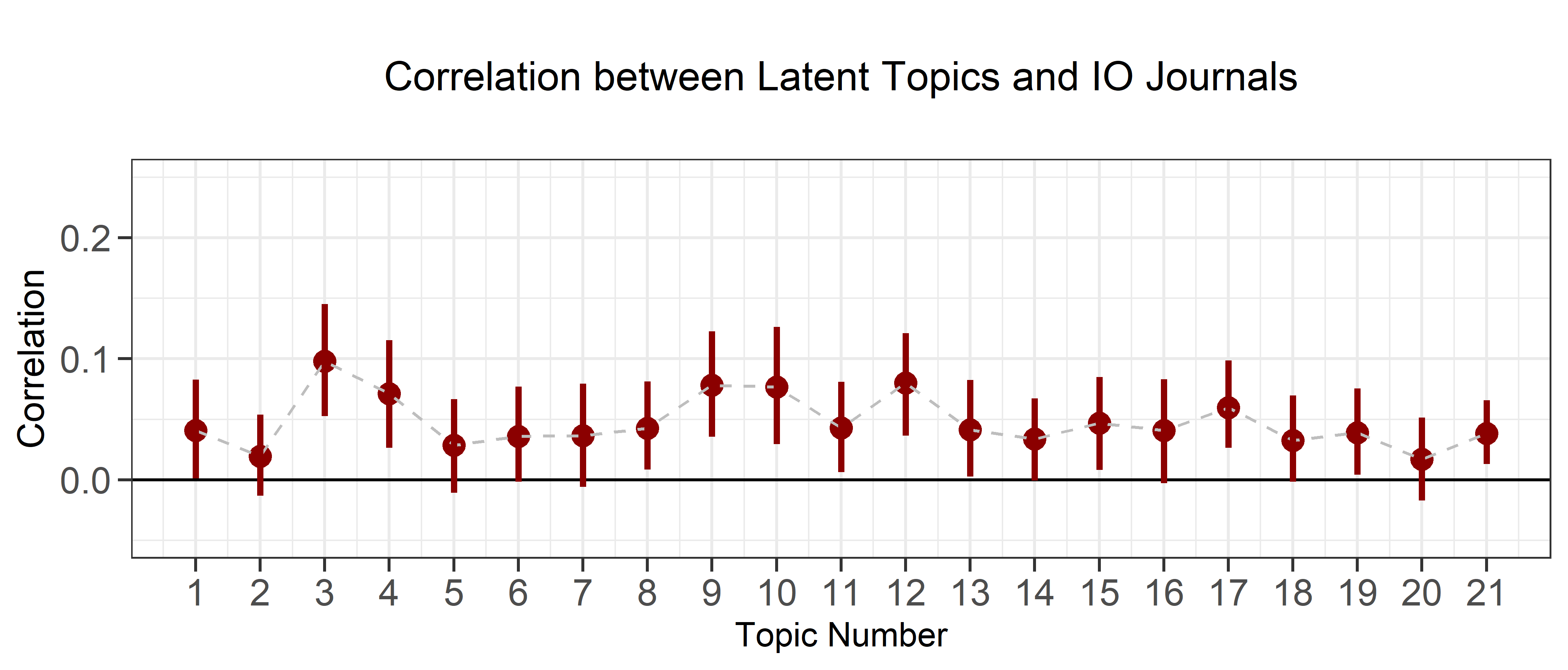}\\
\includegraphics[scale=.7]{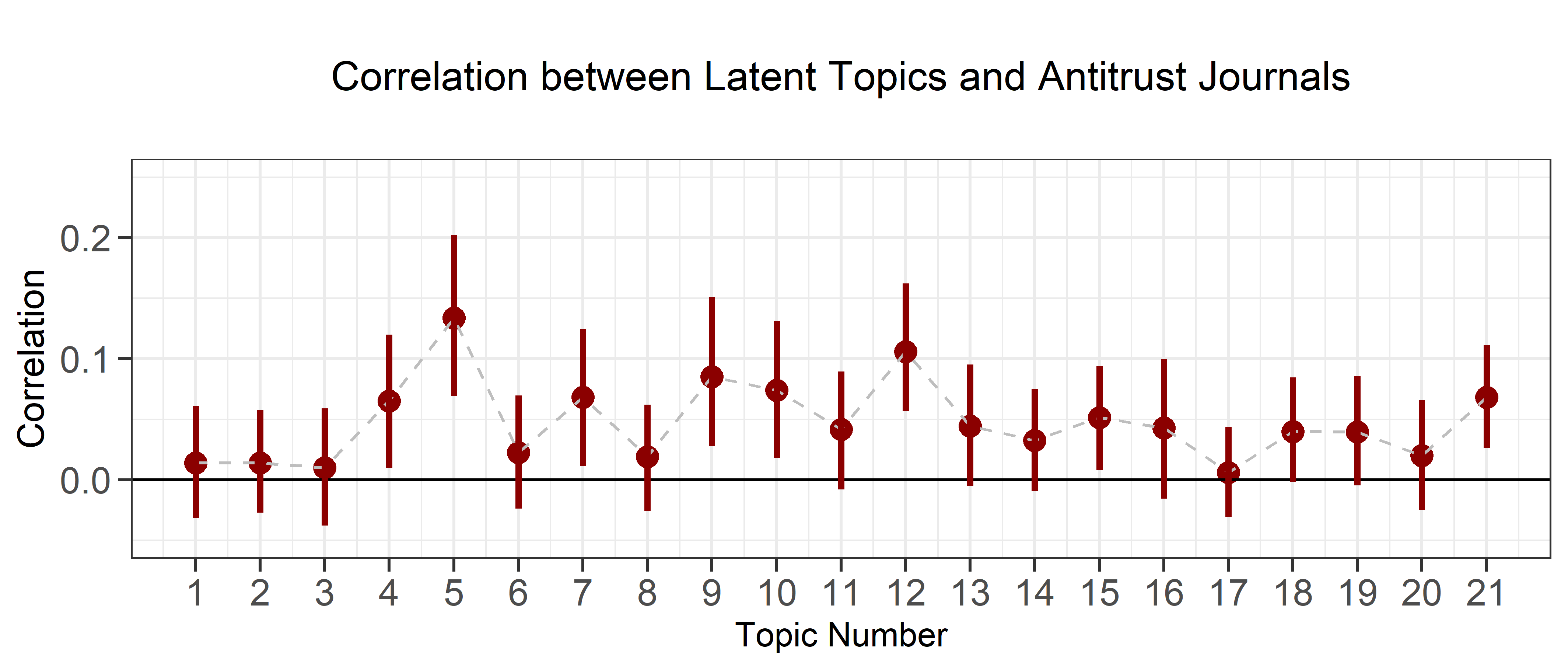}
\caption{Topic Correlation with journal types}
\label{fig.a.corr}
\end{figure}

\newpage

\begin{figure}[H]
\centering
\includegraphics[scale=.75]{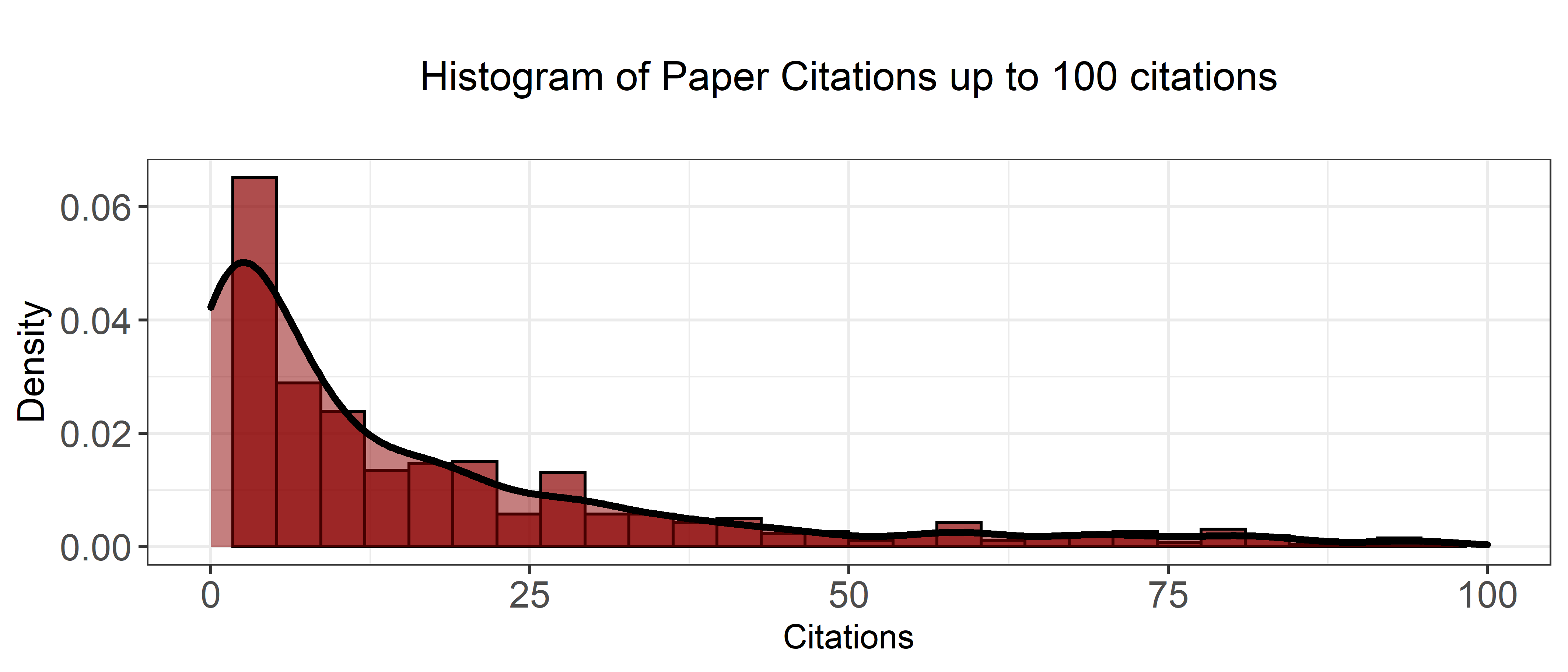}\\
\caption{Histogram of Citations $\le$ 100}
\label{fig.citecount}
\end{figure}

\begin{figure}[H]
\centering
\includegraphics[scale=.75]{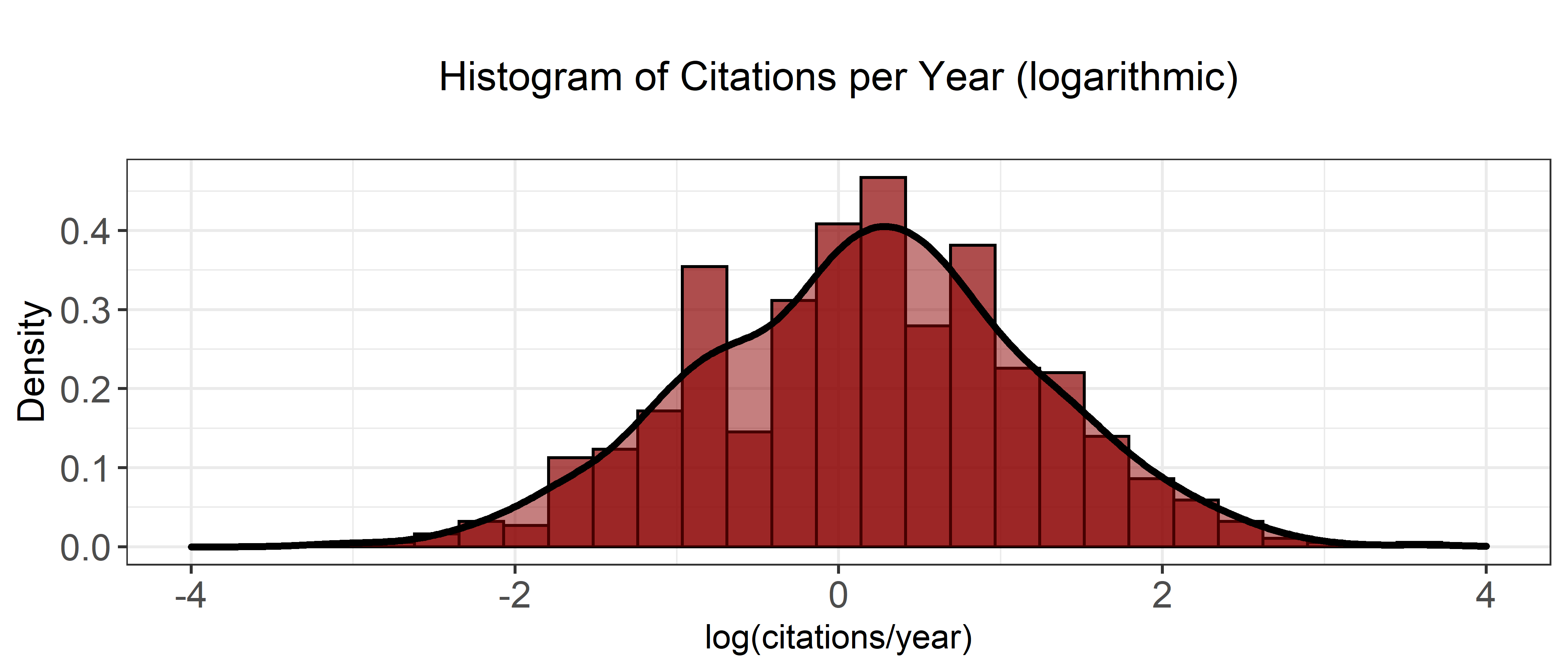}\\
\caption{Histogram of logarithmic citations per year}
\label{fig.citeyear}
\end{figure}

\begin{table}

\begingroup  
\centering
\begin{tabular}{lccc}
   \tabularnewline \midrule \midrule
& \textbf{$log(c/y)$}      & \textbf{$log[(c/y)+1]$}          & \textbf{$hsin\:transf.$} \\    \midrule
   \textbf{log(Topic 3)}                  & -0.116 (0.153) & -0.119$^{**}$ (0.050) & -0.153$^{**}$ (0.066)\\   
   \midrule
   \emph{Fit statistics}\\
   Observations         & 675            & 777                   & 777\\  
   R$^2$                & 0.994          & 0.992                 & 0.992\\  
   Within R$^2$         & 0.986          & 0.980                 & 0.980\\  
   \midrule
 \end{tabular}
\par\endgroup

\begingroup 
\centering
\begin{tabular}{lccc}
   \tabularnewline \midrule \midrule
  \textbf{ log(Topic 9) }                 & -0.128 (0.136) & -0.146$^{*}$ (0.076) & -0.188$^{*}$ (0.102)\\   
   \midrule
   \emph{Fit statistics}\\
   Observations         & 675            & 777                  & 777\\  
   R$^2$                & 0.994          & 0.992                & 0.992\\  
   Within R$^2$         & 0.986          & 0.980                & 0.980\\  
   \midrule
 \end{tabular}

\par\endgroup

\begingroup 
\centering
\begin{tabular}{lccc}
   \tabularnewline \midrule \midrule
  \textbf{ log(Topic 10)}                 & -0.030 (0.109) & -0.126$^{**}$ (0.051) & -0.161$^{**}$ (0.067)\\   
   \midrule
   \emph{Fit statistics}\\
   Observations         & 675            & 777                   & 777\\  
   R$^2$                & 0.994          & 0.992                 & 0.992\\  
   Within R$^2$         & 0.985          & 0.979                 & 0.979\\  
   \midrule 
 \end{tabular}
\par\endgroup

\begingroup 
\centering
\begin{tabular}{lccc}
   \tabularnewline \midrule \midrule
  \textbf{ log(Topic 15)}                 & -0.021 (0.087) & -0.075$^{**}$ (0.038) & -0.100$^{**}$ (0.050)\\   
   \midrule
   \emph{Fit statistics}\\
   Observations         & 675            & 777                   & 777\\  
   R$^2$                & 0.994          & 0.991                 & 0.991\\  
   Within R$^2$         & 0.985          & 0.977                 & 0.977\\  
   \bottomrule
 \end{tabular}
\begin{threeparttable}
\vspace{-5mm}
\begin{tablenotes}
\begin{singlespace}
\item
\footnotesize{\emph{Signif. Codes: ***: 0.01, **: 0.05, *: 0.1}. Standard errors clustered on year$\times$journal level and small sample corrected in parentheses. Further variables: open access (Y/N), corresponding author, see eq.~(\ref{eq.panel}). Plot of the results in column 2 and 3 to be found in Figure \ref{fig.cit1}.}
\end{singlespace}
\end{tablenotes}
\end{threeparttable}
\par\endgroup
\caption{Regression Tables for the relationship between Topic Prevalence and Citations}
\label{tab.reg1}
\end{table}

\end{small}

\begin{figure}[H]
\centering
\includegraphics[scale=.6]{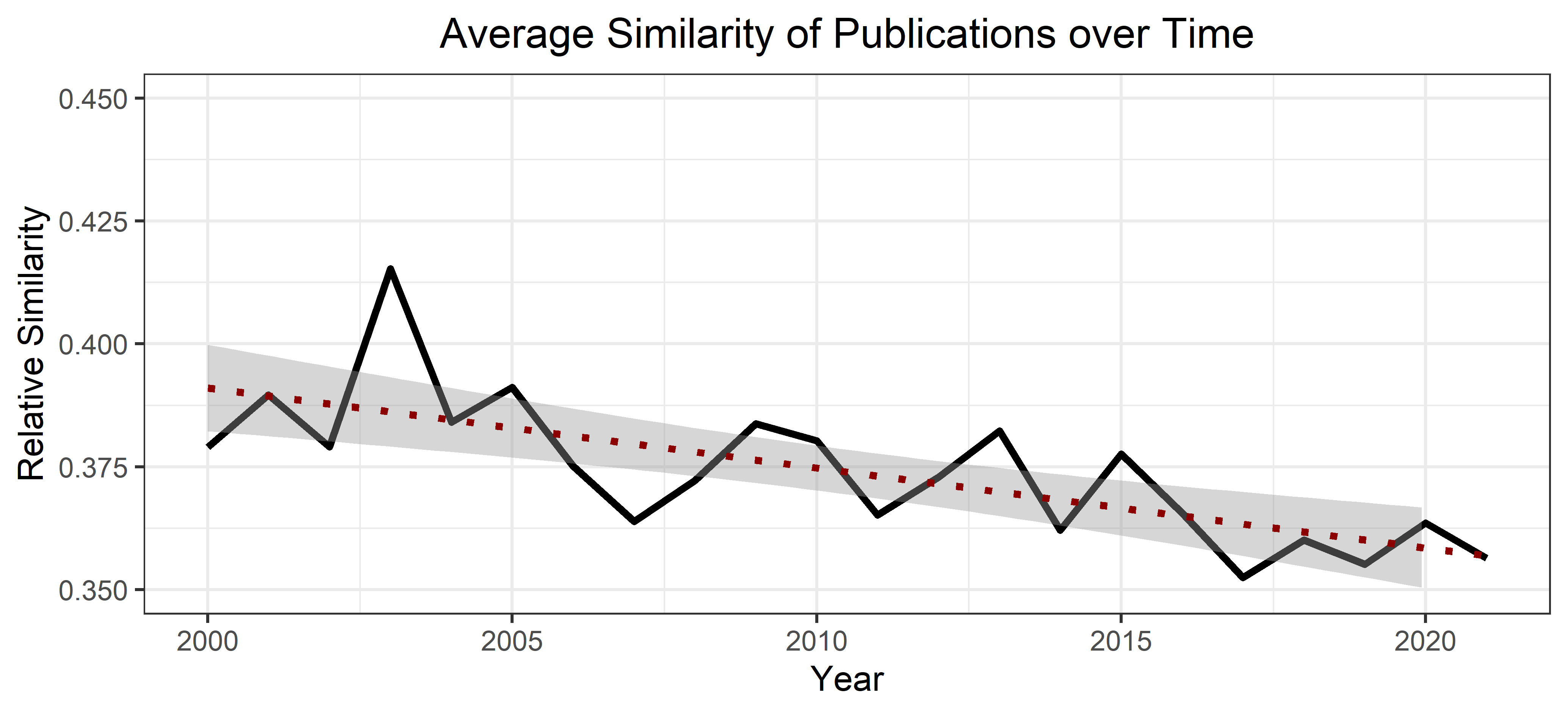}\\
\includegraphics[scale=.6]{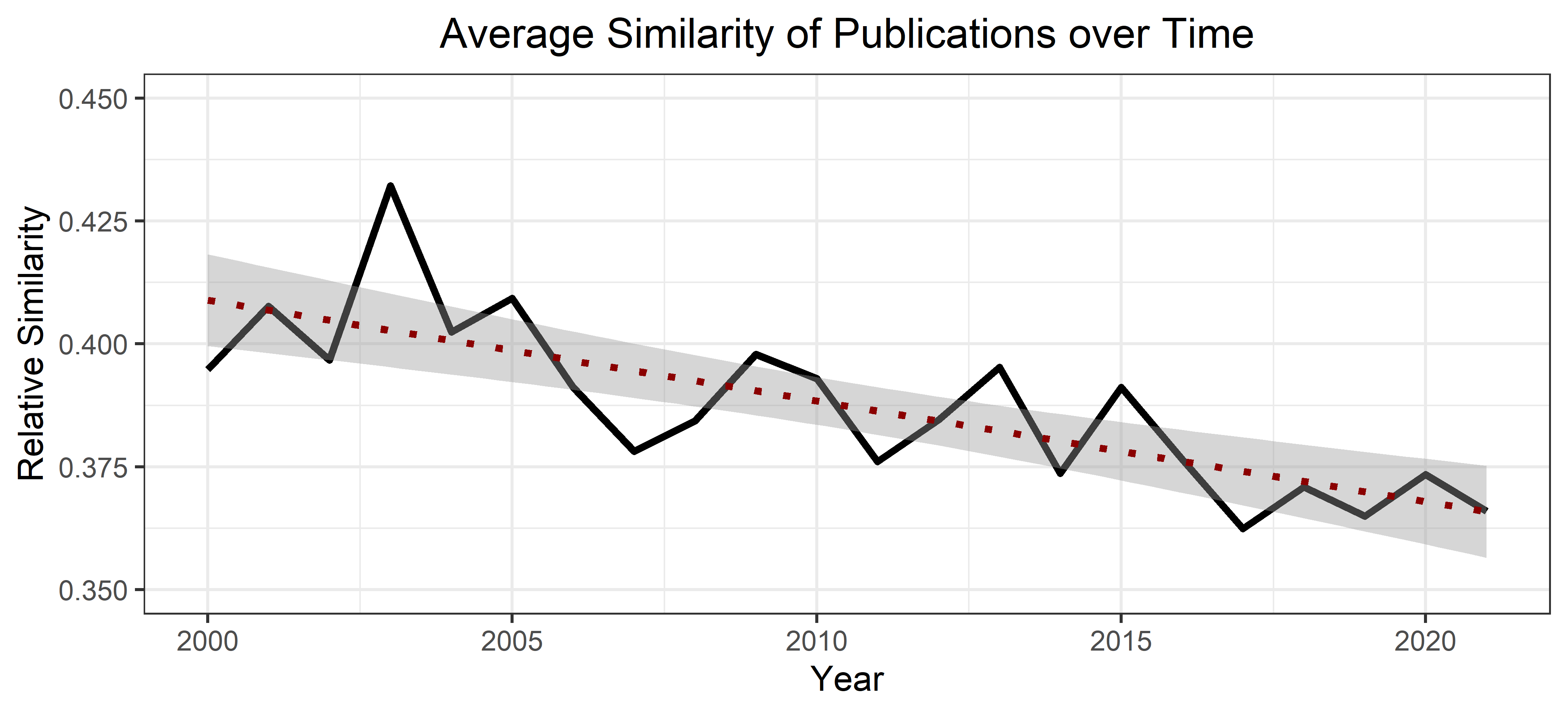}
\fnote{Upper plot:~Paragraph Vector dimension: 100. Lower plot:~Vector dimension: 500}
\caption{Publication Similarity over time using different neural network specifications}
\label{fig.app.neural}
\end{figure}
\vspace{-2mm}

\begin{table}[ht]
\centering
\begin{tabular}{ccrr}
  \toprule
  \toprule
$Non$-$stationarity\:type$ & $lag$ & $statistic$ & $p$-$value$ \\ 
\midrule
\midrule
  \multicolumn{4}{c}{\emph{Augmented Dickey-Fuller Test}}\\
  \midrule
type1 & 0 & -0.2669 & 0.5584 \\ 
  type1 & 1 & -0.4802 & 0.4964 \\ 
  type1 & 2 & -0.4658 & 0.5012 \\ 
  type2 & 0 & -3.4869 & 0.0194 \\ 
  type2 & 1 & -1.7716 & 0.4129 \\ 
  type2 & 2 & -1.3820 & 0.5510 \\ 
  type3 & 0 & -5.4514 & 0.0100 \\ 
  type3 & 1 & -3.4341 & 0.0730 \\ 
  type3 & 2 & -4.1344 & 0.0186 \\ 
    \midrule
   \multicolumn{4}{c}{\emph{Phillips-Perron Test}}\\
   \midrule 
  type 1 & 2 & -0.0389 & 0.6707 \\ 
  type 2 & 2 & -18.3826 & 0.0100 \\ 
  type 3 & 2 & -23.4225 & 0.0100 \\ 
   \midrule 
    \multicolumn{4}{c}{\emph{KPSS Test}}\\
    \midrule
  -- & 2 & 0.5281 & 0.0353 \\ 
   \bottomrule
\end{tabular}
\caption{Unit root tests for non-stationarity of publication similarity over time} 
\label{tab.ur_neural}
\end{table}

\pagebreak

\begin{table}[H]
\centering
\textbf{Johansen Test}\\
\vspace{3mm}
\begin{tabular}{lcccccc}
\toprule
\toprule
Coint.  & param.   & LL  & eigenvalue   & trace stat. & 5\% crit. value \\
\midrule
0               & 12              & -34.5435 &              & 51.6992         & 29.68                      \\
1               & 17              & -19.0674   & 0.7873      & 20.7470         & 15.41  \\
2$^\mathparagraph$  & 20   & -9.0927  & 0.6312  & 0.7977  & 3.76  \\
3               & 21              & -8.6939 & 0.0391      &                 &                   \\
\bottomrule            
\end{tabular}\\
\vspace{10mm}
{
\textbf{Engle-Granger Test}\\
\vspace{2mm}
Regression Results First Step\\
\vspace{2mm}
\def\sym#1{\ifmmode^{#1}\else\(^{#1}\)\fi}
\begin{tabular}{lll}
\toprule
\toprule
            &\multicolumn{2}{c}{External Topics} \\
\midrule
Rules      &      -2.328\sym{***} & (0.568)\\
[1em]
Attr. of the market    &      -0.749   & (0.847)      \\
[1em]
Constant     &      -6.997\sym{***} & (1.771)\\
\midrule
\(N\)       &          22         \\
$F(2, 19)$  &     8.50			\\
\(R^{2}\)   &       0.472         \\
adj. \(R^{2}\)&       0.417         \\
\bottomrule
\multicolumn{2}{l}{\footnotesize Standard errors in parentheses}\\
\multicolumn{2}{l}{\footnotesize \sym{*} \(p<0.05\), \sym{**} \(p<0.01\), \sym{***} \(p<0.001\)}\\
\end{tabular}\\
}
{
\vspace{6mm}
Second Step: Dickey-Fuller Test on Residuals\\
\begin{tabular}{cccccc}
  \toprule
  \toprule
$Non$-$stationarity\:type$ & $lag$ & $statistic$ & \multicolumn{3}{c}{$adjusted\:DF\:statistic$} \\ 
&&& 1\% & 5\% & 10\% \\
  \hline
  type 1 & 1 & -4.501 & -4.29 & -3.74 & -3.45  \\ 
   \bottomrule
\end{tabular}
}
\fnote{Johansen Test: Constant trend; Lag length$=2$, $N=20$, time: 2000-2021. Engle-Granger test: $N=22$, adjusted DF-statistic critical values based on \textcite[][p.~722]{Davidson.1993} since the standard DF-critical values cannot be used as the OLS estimation in the first step distorts the variance.  $^\mathparagraph$ marks the cointegration level at which the critical value exceeds the trace statistic.}
\caption{Tests for Cointegration}
\label{tab.vec.coint}
\end{table}

\begin{table}[H]
\centering
\begin{tabular}{cccc}
\toprule
\toprule
lag                    & $\chi^2$    & df & Prob \textgreater $\chi^2$  \\
\midrule
1                        &  5.5564  & 9  &  0.7834                   \\
2                        &  8.5581   & 9  &  0.4790      \\
\bottomrule             
\end{tabular}
\fnote{ H0: no autocorrelation at lag order $i$.}
\caption{Lagrange-Multiplier Test for autocorrelation of residuals in the VECM}
\label{tab.vec.autocorr}
\end{table}

\begin{table}[H]
\centering
\begin{tabular}{lllll}
\toprule
\toprule
\emph{Jarque-Bera test} &          &       &    &                        \\
Equation         &          & $\chi^2$  & df & Prob \textgreater $\chi^2$ \\
External       &          & 1.062 & 2  & 0.5881                \\
Rules     &          & 0.432 & 2  & 0.8057                \\
Attr. of the market     &          & 0.167 & 2  & 0.9198                \\
All             &          & 1.661 & 6  & 0.9481               \\
\midrule
\emph{Skewness test }   &          &       &    &                        \\
Equation         & Skewness & $\chi^2$  & df & Prob \textgreater $\chi^2$ \\
External      & -0.5623  & 1.054 & 1  & 0.3046                \\
Rules       & -0.0193  & 0.001 & 1  & 0.9719                \\
Attr. of the market    & -0.0441  & 0.006 & 1  & 0.9359                \\
All             &          & 1.062 & 3  & 0.7863                \\
\midrule
\emph{Kurtosis test}    &          &       &    &                        \\
Equation         & Kurtosis & $\chi^2$  & df & Prob \textgreater $\chi^2$ \\
External       & 3.0964   & 0.008 & 1  & 0.9299                \\
Rules      & 2.281    & 0.431 & 1  & 0.5116                \\
Attr. of the market     & 2.5608   & 0.161 & 1  & 0.6885                \\
All              &          & 0.599 & 3  & 0.8966    \\          
\bottomrule
\end{tabular}
\caption{Test for normally distributed residuals of the VECM}
\label{tab.vec.norm}
\end{table}

\begin{figure}[H]
\centering
\includegraphics[width=.8\linewidth]{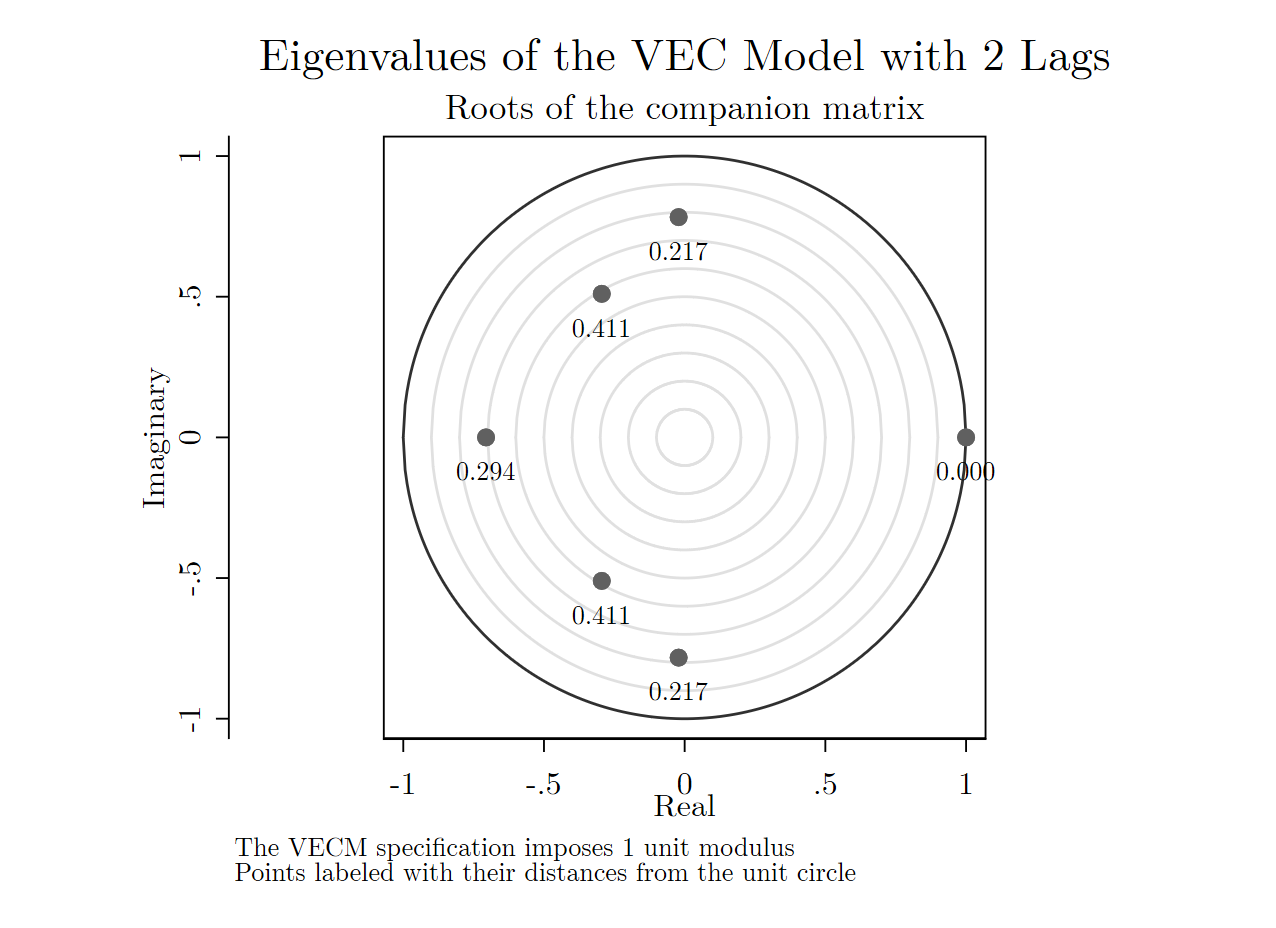}
\vspace{-2mm}
\caption{Unit Circle of the VECM}
\label{fig.unitcircle_vec}
\end{figure}

\begin{table}[H]
\centering
\begin{tabular}{rlccc}
\toprule
\toprule
\multicolumn{2}{c}{Eigenvalue} &      Modulus   &  &  \\
real part & imaginary part & \\
\midrule
+1.0                            &          &   1.0      &  &  \\
-0.0214                       & $+$0.7829$i$ & 0.7832 &  &  \\
-0.0214                          & $-$0.7829$i$ & 0.7832 &  &  \\
-0.7059                             &    &    0.7059     &  &  \\
-0.2945                           & $+$0.5102$i$ & 0.5891 &  &  \\
-0.2945                           & $-$0.5102$i$ & 0.5891 &  &  \\
\bottomrule
\end{tabular}
\fnote{The VECM specification imposes a unit modulus.}
\caption{Eigenvalues for test of the Eigenvalue stability condition.}
\label{tab.vec.ev}
\end{table}

\begin{table}[H]
\begin{tabular}{lllllllllll}
\toprule
\toprule
lag & LL       & LR      & df & p     & FPE      & AIC      & HQIC     & SBIC   &  &  \\
\midrule
0   & -31.148  &         &    &       & 0.0073  & 3.5945  & 3.6198  & 3.7437$^\mathparagraph$ &  &  \\
1   & -19.9459 & 22.404  & 9  & 0.008 & 0.0059  & 3.3627  & 3.4637 & 3.9592  &  &  \\
2   & -7.0418  & 25.808$^\mathparagraph$ & 9  & 0.002 & 0.0043$^\mathparagraph$ & 2.9518$^\mathparagraph$ & 3.1284$^\mathparagraph$ & 3.9956 &  &  \\
3   & -0.3056 & 13.472  & 9  & 0.142 & 0.0069  & 3.1901  & 3.4424  & 4.6813  &  &  \\
\midrule
    &          &         &    &       &          &          &          &          &  &  \\
0   & -35.4609 &         &    &       & 0.0094  & 3.8461  & 3.8753  & 3.9955$^\mathparagraph$ &  &  \\
1   & -25.1956 & 20.531  & 9  & 0.015 & 0.0084  & 3.7196  & 3.8362  & 4.317    &  &  \\
2   & -8.6939 & 33.003$^\mathparagraph$ & 9  & 0.0 & 0.0043$^\mathparagraph$  & 2.9694$^\mathparagraph$ & 3.1735$^\mathparagraph$ & 4.0149 &  &\\
 \bottomrule
\end{tabular}
\fnote{\footnotesize $^\mathparagraph$ marks the best outcome in each column. For two lags: $N=20$, for three lags: $N=19$.}
\caption{Lag selection statistics for the VECM}
\label{tab.vec.lags}
\end{table}

\begin{figure}[H]
\includegraphics[width=.49\linewidth]{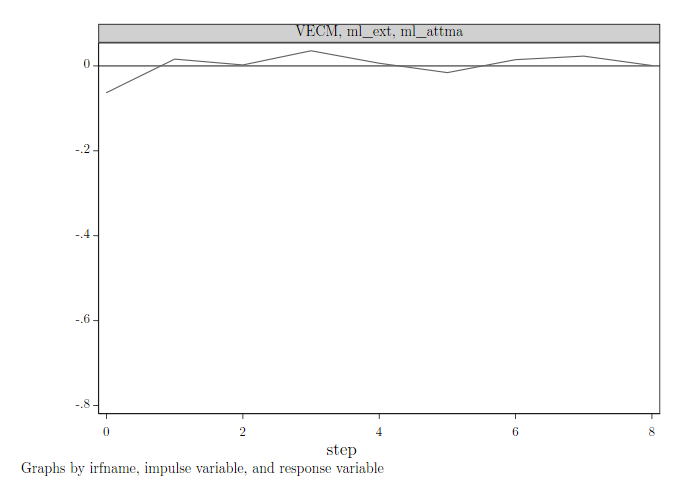}
\includegraphics[width=.49\linewidth]{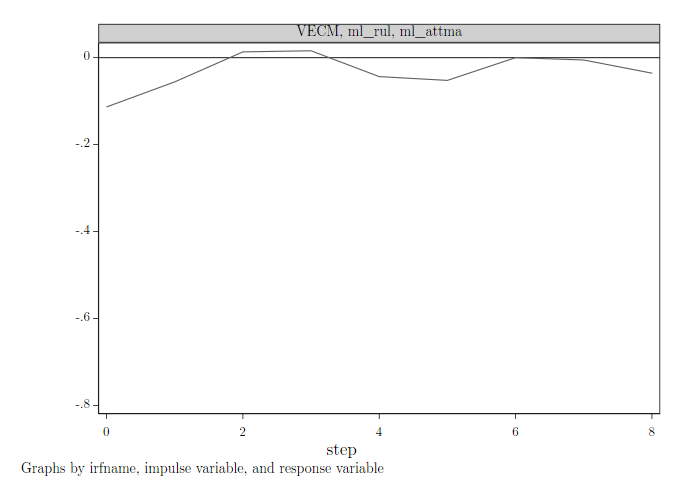}\\
\includegraphics[width=.49\linewidth]{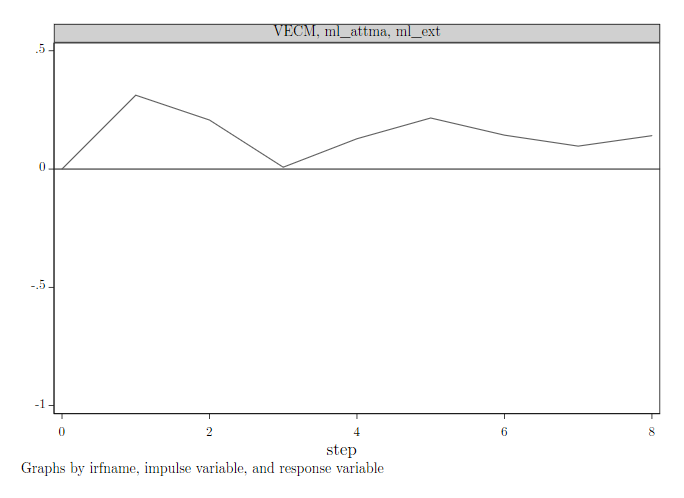}
\includegraphics[width=.49\linewidth]{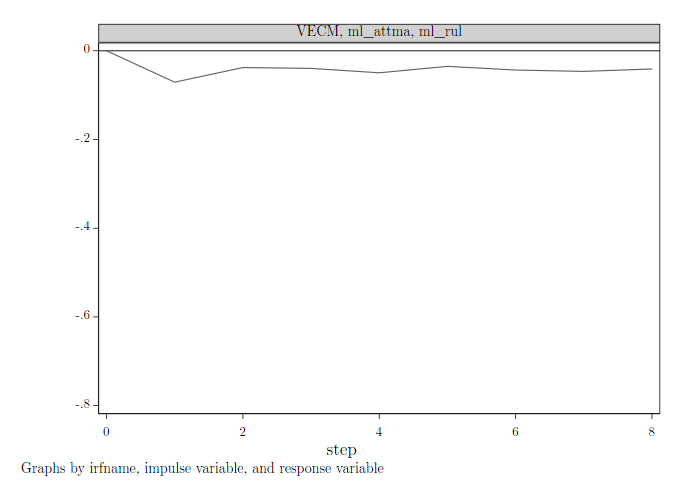}
\caption{Further Impulse Response Functions VEC model}
\label{fig.vecm.irf_add}
\end{figure}


\begin{table}[H]
\centering
\textbf{Johansen Test}\\
\vspace{3mm}
\begin{tabular}{lcccccc}
\toprule
\toprule
Coint.  & param.   & LL  & eigenvalue   & trace stat. & 5\% crit. value \\
\midrule
0               & 10 & -19.8128 &              & 16.4807  &  15.41     \\
1$^\mathparagraph$               & 13  & -12.1547   &  0.5534      & 1.1645 & 3.76  \\
2               & 14  & -11.5724 & 0.0595      &                 &                   \\
\bottomrule            
\end{tabular}
\fnote{\footnotesize Johansen Test: Constant trend; Lag length$=3$, $N=19$, time: 2000-2021. $^\mathparagraph$ marks the cointegration level at which the critical value exceeds the trace statistic.}
\caption{Tests for Cointegration for the two variable specification with three lags}
\label{tab.var.coint}
\end{table}

\begin{table}[H]
\begin{tabular}{lllllllll}
\toprule
\toprule
lag                      & LL          & LR        & df      & p       & FPE      & AIC      & HQIC     & SBIC     \\
\midrule

0                        & -34.6217  & & &  & 0.1335   & 3.6622 & 3.6816  & 3.7617    \\
1                        & -30.348     & 8.5474    & 4       & 0.073   & 0.1305  & 3.6348   & 3.6931  & 3.9335  \\
2                        & -16.1596    & 28.377$^\mathparagraph$   & 4       & 0.000   & 0.0479$^\mathparagraph$ & 2.616$^\mathparagraph$ & 2.7132$^\mathparagraph$ & 3.1138$^\mathparagraph$ \\
\midrule
 &  &     &         &         &          &          &          &          \\
0                        & -29.8873    &&& & 0.0984   & 3.3566 & 3.3734 & 3.4560\\
1                        & -24.9648    & 9.8451    & 4       & 0.043   & 0.0897  & 3.2595  & 3.3099  & 3.5577  \\
2                        & -13.873     & 22.183$^\mathparagraph$   & 4       & 0.000   & 0.0434$^\mathparagraph$ & 2.513$^\mathparagraph$ & 2.5971$^\mathparagraph$ & 3.01$^\mathparagraph$ \\
3                        & -11.5724    & 4.6011    & 4       & 0.331   & 0.0544 & 2.6918  & 2.8096  & 3.3877 \\
\bottomrule
\end{tabular}
\fnote{\footnotesize $^\mathparagraph$ marks the best outcome in each column. For two lags: $N=20$, for three lags: $N=19$.}
\caption{Selection-order criteria VAR}
\label{tab.var.lags}
\end{table}

\begin{table}[H]
\centering
\begin{tabular}{lllll}
\toprule
\toprule
\emph{Jarque-Bera test} &          &       &    &                        \\
Equation         &          & $\chi^2$  & df & Prob \textgreater $\chi^2$ \\
External       &          & 1.058 &  2  &  0.5892               \\
Rules     &          &  1.276 &  2  &  0.5283                 \\ 
All             &          & 2.334 &  4  &  0.6745                \\
\midrule
\emph{Skewness test }   &          &       &    &                        \\
Equation         & Skewness & $\chi^2$  & df & Prob \textgreater $\chi^2$ \\
External      & -0.4217 &   0.563 &  1  &  0.453              \\
Rules       & -0.468 &   0.694 &  1   & 0.405                \\
All             &          &  1.257  & 2  &  0.5335               \\
\midrule
\emph{Kurtosis test}    &          &       &    &                        \\
Equation         & Kurtosis & $\chi^2$  & df & Prob \textgreater $\chi^2$ \\
External       & 2.2092 &   0.495  & 1   & 0.4817                \\
Rules      & 2.1421 &   0.583  & 1   & 0.4453                \\
All              &          &  1.078  & 2  &  0.5834     \\          
\bottomrule
\end{tabular}
\caption{Test for normally distributed residuals of the VAR model}
\label{tab.var.norm}
\end{table}

\begin{table}[H]
\centering
\begin{tabular}{cccc}
\toprule
\toprule
lag                    & $\chi^2$    & df & Prob \textgreater $\chi^2$  \\
\midrule
1                        &   9.2354  & 4  &  0.0555                   \\
2                        &   6.6933   & 4  &  0.153      \\
\bottomrule             
\end{tabular}
\fnote{\footnotesize H0: no autocorrelation at lag order $i$.}
\caption{Lagrange-Multiplier Test for autocorrelation of residuals in the VAR model}
\label{tab.var.autocorr}
\end{table}

\begin{table}[H]
\centering
\begin{tabular}{rcccc}
\toprule
\toprule
\multicolumn{2}{c}{Eigenvalue}                            &      Modulus   &  &  \\
real part & imaginary part & \\
\midrule
0.8888 &	&	0.8888  \\
0.0813 &	$+$  0.8141$ i $ &	0.8181	  \\
0.0813 &	$-$  0.8141$ i $ &	0.8181	  \\
$-$0.6004 &	$+$  0.2066$ i $ &  0.6350	  \\
$-$0.6004 &	$-$  0.2066$ i $ &	0.6350	  \\
0.1924 &	&	0.1924 \\
\bottomrule
\end{tabular}
\fnote{\footnotesize All eigenvalues have a modulus \textless 1. The VAR fulfills the stability condition.}
\caption{Eigenvalues for test of the Eigenvalue stability condition.}
\label{tab.var.ev}
\end{table}

\begin{figure}[H]
\centering
\includegraphics[width=.8\linewidth]{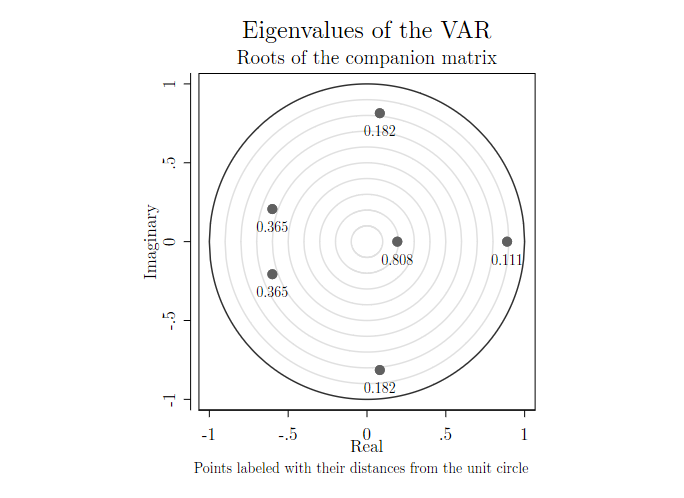}
\vspace{-2mm}
\caption{Unit Circle of the VAR model}
\label{fig.unitcircle_var}
\end{figure}

\begin{table}[H]
\centering
\begin{tabular}{lllll}
\toprule
\toprule
Step & FEVD    & Std. Err.   & \multicolumn{2}{c}{95\% CI}  \\
\midrule
0    & 0       & 0       & 0       & 0       \\
1    & 0       & 0       & 0       & 0       \\
2    & 0.3119 & 0.1575 & 0.6206 & 0.0031 \\
3    & 0.4956 & 0.1615 & 0.8122 & 0.179 \\
4    & 0.4914  & 0.1631 & 0.8112 & 0.1717 \\
5    & 0.45 & 0.1595 & 0.7726 & 0.1474  \\
6    & 0.4584 & 0.1606 & 0.7731 &  0.1436 \\
7    & 0.4915 & 0.1665  & 0.8178 & 0.1653 \\
8    & 0.4926 & 0.1694 & 0.8247 & 0.1605\\
\bottomrule
\end{tabular}
\fnote{Impulse variable: rules, response variable: External topics. FEVD Computation for Fig. \ref{fig.var.fevd}.}
\caption{Forecast Error Variance Decomposition -- Table}
\label{tab.var.fevd}
\end{table}

\subsection*{Appendix B: Selection of latent topics}
\label{app.B}

\begin{figure}[H]
\centering
\includegraphics[width=.8\linewidth]{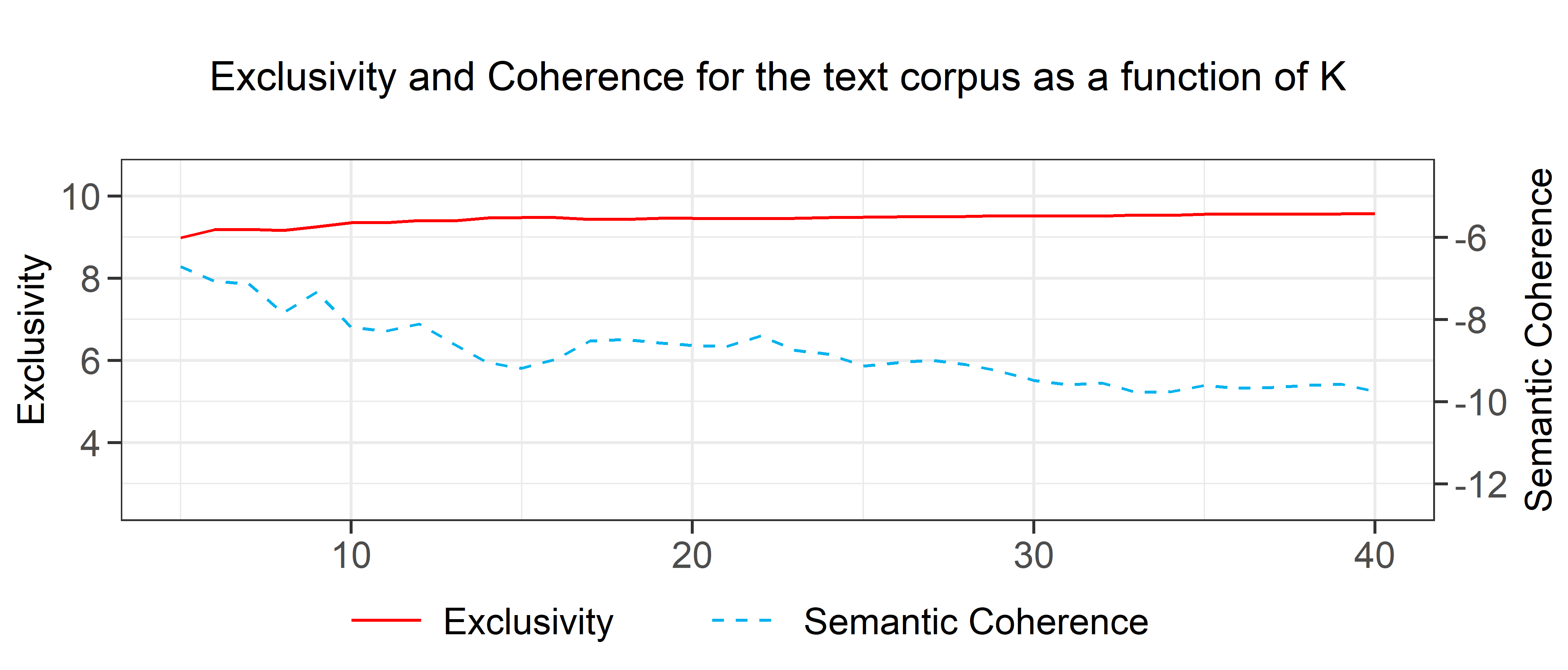}
\begin{threeparttable}
\vspace{-5mm}
\begin{tablenotes}
\item \footnotesize{$K \in [5,40]$, Semantic Coherence adjusted: $\hat{SC}=\frac{SC}{10}+16$ to be comparable with exclusivity.}
\end{tablenotes}
\end{threeparttable}
\caption{Measures to identify the optimal $K$ for the text corpus.}
\label{fig.k1}
\end{figure}

Figure \ref{fig.k1} presents the two selection measures exclusivity ($EX$) and semantic coherence ($SC$) for the text corpus. One can easily see in all four plots a negative slope for semantic coherence and a positive slope for exclusivity as suggested in theory. The choice of the optimal $K^*$ is subject to individual reasoning as mentioned beforehand. Still, I propose a formalization of the decision based on $EX$ and $SC$ as follows:
\vspace{-3mm}
\begin{align}
SER_k &= \frac{SC_k}{EX_k} \label{SEratio} \\
\Delta SER_k &= SER_k - SER_{k-1} \label{DSE} \\
wSER_k &= \frac{\Delta SER_k}{SER_k} \label{weightSE}
\end{align}

\begin{figure}[H]
\centering
\includegraphics[width=.8\linewidth]{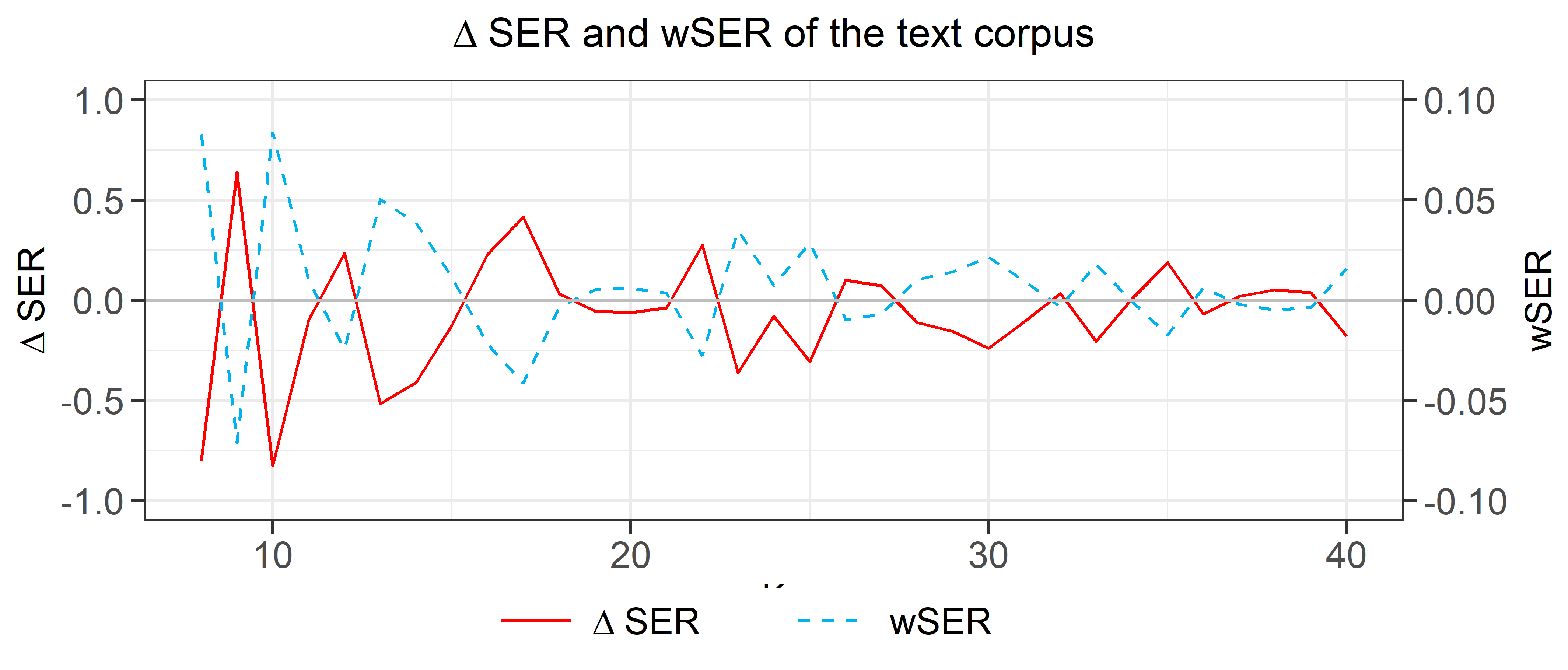}
\begin{threeparttable}
\vspace{-5mm}
\begin{tablenotes}
\item \footnotesize{$\Delta SER_k$ (red solid line) and $wSER_k$ (blue dashed line) for the Text Corpus. $K \in [5,40]$}
\end{tablenotes}
\end{threeparttable}
\caption{$\Delta SER_k$ and $wSER_k$ for the Text Corpora}
\label{fig.ser}
\end{figure}

Equation (\ref{SEratio}) describes the ratio of semantic coherence and exclusivity for each $K=k$. In a second step, I use the change in the $SER$ to quantify the effect of a marginal (discrete) change in $k$ to $k+1$ as shown in eq. (\ref{DSE}). This approach captures changes in both measures at the same time. As coherence and exclusivity are not stationary, eq. (\ref{weightSE}) weights the change in $SER$ by the absolute value of the $SER$. Due to the nature of the data $EX >0$ and $SC<0$ holds, such that $SER_k<0$ always holds. However, the change in $\Delta SER_k$ depends on how both measures jointly change. A sole decrease in exclusivity would imply $\Delta SER_k>0$ as well as an increase in semantic coherence. As $SER_k <0$ always holds, $wSER_k<0$ corresponds to an increase in one of the measures. Figure \ref{fig.ser} plots the measures in eqs. (\ref{DSE}) and (\ref{weightSE}). 

Eventually, I limit the number of possible topics to $K \in [5,40]$ as due to the comparatively narrow scope of the paper selection I consider a sufficiently larger amount of separate topics captured in paper abstracts as unreasonable. On the other hand, it is unreasonable that only three or four underlying topics in collusion research exist. The absolute minimum of topics would be two as one topic would essentially cover everything. 

Stepping back, one should reconsider that the choice of $K$ is up to the researcher and existing literature applying topic modeling algorithms to the best of my knowledge applies less sophisticated decision rules. Hence, the proposed decision rule attempts to contribute a new metric to find $K^*$ but is still not independent from my personal understanding.

Looking at Figure \ref{fig.ser}, one can see that $wSER_k<0$ often occurs, which corresponds to stronger fluctuations in semantic coherence across $K$ values (see Figure \ref{fig.k1}). Hence, semantic coherence increases occasionally with an increasing number of $K$. As one can see in Figure \ref{fig.ser}, especially for $K<20$ exist several values at which semantic coherence improves. At the same, one can see in Figure \ref{fig.k1} that the absolute value of semantic coherence increases at $K=21$ that is higher than for $K\in[15,20]$ and for all $K>21$, semantic coherence is lower. Furthermore, $K=21$ appears as a reasonable amount of topics for the temporal range of 22 years and the chosen focus on collusive practices.

\begin{table}[H]
\centering
\begin{singlespace}
\begin{footnotesize}
\begin{tabular}{l|l}
\toprule
\toprule
\textbf{Topic 1: } & \\
 	 Highest Prob:& competit, articl, author, enforc, law, agreement, effect   \\
 	 FREX:& articl, programm, remedi, applic, author, jurisdict, review \\
\textbf{Topic 2: } & \\
 	 Highest Prob:& damag, court, case, decis, law, competit, european  \\
 	 FREX:& settlement, damag, court, rpm, infring, suprem, appeal \\
\textbf{Topic 3: } & \\
 	 Highest Prob:& market, competit, industri, firm, effect, model, find \\
 	 FREX:& market, industri, concentr, entri, experi, predict, size \\
\textbf{Topic 4: } &\\
 	 Highest  & auction, bid, bidder, use, procur, revenu, two  \\
 	 FREX:& auction, bid, bidder, first-pric, ring, procur, revenu \\
\textbf{Topic 5: } &\\
 	 Highest Prob:& equilibrium, inform, communic, game, privat, signal, player \\
 	 FREX:&  signal, communic, equilibrium, player, payoff, game, perfect \\
\textbf{Topic 6: } & \\
 	 Highest Prob:&  competit, antitrust, trade, econom, polici, athlet, ncaa  \\
 	 FREX:& athlet, ncaa, restraint, sport, amateur, wto, trade  \\
\textbf{Topic 7: } &\\
 	 Highest Prob:& agent, contract, incent, princip, organ, optim, cost  \\
 	 FREX:&  princip, agent, deleg, corrupt, team, effort, contract  \\
\textbf{Topic 8: } &\\
 	 Highest Prob:&  lenienc, effect, polici, firm, fine, enforc, antitrust  \\
 	 FREX:&  lenienc, program, deterr, fine, penalti, crimin, deter   \\
\textbf{Topic 9: } &\\
 	 Highest Prob:& price, cost, consum, retail, demand, firm, use   \\
 	 FREX:& retail, price, gasolin, consum, guarante, demand, deviat   \\
\textbf{Topic 10: } &    \\
 	 Highest Prob:& firm, profit, merger, product, effect, vertic, sustain \\
 	 FREX:&  invest, merger, integr, capac, firm, downstream, rival   \\ 	 	 	
 	 \bottomrule
 \end{tabular}
 \begin{threeparttable}
\begin{tablenotes}
\item \footnotesize{Alternative value of $K = 10$. Text corpus contains 777 documents, 2590 terms and 37746 tokens. EM-Algorithm iterations: $\le 75$. Spectral initialization.}
\end{tablenotes}
\end{threeparttable}
\caption{Latent Topics in joint 21st Century Antitrust \& Collusion Text Corpus: Alternative choice of $K$}
\label{tab.words_alt_K10}
\end{footnotesize}
\end{singlespace}
\end{table}

\end{singlespace}

\end{onehalfspace}
\end{document}